%% file: paper.tex
\documentclass[acmtog]{acmart}

\usepackage{booktabs} 
\usepackage{subcaption}

\usepackage[ruled]{algorithm2e} 

\SetAlFnt{\small}
\SetAlCapFnt{\small}
\SetAlCapNameFnt{\small}
\SetAlCapHSkip{0pt}

\newcommand{\mydraft}{false}
\newcommand{\mb}{\mathbf}
\newcommand{\xx}{\mb{x}}

\newcommand{\oo}{\mb{o}}

\newcommand{\mm}{\mb{m}}

\newcommand{\qq}{\mb{q}}

\ifdef{\vv}{\renewcommand{\vv}{\mb{v}}}{\newcommand{\vv}{\mb{v}}}

\newcommand{\pp}{\mb{p}}
\renewcommand{\ss}{\mb{s}}

\newcommand{\ff}{\mb{f}}

\newcommand{\ii}{\mb{i}}
\newcommand{\jj}{\mb{j}}

\newcommand{\fm}{\boldsymbol \phi}

\newcommand{\Surf}{\mathcal{S}}
\newcommand{\TSurf}{\tilde{\mathcal{S}}}
\newcommand{\Vs}{\mathcal{V}^S}
\newcommand{\Gdx}{\mathcal{G}_{\Delta x}}

\newcommand{\fms}{\fm^S_{\tilde{S}}}

\acmJournal{TOG}

\begin{document}
\title{A Robust Grid-Based Meshing Algorithm for Embedding Self-Intersecting Surfaces}
\begin{teaserfigure}
    \centering
    \includegraphics[draft=\mydraft,width=\textwidth]{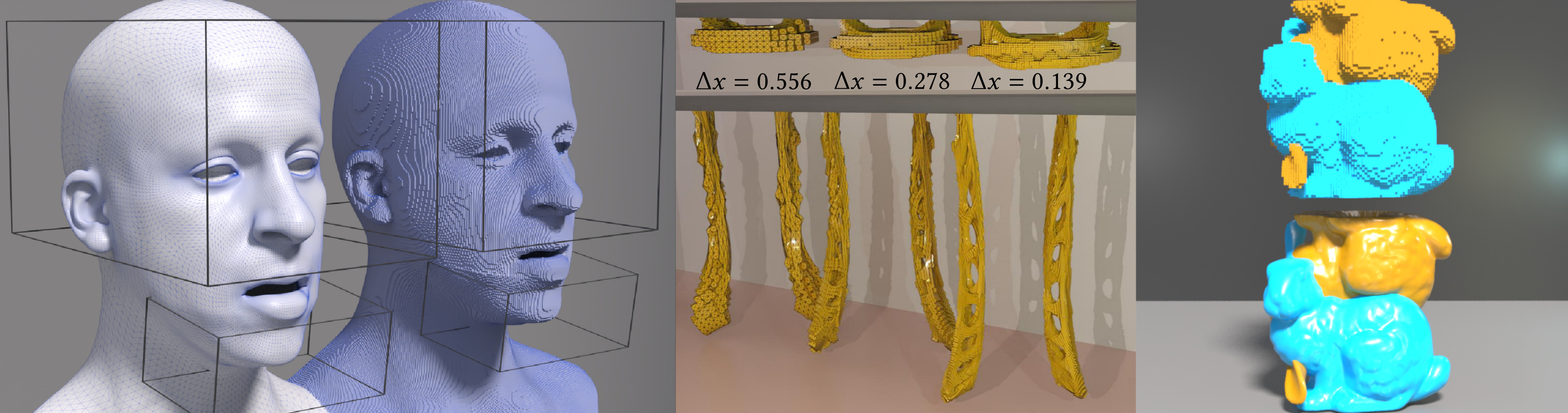}
    \caption{\textit{(Left)} Our method can generate a consistent volumetric mesh for a facial geometry that contains self-intersections e.g.\ around the lips.  \textit{(Middle)} Two interlocking M\"obius-strip-like bands separate freely at various spatial resolutions of the background grid, despite many near self-intersections in the surface geometry.  \textit{(Right)} Two bunny geometries can naturally separate despite significant initial overlaps.}
\end{teaserfigure}

\author{Steven W. Gagniere}
\affiliation{%
    \institution{UCLA}
    \streetaddress{Box 951555}
    \city{Los Angeles}
    \state{CA}
    \postcode{90095-1555}
    \country{USA}}
\email{sgagniere@math.ucla.edu}
\author{Yushan Han}
\affiliation{%
    \institution{UCLA}
    \streetaddress{Box 951555}
    \city{Los Angeles}
    \state{CA}
    \postcode{90095-1555}
    \country{USA}}
\email{yushanh1@math.ucla.edu}
\author{Yizhou Chen}
\affiliation{%
    \institution{UCLA}
    \streetaddress{Box 951555}
    \city{Los Angeles}
    \state{CA}
    \postcode{90095-1555}
    \country{USA}}
\email{chenyizhou@math.ucla.edu}
\author{David A. B. Hyde}
\affiliation{%
    \institution{Vanderbilt University}
    \streetaddress{PMB 351679, 2301 Vanderbilt Place}
    \city{Nashville}
    \state{TN}
    \postcode{37235-1679}
    \country{USA}}
\email{david.hyde.1@vanderbilt.edu}
\author{Alan Marquez-Razon}
\affiliation{%
    \institution{UCLA}
    \streetaddress{Box 951555}
    \city{Los Angeles}
    \state{CA}
    \postcode{90095-1555}
    \country{USA}}
\email{marqueza04@g.ucla.edu}
\author{Joseph Teran}
\affiliation{%
    \institution{UCLA}
    \streetaddress{Box 951555}
    \city{Los Angeles}
    \state{CA}
    \postcode{90095-1555}
    \country{USA}}
\email{jteran@math.ucla.edu}
\author{Ronald Fedkiw}
\affiliation{%
    \institution{Stanford Unviersity}
    \streetaddress{Gates Computer Science Building, 353 Jane Stanford Way}
    \city{Stanford}
    \state{CA}
    \postcode{94305-9025}
    \country{USA}}
\email{fedkiw@cs.stanford.edu}

\begin{abstract}
The creation of a volumetric mesh representing the interior of an input polygonal mesh is a common requirement in graphics and computational mechanics applications.
Most mesh creation techniques assume that the input surface is not self-intersecting.
However, due to numerical and/or user error, input surfaces are commonly self-intersecting to some degree.
The removal of self-intersection is a burdensome task that complicates workflow and generally slows down the process of creating simulation-ready digital assets.
We present a method for the creation of a volumetric embedding hexahedron mesh from a self-intersecting input triangle mesh.
Our method is designed for efficiency by minimizing use of computationally expensive exact/adaptive precision arithmetic.
Although our approach allows for nearly no limit on the degree of self-intersection in the input surface, our focus is on efficiency in the most common case: many minimal self-intersections.
The embedding hexahedron mesh is created from a uniform background grid and consists of hexahedron elements that are geometrical copies of grid cells.
Multiple copies of a single grid cell are used to resolve regions of self-intersection/overlap.
Lastly, we develop a novel topology-aware embedding mesh coarsening technique to allow for user-specified mesh resolution as well as a topology-aware tetrahedralization of the hexahedron mesh.
\end{abstract}

%
%

\ccsdesc[500]{Computing methodologies~Computer graphics}
\ccsdesc[500]{Mathematics of computing~Mesh generation}

%
%


\maketitle

\input{paper_body}

\end{document}

%% file: paper_body.tex
\begin{figure*}[t]
    \includegraphics[draft=\mydraft,width=\textwidth]{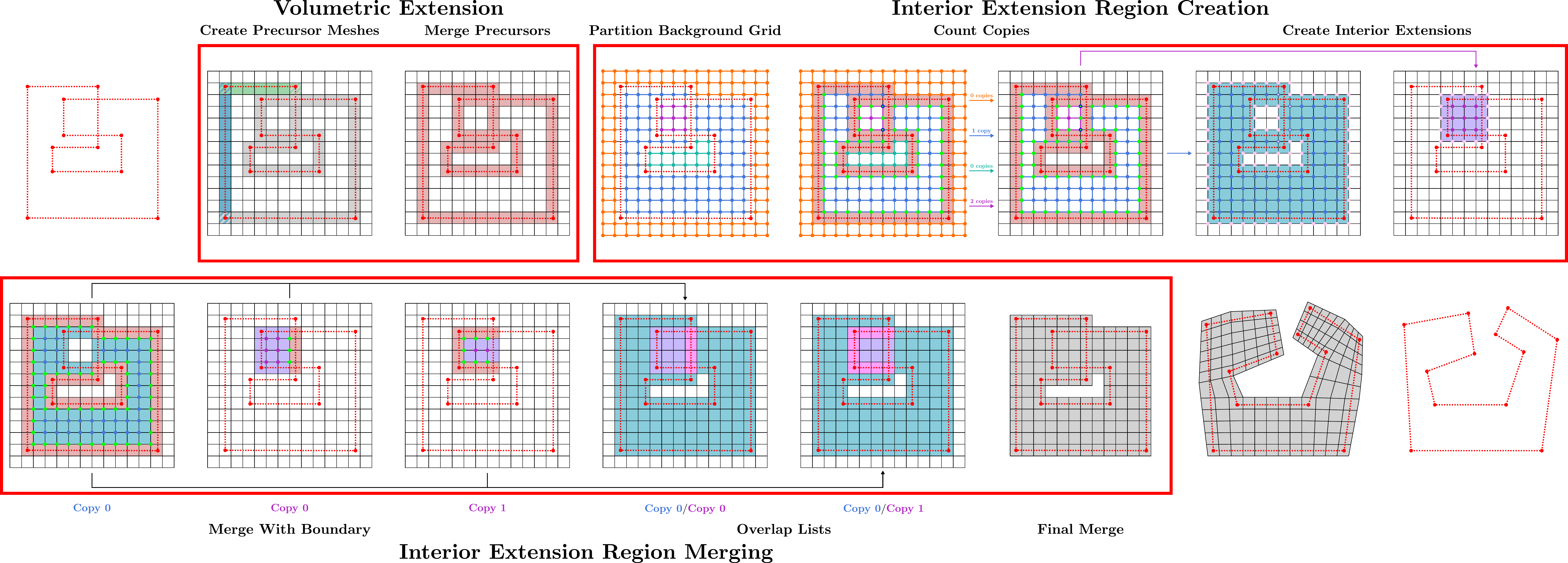}
    \caption{
        {\textbf{Algorithm overview.}}
        Given an initial input surface mesh $\mathcal{S}$, there are three major steps in the computation of the final volumetric extension mesh $\mathcal{V}$: Volumetric Extension, Interior Extension Region Creation, and Interior Extension Region Merging.
        \textit{(Volumetric Extension)} In this step, we create a precursor mesh for each element in $\mathcal{S}$, and compute preliminary signing information for the vertices. We then merge the precursor meshes to create the volumetric extension $\mathcal{V}^S$ and correct the signing information where necessary.
        \textit{(Interior Extension Region Creation)} In preparation for growing the volumetric extension into the interior, we first partition the nodes of the background grid using the edges cut by $\mathcal{S}$. We decide which regions are interior and count the copies of each region using the vertices of $\mathcal{V}^S$ which have negative sign. For each interior region $j^I$ with at least one copy, we then create a hexahedron mesh $\mathcal{V}^{j^I,c}$ for each copy $c$.
        \textit{(Interior Extension Region Merging)} The merging process begins with copying relevant hexahedra from $\mathcal{V}^S$ into $\mathcal{V}^{j^I,c}$. First, certain vertices of $\mathcal{V}^{j^I,c}$ are replaced by corresponding vertices from $\mathcal{V}^S$. Hexahedra to be replaced are then removed from $\mathcal{V}^{j^I,c}$ before the boundary hexahedra are copied in. We then merge the various meshes $\mathcal{V}^{j^I,c}$ by first determining where different meshes overlap, and then using these hexahedra overlap lists to perform the final merge.
    }
    \label{fig:overview}
\end{figure*}

\section{Introduction}\label{section:introduction}
In many computer graphics and computational mechanics applications, it is necessary to create a volumetric mesh associated with the interior of an input polygonal surface mesh. Most commonly a volumetric tetrahedron mesh is created whose boundary coincides topologically and/or geometrically with an input triangle mesh \cite{molino:2003:mesh,labelle:2007:mesh,hu:2018:tetwild,si:2015:tetgen}.
A volumetric embedding mesh that contains the input surface but whose boundary is different than an input triangle mesh is also commonly used \cite{sifakis:2007:arbitrary,tao:2019:mandoline,koschier:2017:xfem,teran:2005:muscle}.
It is generally required that the surface mesh be closed and orientable.
It is also generally required that the surface mesh is free of self-intersection or overlap. 
While the closed and orientable requirements are relatively easy to satisfy in practice, the self-intersection constraint is more challenging, particularly near regions of high-curvature.  
In many computer graphics applications, this constraint can be violated without any artifacts since the overlap regions are not visible, however most volumetric mesh creation techniques either break down or give numerically ``glued'' meshes if the constraint is violated. 
Even intersection free, but nearly intersecting meshes can cause problems for many volumetric mesh creation techniques.

While many surface geometry creation techniques address the importance of its prevention \cite{harmon:2011:interference,vonfunck:2006:vector,attene:2010:lightweight,angelidis:2006:swirling,gain:2001:preventing}, as noted in e.g. \cite{sacht:2013:consistent,li:2018:immersion}, self-intersecting surface meshes are common in practice. 
Often those involved in the surface geometry creation process are not involved in volumetric simulation or similar down-stream portions of the production pipeline and introduction of self-intersecting regions arises from a lack of communication. 
Furthermore, completely removing all regions of self-intersection is often deemed not worthy of the effort since it can significantly increase modeling time.
In some cases it is even desirable to have an overlapping input surface. E.g. it is desirable to have overlapping lips in the neutral pose of a deformable volumetric face mesh since lips resting in non-overlapping contact are not in a stress free state \cite{cong:2015:fully,cong:2016:face}.
It should be noted that although in practice a non-negligible number of slightly overlapping or nearly overlapping regions are common, generally the intersection-free constraint is not violated to an extreme degree with overlap regions typically having minimal volume.

Various approaches have developed volumetric mesh creation techniques specifically designed to be robust to self-intersecting \cite{sacht:2013:consistent,li:2018:immersion} or nearly self-intersecting \cite{teran:2005:muscle,li:2018:immersion} input surfaces.
Sacht et al. \shortcite{sacht:2013:consistent} use conformalized mean-curvature flow (cMCF) to first evolve the surface to a self-intersection-free state from which the flow is reversed, attracting the surface to its original, self-intersecting state but with a collision prevention safeguard.
This defines an intersection free counterpart to the original input surface which can be meshed with standard techniques.
Li and Barbi\v{c} \shortcite{li:2018:immersion} create embedding tetrahedron meshes from unmodified surface meshes with self-intersection by computing locally-injective immersions that can be used to unambiguously duplicate embedded mesh regions near overlaps. 
They sew these duplicated regions together using a technique inspired by the Constructive Solid Geometry (CSG) approaches in \cite{teran:2005:muscle,sifakis:2007:arbitrary} but with reduced use of expensive exact precision arithmetic. 
Teran et al. \shortcite{teran:2005:muscle} use an element duplication/sewing technique to create embedding tetrahedron meshes for nearly intersecting input surfaces meshes.

\begin{figure}[t]
    \includegraphics[draft=\mydraft,width=\columnwidth]{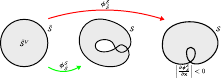}
    \caption{
        {\textbf{Intersection-free mapping.}} Two mappings from a non-self-intersecting region $\tilde{\mathcal{S}}^V$ to self-intersecting boundary $\mathcal{S}$ are shown. The second mapping (right) requires the existence of a negative Jacobian determinant.
    }
    \label{fig:cases}
\end{figure}

We design an approach for the construction of a uniform-grid-based embedding hexahedron mesh counterpart $\mathcal{V}$ to an input triangulated surface mesh $\Surf$ that is well-defined (i.e. free from numerical mesh ``glueing'' artifacts) when the surface is self-intersecting.
As in \cite{sacht:2013:consistent}, we assume there exists a nearby non-self-intersecting mesh $\TSurf$ and a mapping $\fms:\TSurf^V \rightarrow \mathbb{R}^3$ with non-singular Jacobian determinant (see Figure \ref{fig:cases}).
Here $\TSurf^V$ is the unambiguously defined interior of the non-self-intersecting $\TSurf$. 
Intuitively, if we can find a mapping $\fms$ then we can define a volumetric embedding mesh for $\TSurf$ unambiguously with standard techniques and then push it forward under the mapping.
However, unlike Sacht et al. \shortcite{sacht:2013:consistent}, we do not explicitly create $\fms$ or $\TSurf$ but rather use their existence to guide our mesh creation strategy.

We build our embedding hexahedron mesh $\mathcal{V}$ from the intersection of the input surface $\Surf$ with a uniform background grid where cells in contiguous regions are copied to form sub-meshes that are sewn together using techniques inspired by Teran et al. \shortcite{teran:2005:muscle} and Sifakis et al. \shortcite{sifakis:2007:arbitrary} but in a manner designed to mimic the image of $\fms$.
Our approach is ultimately similar to that of Li and Barbi\v{c} \shortcite{li:2018:immersion} in that we create the volumetric embedding mesh without modifying the self-intersecting surface and our region duplication/sewing is equivalent to discovering immersions.
Unlike \cite{li:2018:immersion}, our approach uses nearly no exact and/or adaptive precision arithmetic as
we do not resolve the geometry of intersection from triangles in $\Surf$ with themselves or with cells in the background grid and we do not use CSG operations as in \cite{sifakis:2007:arbitrary}.
We simply require accurate determination of which triangles intersect which grid cells. 
This limits the accuracy of our method for large grid spacing (low-resolution) and we run with smaller grid spacing (high-resolution) when necessary.
To prevent this from causing excessive element counts, we provide a topology-preserving mesh coarsening strategy similar to that of Wang et al. \shortcite{wang:2015:cutting}.
Lastly, we provide a technique for efficiently converting the uniform-grid-based embedding hexahedron mesh to a tetrahedron mesh that robustly handles duplicated regions of the hexahedron mesh near self-intersecting features.
As in \cite{li:2018:immersion}, we use a body-centered cubic (BCC) structure \cite{molino:2003:mesh} for this conversion.

We summarize our novel contributions as:
\begin{itemize}
    \item An efficient technique with reduced use of exact/adaptive precision arithmetic for building an embedding hexahedron mesh for an input self-intersecting triangle mesh from a uniform grid that is equivalent to pushing forward one unambiguously defined from a self-intersection-free state. 
    \item A topology aware embedding mesh coarsening strategy to provide for flexible resolution/element count.
    \item A topology aware BCC approach for converting the embedding hexahedron mesh into an embedding tetrahedron mesh.
\end{itemize}

\section{Related Work}\label{section:related}
We discuss methods in the existing literature that are related to our approach.
We first provide detailed discussion of \cite{li:2018:immersion} and \cite{sacht:2013:consistent} since these works are most relevant to ours.
In addition to techniques that compute a volumetric mesh from an input triangle mesh, we discuss relevant works in the fracture and virtual surgery literature since our approach makes use of grid cutting operations to intersect the input surface mesh with a uniform background grid. 
Lastly, we discuss relevant surface modeling techniques that address prevention of self-intersection and overlap.

\subsection{Volumetric Mesh Creation from a Self-Intersecting Triangle Mesh}
Sacht et al. \shortcite{sacht:2013:consistent} were the first to design an approach that creates an appropriately overlapping tetrahedron mesh from a self-intersecting triangle mesh. As with our approach, they assume the existence of a mapping $\fms$ from a non-self intersecting counterpart $\TSurf$ to the input mesh $\Surf$. Unlike our approach, they explicitly form $\TSurf$ and the mapping $\fms$.
$\TSurf$ is created by a backward process using cMCF followed by a forward process that minimizes distortion-energy and deviation from $\Surf$ subject to collision constraints. 
The cMCF is known to remove self-intersections for sphere-topology surfaces \cite{kazhdan:2012:flow} and accordingly, their method is limited to input surfaces with genus zero.
They create a tetrahedron mesh using the self-intersection free $\TSurf$ and then push it forward under $\fms$ which is created by mapping the boundary of the tetrahedron mesh to $\Surf$ and propagating deformation to the interior.
Our approach is similar in spirit, but we do not explicitly create $\TSurf$ or $\fms$; furthermore, we can support input surfaces with genus larger than zero. 
In addition, since they do not directly generate tetrahedra in world space, they must take care to maintain tetrahedron mesh quality under deformation in $\fms$.

\begin{figure}[t]
    \centering
    \begin{subfigure}[b]{0.49\columnwidth}
        \includegraphics[draft=\mydraft,width=\linewidth,trim={300px 50px 270px 50px},clip]{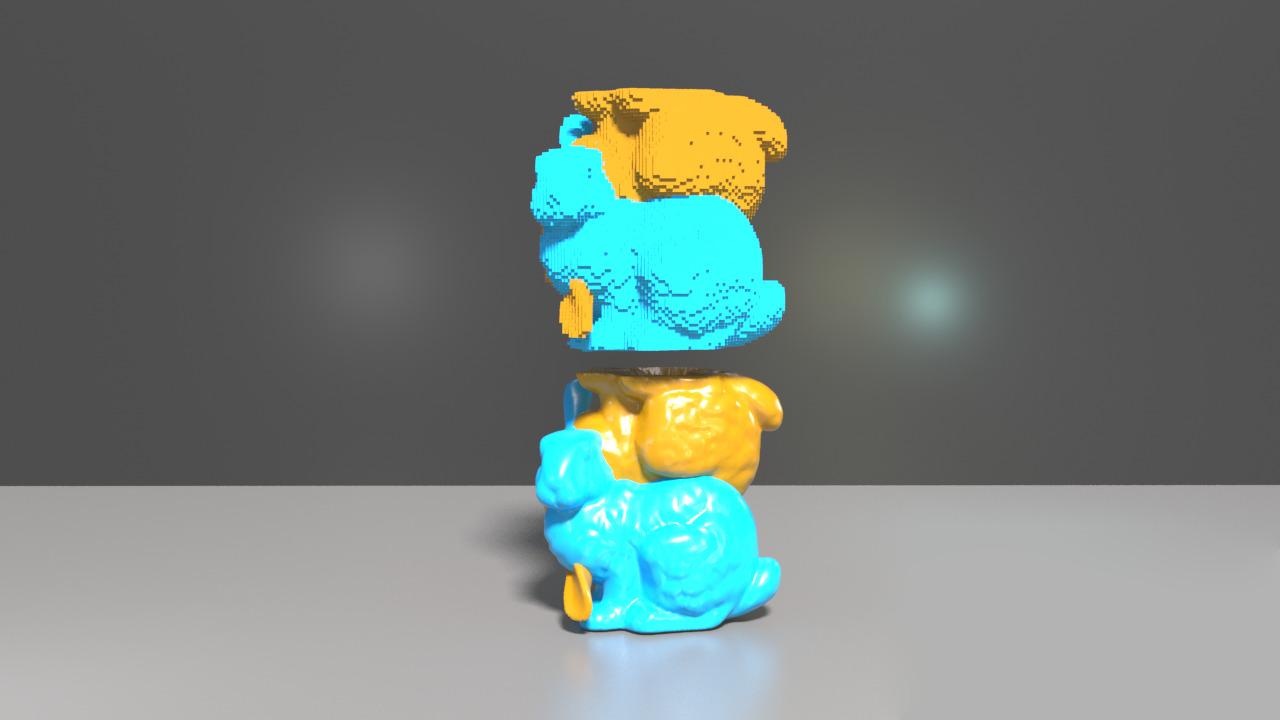}
        \caption{Frame 1}
    \end{subfigure}
    \begin{subfigure}[b]{0.49\columnwidth}
        \includegraphics[draft=\mydraft,width=\linewidth,trim={300px 50px 270px 50px},clip]{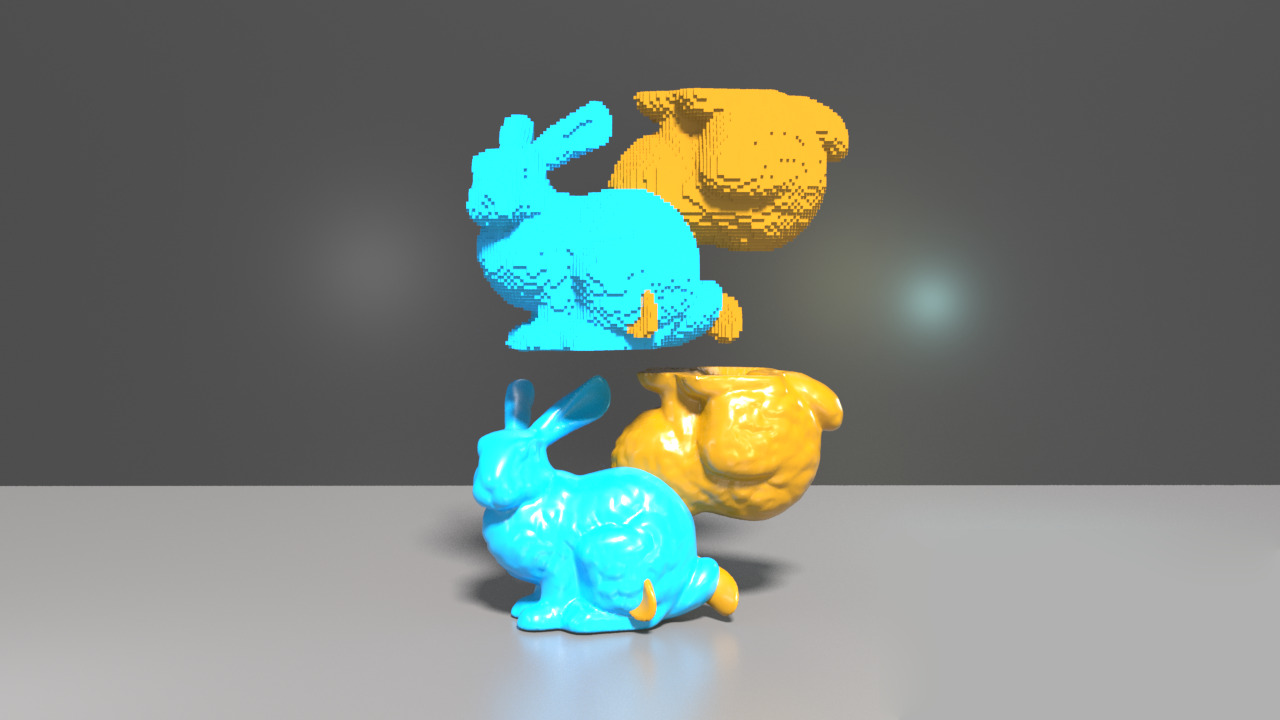}
        \caption{Frame 27}
    \end{subfigure}
    \hfill
    \begin{subfigure}[b]{0.49\columnwidth}
        \includegraphics[draft=\mydraft,width=\linewidth,trim={300px 50px 270px 50px},clip]{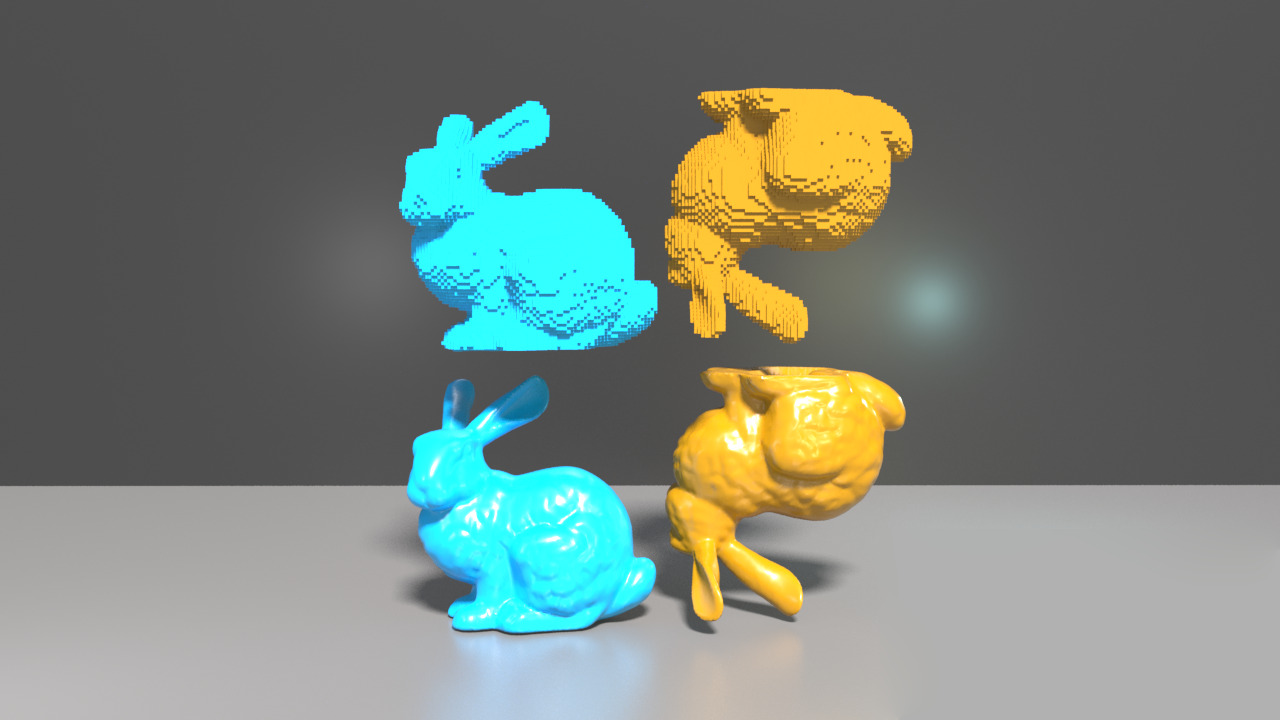}
        \caption{Frame 54}
    \end{subfigure}
    \begin{subfigure}[b]{0.49\columnwidth}
        \includegraphics[draft=\mydraft,width=\linewidth,trim={300px 50px 270px 50px},clip]{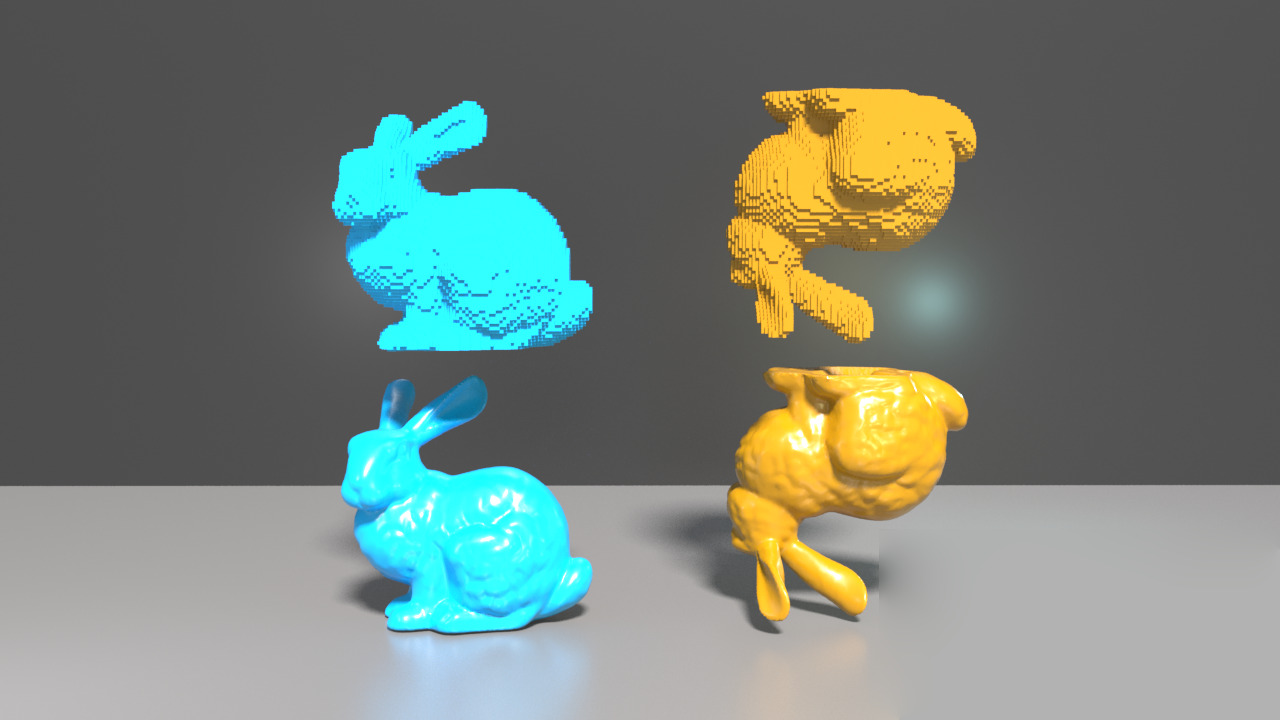}
        \caption{Frame 81}
    \end{subfigure}
    \hfill
    \caption{Two overlapping bunnies naturally separate.  The top part of each subfigure shows the meshes generated by our algorithm, while the bottom part of each subfigure shows the corresponding surface meshes.}
    \label{fig:twin-bunnies}
\end{figure}
Like Li and Barbi\v{c} \shortcite{li:2018:immersion}, we create a volumetric embedding mesh in world space. 
Li and Barbi\v{c} \shortcite{li:2018:immersion} observed that the creation of a volumetric mesh from a self-intersecting surface is related to the geometric and algebraic topological determination of immersions (locally injective mappings) from a compact 3-manifold to a portion of the world space domain.
As in our approach, they start by dividing world space into contiguous regions using the input surface mesh $\Surf$.
However, they use exact/adaptive precision arithmetic to intersect $\Surf$ with itself to achieve this.
We use simplified/less costly intersections of triangles in $\Surf$ with uniform background grid cells and edges.
We only need to know whether an intersection occurs or not; we do not need to resolve the intersection geometry.
Immersions do not always exist, and Li and Barbi\v{c} \shortcite{li:2018:immersion} developed a graph based algorithm to determine if one exists.
Their method for computing these is NP-complete; however, as they note, this is not a bottleneck for most computer graphics applications.
When such an immersion exists, they compute it by duplicating the contiguous regions, intersecting each duplicate with a uniform background tetrahedron lattice to create local tetrahedron meshes that are then sewn together appropriately using their graph structure.
We also duplicate and then sew together contiguous regions, but we use simplified criteria that, while more efficient, can only give accurate results for simple immersions.
Although, as Li and Barbi\v{c} \shortcite{li:2018:immersion} note, the vast majority of applications in computer graphics only require simple immersions.
As with our approach, they also prevent artificial glueing for embedded meshes with nearly intersecting features.
While Li and Barbi\v{c} \shortcite{li:2018:immersion} can accurately compute non-simple immersions, they cannot handle exactly coincident portions $\Surf$ with non-zero measure, which we can handle.
Broadly speaking, the Li and Barbi\v{c} \shortcite{li:2018:immersion} approach is more general than our method, but more costly, primarily due to the comparably large use of exact/adaptive precision arithmetic.

\subsection{Mesh Creation and Mesh Cutting}
The virtual node algorithm (VNA) of \cite{molino:2004:vna} allows cutting a tetrahedron mesh along piecewise-linear paths through the mesh.
As in our approach, duplicates of cut elements are used to resolve necessary topological features.
Teran et al. \shortcite{teran:2005:muscle} built a generalization of this approach to create embedding meshes for nearly overlapping input triangle meshes.
Sifakis et al. \shortcite{sifakis:2007:arbitrary} further extended the VNA to allow for arbitrary cut geometry.
A downside to the geometric flexibility provided by these generalizations is their need for adaptive precision arithmetic and CSG.
Motivated by this, Wang et al. \shortcite{wang:2015:cutting} developed a technique that allows for geometric flexibility without the need for adaptive precision arithmetic.
Their approach allows for arbitrary cut surfaces by generalizing the original VNA \cite{molino:2004:vna} to allow cuts to pass through vertices, edges, or faces of the embedding mesh. 
This alone does not provide sufficient geometric flexibility since cuts cannot pass through facets multiple times. 
To resolve such cuts, the algorithm is run at high-resolution where facets are only intersected once and then coarsened in a topologically-aware manner.

The extended finite element method (XFEM) \cite{belytschko:1999:elastic} is very similar to VNA. 
An XFEM-based but remeshing-free approach for cutting of deformable bodies is presented in \cite{koschier:2017:xfem}.
In a similar spirit, Zhang et al. \shortcite{zhang:2018:stable} utilized the cracking node method \cite{song:2009:cracking}, which is similar to XFEM but uses discontinuous cracks centered at nodes in order to approximate crack paths.
This yields an efficiency advantage over XFEM which in turn allows for simulating materials with many evolving, branching cracks.
The reader is also referred to the survey of Wu et al. \shortcite{wu:2015:cutting} for more discussion of mesh cutting techniques in computer graphics.
\begin{figure}[t]
    \centering
    \begin{subfigure}[b]{0.49\columnwidth}
        \includegraphics[draft=\mydraft,width=\linewidth,trim={0 0 0 0},clip]{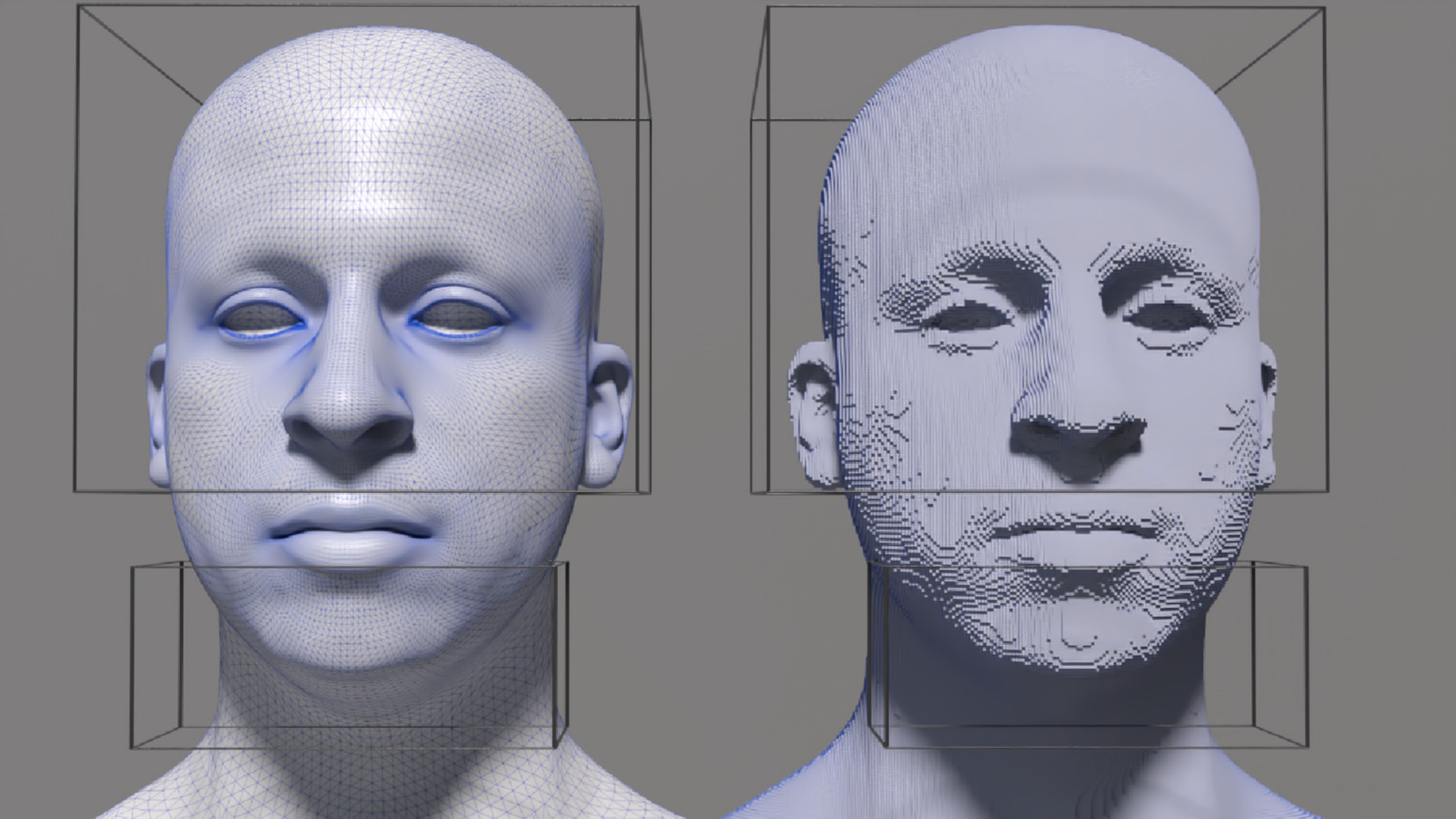}
        \caption{Frame 0}
    \end{subfigure}
    \begin{subfigure}[b]{0.49\columnwidth}
        \includegraphics[draft=\mydraft,width=\linewidth,trim={0 0 0 0},clip]{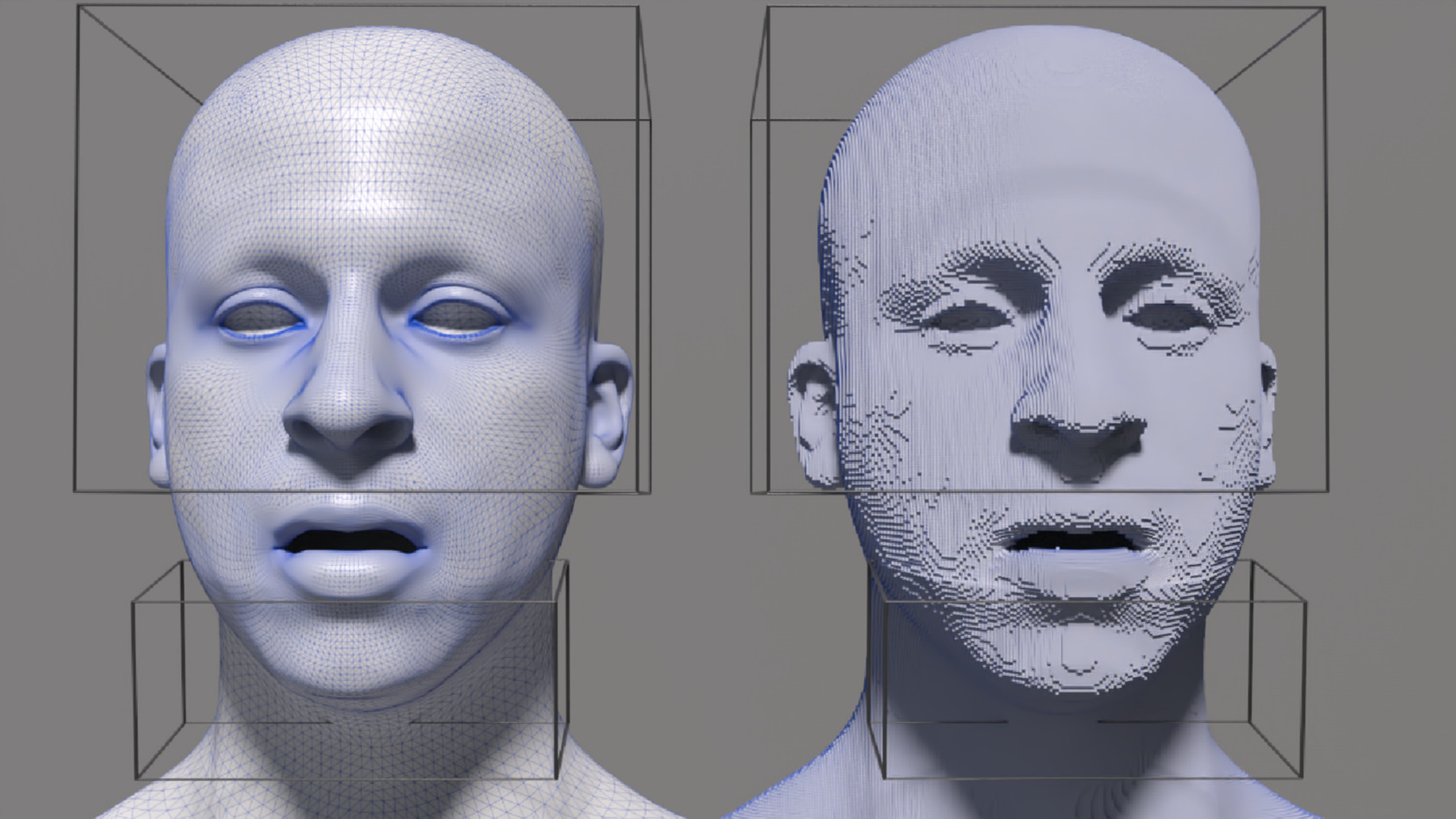}
        \caption{Frame 40}
    \end{subfigure}
    \hfill
    \begin{subfigure}[b]{0.49\columnwidth}
        \includegraphics[draft=\mydraft,width=\linewidth,trim={0 0 0 0},clip]{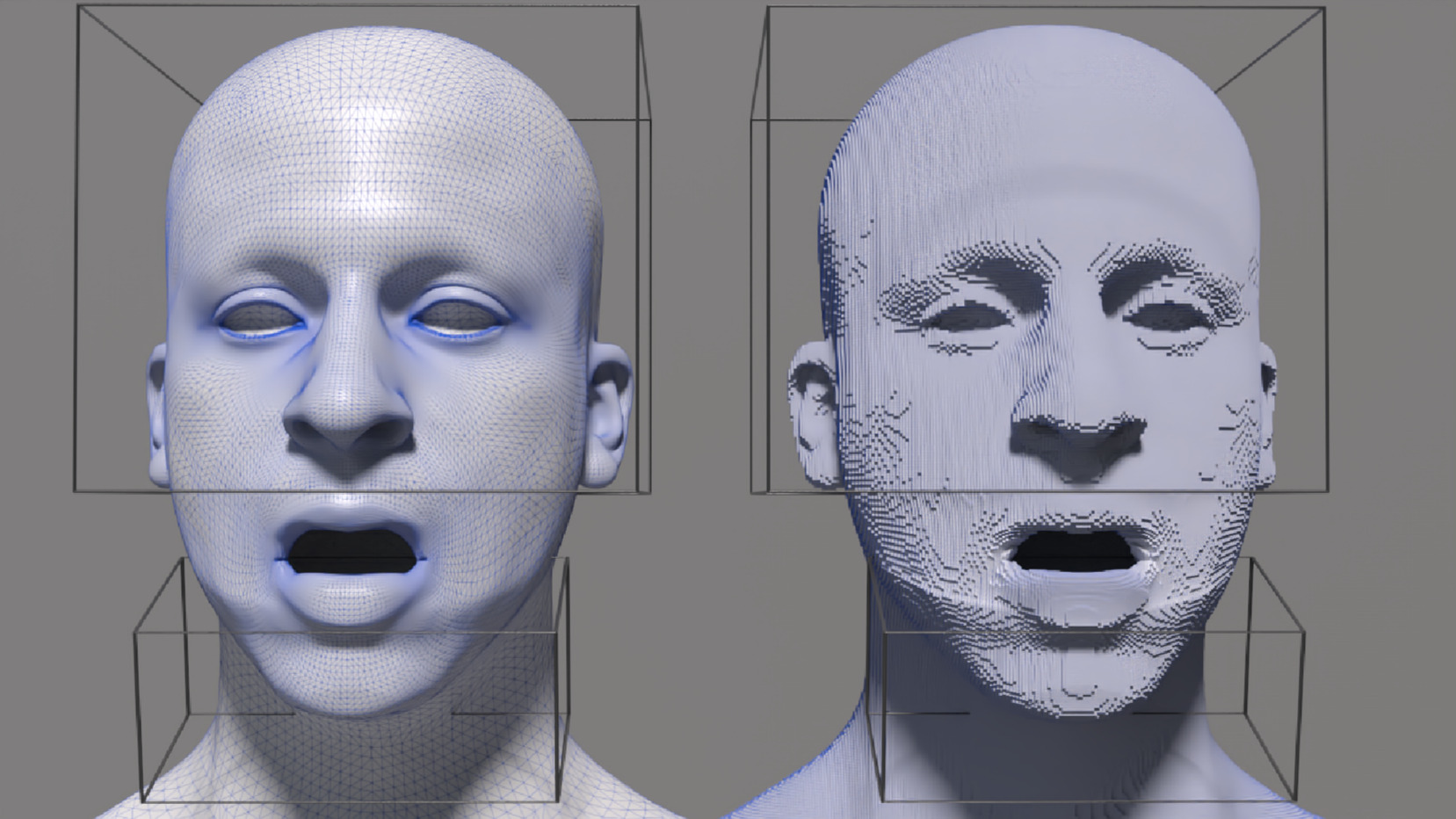}
        \caption{Frame 80}
    \end{subfigure}
    \begin{subfigure}[b]{0.49\columnwidth}
        \includegraphics[draft=\mydraft,width=\linewidth,trim={0 0 0 0},clip]{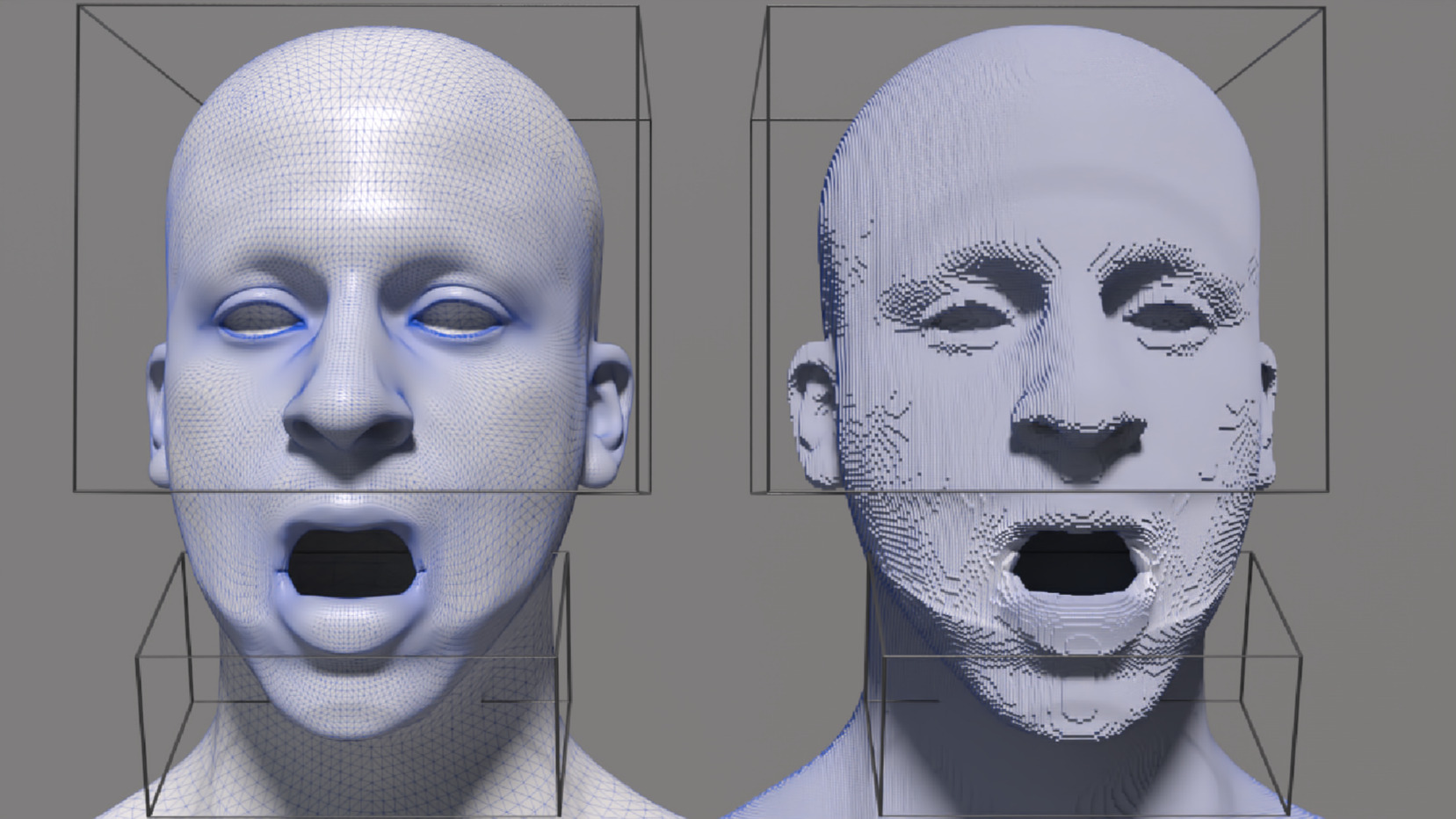}
        \caption{Frame 120}
    \end{subfigure}
    \hfill
    \begin{subfigure}[b]{\columnwidth}
        \includegraphics[draft=\mydraft,width=\linewidth,trim={0 0 0 0},clip]{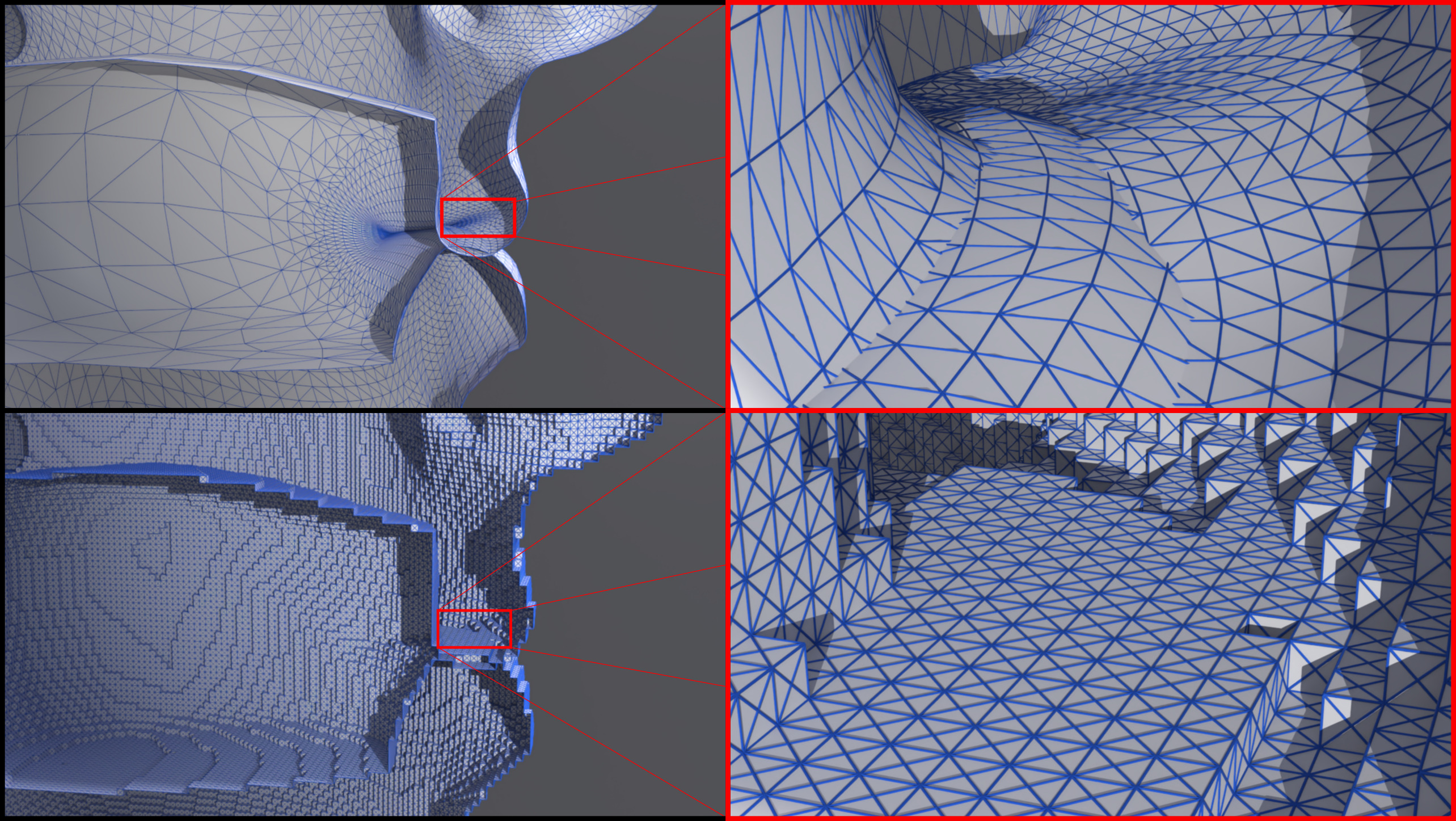}
        \caption{Interior view of lips}
    \end{subfigure}
    \hfill
    \caption{A face surface with self-intersecting lips is successfully meshed.  The right-hand side of each of the first four frames shows the deformed hexahedron mesh, while each left-hand side shows the corresponding surface mesh. The wireframe boxes represent Dirichlet boundary condition regions. In the bottom four subfigures, lip intersection is visualized in the input surface and subsequent hexahedron mesh.}
    \label{fig:head-3d}
\end{figure}

More generally, tetrahedron mesh creation has been robustly addressed by a number of works \cite{si:2015:tetgen,hu:2018:tetwild,labelle:2007:mesh,molino:2003:tetrahedral,doran:2013:isosurface,jamin:2015:cgal}.  For example, Si \shortcite{si:2015:tetgen} pursued a Delaunay refinement strategy in order to provide certain guarantees on tetrahedron quality.  However, sliver tetrahedra are still possible \cite{hu:2018:tetwild}.  The method presented in \cite{hu:2018:tetwild} can handle arbitrary triangle soup as input and returns a high-quality approximated constrained tetrahedron mesh, though performance is hindered to an extent due to prominent usage of exact rational arithmetic.  However, recently, those performance bottlenecks were alleviated and replaced with floating-point computations \cite{hu:2020:fast}.  Notably, researchers have recently presented a successful method for learning high-quality tetrahedron meshes from noisy point clouds or a single image \cite{gao:2020:learning}.



\subsection{Self-Intersecting Curves and Surfaces}
Self-intersecting curves and surface meshes have been considered for many years in both the mathematics and computer science literature.
In two dimensions, algorithms and theorems related to identifying self-intersecting curves date back to  \cite{titus:1961:combinatorial}, with many more recent contributions \cite{blank:1967:extending,marx:1974:extensions,shor:1992:detecting,hu:1995:geometric,graver:2011:does,evans:2020:combinatorial}.
Notably, many problems related to identifying self-intersections are NP-complete \cite{eppstein:2009:self}.
Despite this, efficient algorithms frequently exist; for example, Mukherjee \shortcite{mukherjee:2014:self} gave a quadratic algorithm (in the number of points on the discrete curve) to determine the mapping from a disk to an arbitrarily stretched, potentially self-overlapping curve, also known as computing an immersion of the disk.
In another vein, Li \shortcite{li:2011:detecting} used Gauss diagrams from knot theory to characterize self-intersecting two-dimensional projections of three-dimensional polygons, in order to understand whether there are one or multiple ways to perform mesh repair algorithms like \cite{brunton:2009:filling}.

In the context of three-dimensional mesh generation and animation, self-intersections are typically treated as degeneracies to be avoided or removed.
For example, Von Funck et al. \shortcite{vonfunck:2006:vector} provided a method for deforming surfaces that prevents new self-intersections from occurring, due to the smoothness requirements they place on the vector fields governing the deformation.
The tool devised in \cite{angelidis:2006:swirling} allows for local prevention of self-intersections when deforming a mesh.
A method for avoiding introducing self-intersections within the free-form deformation (FFD) modeling scheme \cite{bezier:1970:numerical,sederberg:1986:free} was presented in \cite{gain:2001:preventing}.
The space-time interference volumes introduced in \cite{harmon:2011:interference} can be used to eliminate self-intersections in meshes, although this method is not always guaranteed to work (the method is primarily intended for interacting with non-self-intersecting input geometry).
Shen et al. \shortcite{shen:2004:interpolating} built an implicit surface from polygon soup, resulting in a watertight mesh that approximates the input surface data.
Attene \shortcite{attene:2010:lightweight} deleted overlapping triangles and subsequently performed a gap-filling procedure in the resulting holes.
Similarly, Jacobson et al. \shortcite{jacobson:2013:robust} presented a method based on the generalized winding number (which, notably, is still applicable to triangle soups and point clouds \cite{barill:2018:fast}, unlike the standard winding number).  Their method results in fusing together self-intersecting parts of the mesh.
Recently, Tao et al. \shortcite{tao:2019:mandoline} demonstrated a method for accurately and efficiently generating cut cell meshes for arbitrary triangulated surfaces, including those with degeneracies.
However, again, they treat self-intersections as flaws to be removed, unlike in our method where self-intersections are valid features of our inputs and outputs.
Nonetheless, an attractive aspect of their algorithm is robust resolution of mesh degeneracies and singularities, unlike methods like \cite{edwards:2014:adaptive-dg,kim:2010:ebmesh} which require random numerical perturbations of the background cut cell grid.  Finally, we also highlight \cite{mitchell:2015:nonmanifold}, which describes a method for representing self-intersecting surfaces using implicit functions sampled on a specialized hexahedron mesh.

\section{Algorithm Overview}\label{section:overview}

The input to our algorithm is a triangulated surface mesh $\Surf$.
The output is a uniform-grid-based embedding hexahedron mesh counterpart $\mathcal{V}$ to $\Surf$ that is well-defined (i.e., free from numerical mesh "glueing" artifacts) even when $\Surf$ is self-intersecting (see Section \ref{section:examples} for examples).    

We briefly summarize the three main stages of our algorithm, as detailed in Figure \ref{fig:overview}.
In the first stage, volumetric extension (Section \ref{section:boundary}), we create a hexahedron mesh $\Vs$ from the background grid that only covers the input surface $\Surf$ with connectivity designed to mimic it.
We sign its vertices depending on inside/outside information derived from the hypothetical self-intersection-free counterpart $\TSurf$.  
We emphasize that this volumetric extension mesh only surrounds $\Surf$.  
Accordingly, the second stage of the algorithm is interior extension region creation (Section \ref{section:regioncopy}).  
Nodes of the background grid are partitioned using the edges cut by $\mathcal{S}$, and then we decide which regions are interior.
Interior regions will be copied to approximate the number of times portions of the interior of the hypothetical self-intersection-free counterpart $\TSurf^V$ will need to overlap after being pushed forward by the hypothetical mapping $\fms$.
For each interior region $j^I$ with at least one copy, we create a hexahedron mesh $\mathcal{V}^{j^I,c}$ for each copy $c$.  
In the third stage of the algorithm (Section \ref{section:merging_boundary}), interior extension regions meshes $\mathcal{V}^{j^I,c}$ are sewn together and into the volumetric extension $\Vs$ to produce the final output mesh. 
We additionally provide a coarsening approach in Section~\ref{section:coarsening} to provide user control over the embedding mesh resolution as well as a topologically-aware technique for converting the hexahedron mesh $\mathcal{V}$ into a tetrahedron mesh $\mathcal{T}$.

\begin{figure}[h]
    \includegraphics[draft=\mydraft,width=\columnwidth]{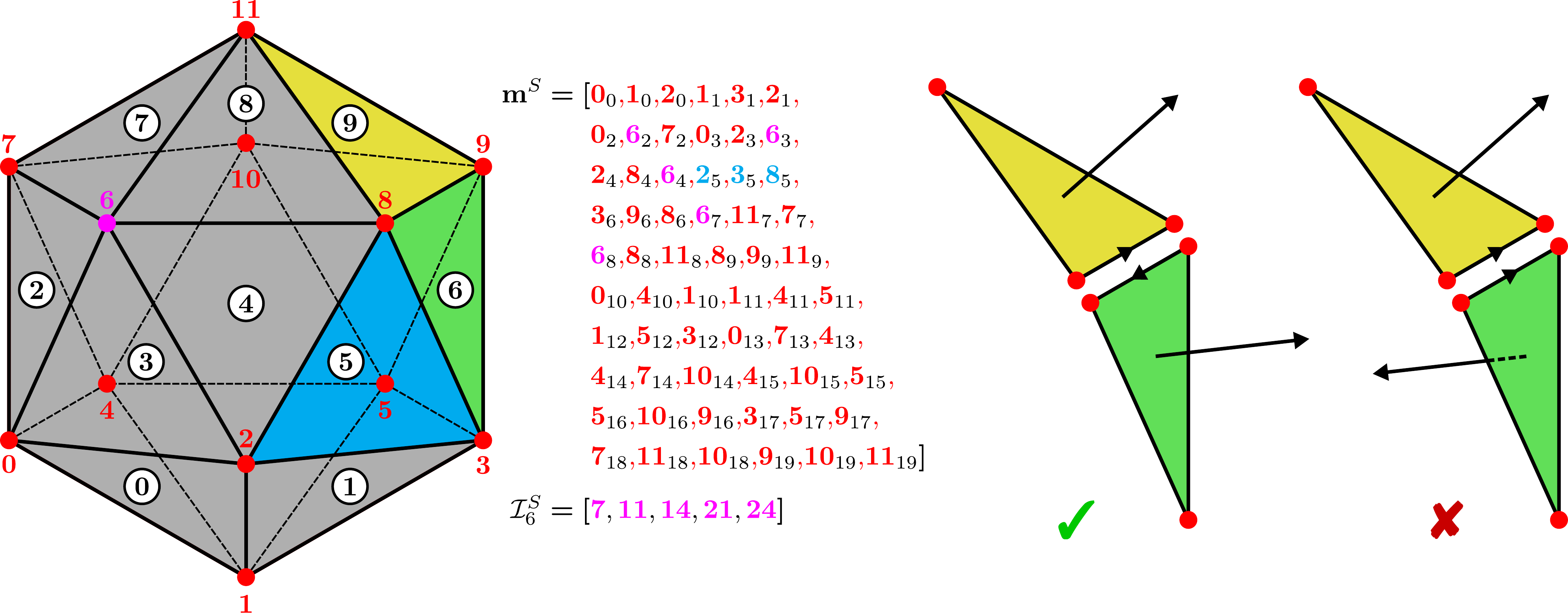}
    \caption{
        {\textbf{Mesh conventions.}} \textit{(Left)} A sample triangle mesh is shown, along with the vector $\mm^S$.
        The incident elements $\mathcal{I}^S_6$ for vertex $6$ are also shown. The first 10 faces, visible from the front, have been labeled on the mesh.
        \textit{(Right)} The left pair of triangles are consistently oriented; the orientations of the edge induced by the normals point in opposite directions.
        For the right pair, the orientations on the common edge point in the same direction; this is not consistent.
    }
    \label{fig:mesh_conventions}
\end{figure}
\section{Definitions and Notation}\label{section:def}
We take a triangle mesh $\mathcal{S}=(\xx^S,\mm^S)$ as input. 
We use $\xx^S=[{\xx^S_0},\hdots,{\xx_{N^S_v-1}^S}]\in\mathbb{R}^{3 N^S_p}$ to denote the vector of triangle vertices $\xx^S_i\in\mathbb{R}^3$ and $\mm^S\in\mathbb{N}^{3N^S_e}$ to denote the vector of indices $m^S_j$ for vertices in $\xx^S$ corresponding triangles $t^S_{\lfloor \frac{j}{3} \rfloor}$, $0\leq {\lfloor \frac{j}{3} \rfloor} <N^S_e$.
For example, for the mesh $\Surf$ in Figure~\ref{fig:mesh_conventions}, triangle $t^S_5$ is made up of vertices $\xx^S_{m^S_j}$ with $j=2,3,8$.
We assume that $\Surf$ is closed (every edge in the mesh has two incident triangles) and consistently oriented (each edge appears with opposite orientations in its two incident triangles).
For each vertex $\xx^S_i$ of $\Surf$, we use $\mathcal{I}^S_i$ to denote the set of incident mesh indices $j$  such that $i=m^S_j$. Figure~\ref{fig:mesh_conventions} demonstrates these conventions. We output a hexahedron mesh $\mathcal{V}=(\xx^V,\mm^V)$ with $\xx^V\in\mathbb{R}^{3 N^V_p}$ denoting the vector of hexahedron vertices and $\mm^V\in\mathbb{N}^{8N^V_e}$ denoting the vector of indices in $\xx^V$ corresponding to vertices in hexahedron $h^V_e$, $0\leq e<N^V_e$.
Each hexahedron in the mesh is geometrically coincident with one grid cell in a background uniform grid $\Gdx$.
We denote the spacing of this grid as $\Delta x$ (uniformly in each direction). For ease of visualization, we use 2D counterparts to $\Surf$ and $\mathcal{V}$ in illustrative figures. In this case, $\Surf$ is a segment mesh and $\mathcal{V}$ is a quadrilateral mesh. 
\begin{figure}[t]
    \includegraphics[draft=\mydraft,width=\columnwidth]{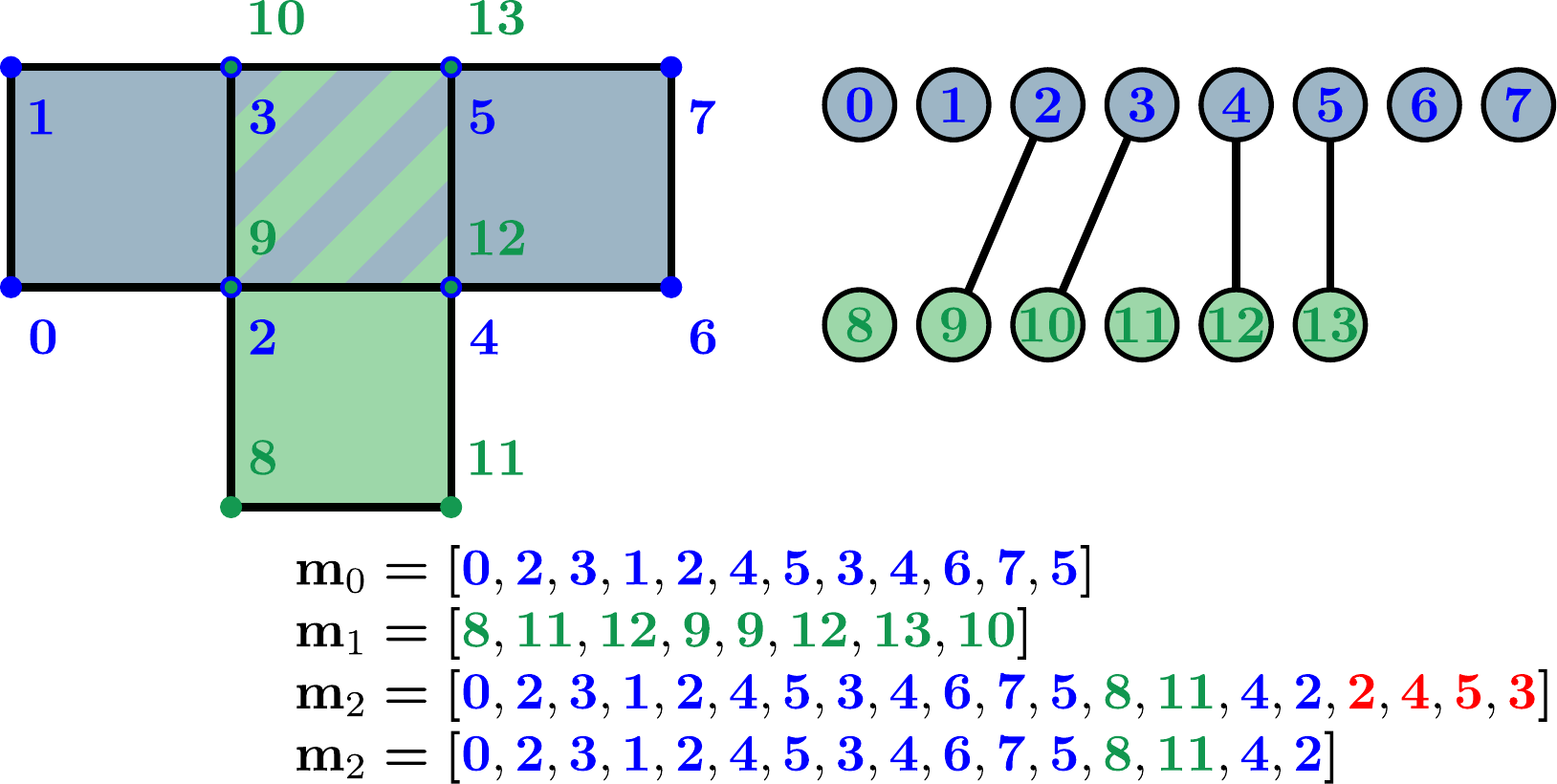}
    \caption{
        {\textbf{Mesh merge.}} An example of two meshes merging together. Vertices 2, 3, 4 and 5 merge with vertices 9, 10, 12 and 13, respectively. A new vector $\mm_2$ is created to hold all of the hexahedron vertices post-merge, and the extra hexahedron (in red) is then removed.
    }
    \label{fig:merge_example}
\end{figure}

\subsection{Merging}\label{subsection:merging}
We construct the final hexahedron mesh $\mathcal{V}$ by merging portions of various precursor hexahedron meshes in a manner similar to techniques used in \cite{teran:2005:muscle,wang:2019:fracture,wang:2015:cutting,li:2018:immersion}. 
As with $\mathcal{V}$, each hexahedron in a precursor mesh is geometrically coincident with background grid cells.  
All precursor meshes share the same vertex array $\xx^V$, although its size will change as we converge to the final $\mathcal{V}$. 
At various stages of the algorithm, we will merge certain geometrically coincident precursor hexahedra.
To perform a merge, we view the set of all vertices in $\xx^V$ as nodes in a single undirected graph and introduce graph edges between nodes corresponding to geometrically coincident vertices. 
In subsequent sections, we refer to such edges in the undirected graph as adjacencies to distinguish them from edges in the various meshes. 
Once all adjacencies are defined, we compute the connected components of the graph using depth-first search. 
All vertices in a connected component are considered to be the same and we choose one representative for all mesh entries. We note that this operation may be carried out on more than two meshes at once and that it can lead to duplicate hexahedra and in this case we remove all but one. Furthermore, replacing all vertices in a connected component with one representative results in unused vertices in $\xx^V$. We remove all unused vertices in a final pass, changing indexing in $\mm^V$ accordingly.
We illustrate the connected component calculation, vertex replacement and unused vertex removal in Figure \ref{fig:merge_example}.

\section{Volumetric Extension}\label{section:boundary}
We first create a volumetric extension $\Vs$ of the surface $\Surf$. 
It is a hexahedron mesh that contains the input surface $\Surf$ and is designed to have topological properties analogous to $\Surf$. 
Since it is an extension of $\Surf$, we can sign the vertices of $\Vs$ depending on which side of the surface they lie on. Overlapping regions in $\Surf$ complicate this process, but it can be disambiguated by considering the pre-image of the surface to its overlap-free counterpart $\tilde{S}$ under the mapping $\fms$. Signing points in $\mathbb{R}^3$ depending on whether or not they are inside $\tilde{S}$ is well-defined and our procedure for signing the vertices in the volumetric extension $\Vs$ is designed considering its pre-image under $\fms$.
\begin{figure}[h]
    \includegraphics[draft=\mydraft,width=\columnwidth]{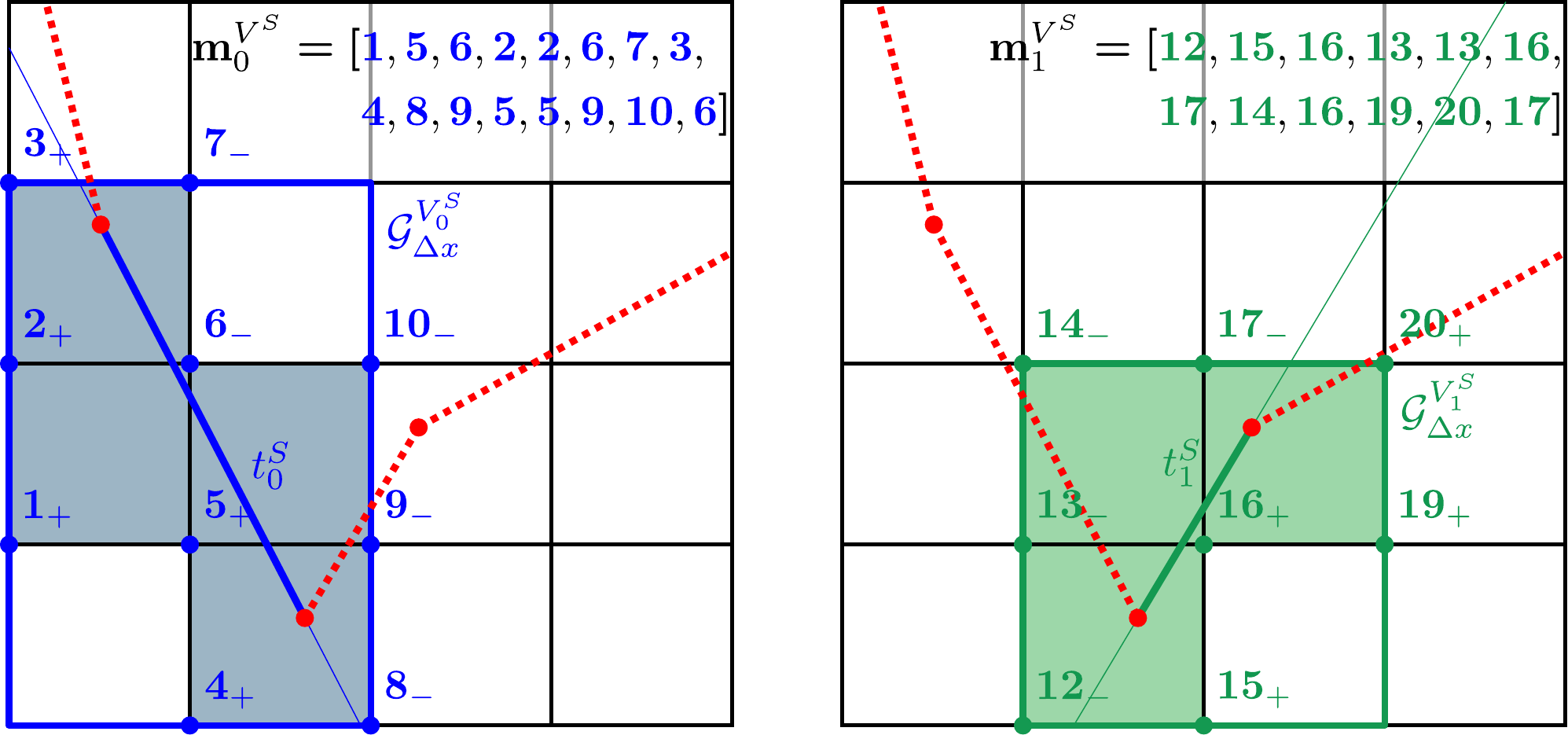}
    \caption{
        {\textbf{Precursor meshes.}} \textit{(Left)} Surface element $t^S_0$ creates quadrilateral mesh $\mathcal{V}^S_0$. \textit{(Right)} Surface element $t^S_1$ creates quadrilateral mesh $\mathcal{V}^S_1$.
        Each element creates copies of the grid cells it intersects by introducing new vertices which are geometrically coincident to grid nodes.
    }
    \label{fig:local_boundary}
\end{figure}
\subsection{Surface Element Precursor Meshes}
In order to mimic the topology of the $\Surf$, we create its volumetric extension $\Vs$ from precursor meshes $\Vs_e=(\xx^V,\mm^{V^S}_e)$ associated with each triangle $t^S_e$ in $\Surf$.
Note that all precursor meshes share the common vertex array $\xx^V$ and that this process begins its evolution to the final $\mathcal{V}$ vertex array.
For each triangle $t^S_e$ in $\Surf$, we define a hexahedron mesh from the subgrid $\mathcal{G}^{V^S_e}_{\Delta x}$ of $\Gdx$ defined by the grid-cell-aligned bounding box of $t^S_e$. 
We add a new hexahedron to $\Vs_e$ corresponding to each background grid cell in $G^{V^S_e}_{\Delta x}$ intersected by $t^S_e$.
We perform this operation using the intersection function from CGAL's 2D/3D Linear Geometry Kernel \cite{cgal:2020:cgal,bronnimann:2020:kernel}.
The hexahedron is geometrically coincident to the intersected grid cell in $\Gdx$, however the vertices introduced into the vertex vector $\xx^V$ are copies of the background grid nodes associated with the sub grid $\mathcal{G}^{V^S_e}_{\Delta x}$.
Note that even though different triangles may intersect the same grid cells, their respective hexahedra correspond to distinct vertices in $\xx^V$. 
Further note that mesh elements in $\Vs_e$ inherit the connectivity of the sub grid $\mathcal{G}^{V^S_e}_{\Delta x}$, that is, hexahedra share common vertices if they are neighbors in $\mathcal{G}^{V^S_e}_{\Delta x}$. We sign the vertices in each $\Vs_e$ depending on which side of the plane containing the triangle $t^S_e$ that they lie on. We illustrate this process in Figure~\ref{fig:local_boundary}. Lastly, we note that these signs are low-cost preliminary approximations to the signs in the final volumetric extension $\Vs$.
In some cases the signs computed in this phase will not be accurate in the volumetric extension, and we provide a more accurate but costly signing when this occurs (discussed in Section \ref{section:merge-meshes}; however, in many cases, they are equal to the final signs, and their comparably-low computational cost improves overall algorithm performance
\subsection{Merge Surface Element Meshes}\label{section:merge-meshes}
\begin{figure}[h]
    \includegraphics[draft=\mydraft,width=\columnwidth]{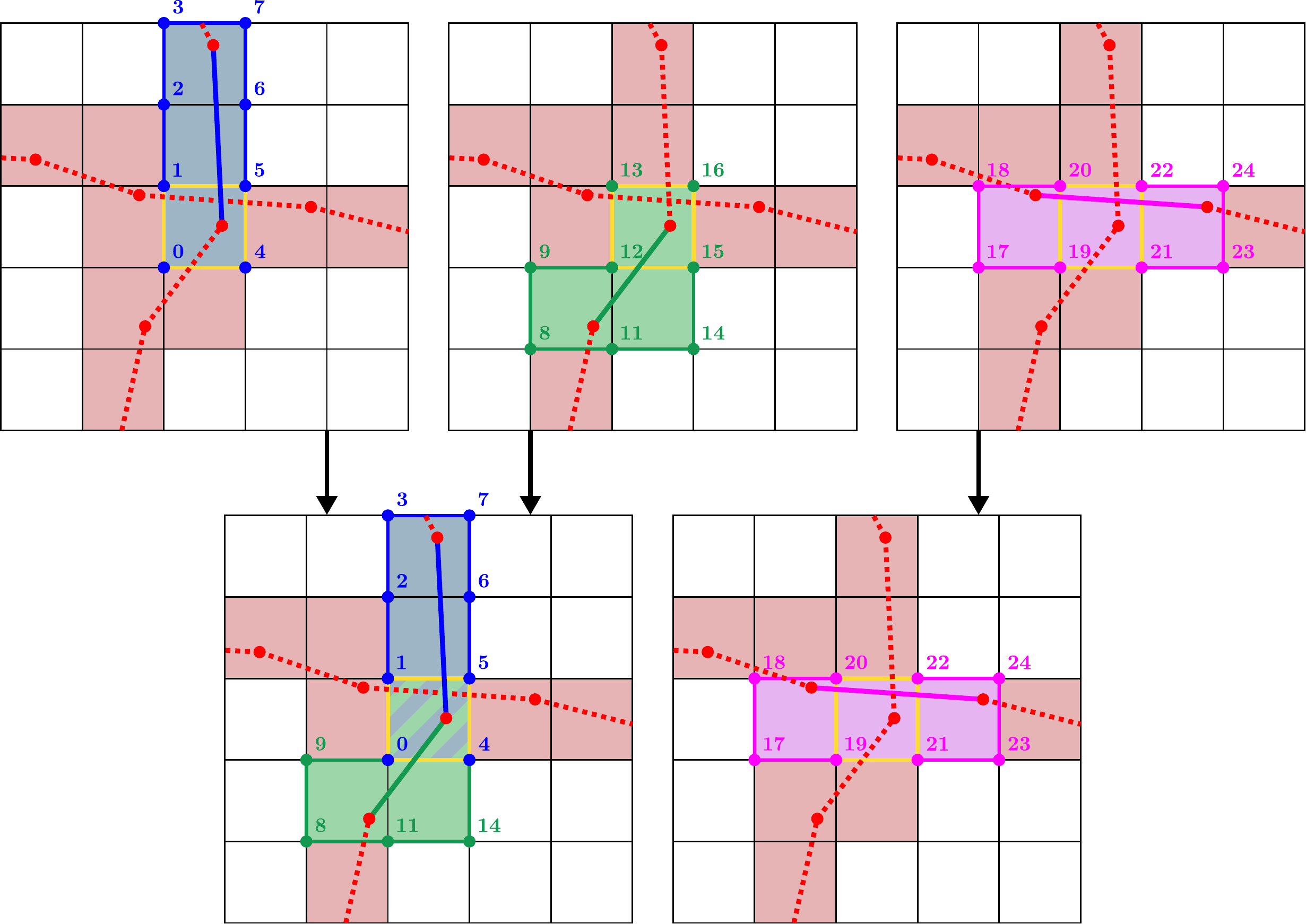}
    \caption{
        {\textbf{Precursor merge.}} The 12 vertices bordering the cell marked in yellow are merged into 8 resulting vertices. Blue vertices 0, 1, 4, 5 and green vertices 12, 13, 15, 16 are merged, respectively. However, magenta vertices 19, 20, 21, 22 do not merge with the blue or green vertices since their associated surface element is topologically distant.
    }
    \label{fig:precursor_merge}
\end{figure}

\begin{figure}[h]
    \includegraphics[draft=\mydraft,width=\columnwidth]{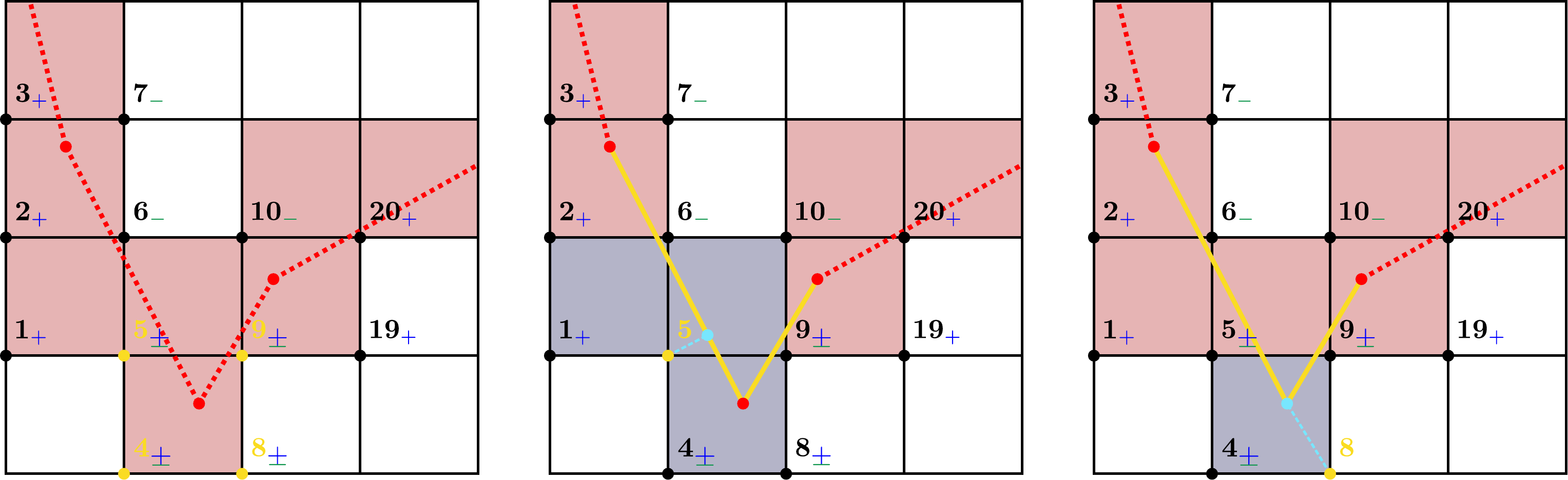}
    \caption{
        {\textbf{Closest facet.}} \textit{(Left)} The four vertices in yellow all have ambiguous signs. \textit{(Middle)} To sign vertex 5, we generate the local patch $S_{5^V}$, which are the segments shown in yellow. The closest facet (indicated in cyan) lies on a face. \textit{(Right)} A similar process is illustrated for vertex 8, but here the closest facet is a vertex.
    }
    \label{fig:sign_conflict}
\end{figure}

We merge portions of the precursor meshes $\Vs_e$ to form the volumetric extension hexahedron mesh $\Vs$ by defining adjacency between vertices in $\xx^V$ as described in Section~\ref{subsection:merging}. 
We define this adjacency from the mesh connectivity of $\Surf$ using its incident elements $\mathcal{I}^S_i$ for each vertex $\xx^S_i$.
Geometrically coincident vertices in $\Vs_{\left \lfloor {j^S_{i,0}/3} \right \rfloor }$ and $\Vs_{\left \lfloor {j^S_{i,1}/3} \right \rfloor }$ for $j^S_{i,0},j^S_{i,1}\in\mathcal{I}^S_i$ are defined to be adjacent if each are on hexahedrons in their respective meshes which are geometrically coincident.
Note in particular that this is different from requiring that geometrically coincident vertices in $\Vs_{\left \lfloor {j^S_i/3} \right \rfloor }$ for $j^S_i\in\mathcal{I}^S_i$ (see the geometry of Figure \ref{fig:node_merging}).
In other words, all geometrically coincident hexahedra in element precursor meshes associated with triangles that share a common vertex are merged (see Figure~\ref{fig:precursor_merge}). Merged vertices retain the sign they were given in $\Vs_e$ when possible. However, if merged vertices have differing signs, e.g. in regions with higher curvature (see Figure~\ref{fig:sign_conflict}), then we must recompute the sign from their geometric relation to $\Surf$.

\begin{figure}[h]
    \includegraphics[draft=\mydraft,width=\columnwidth]{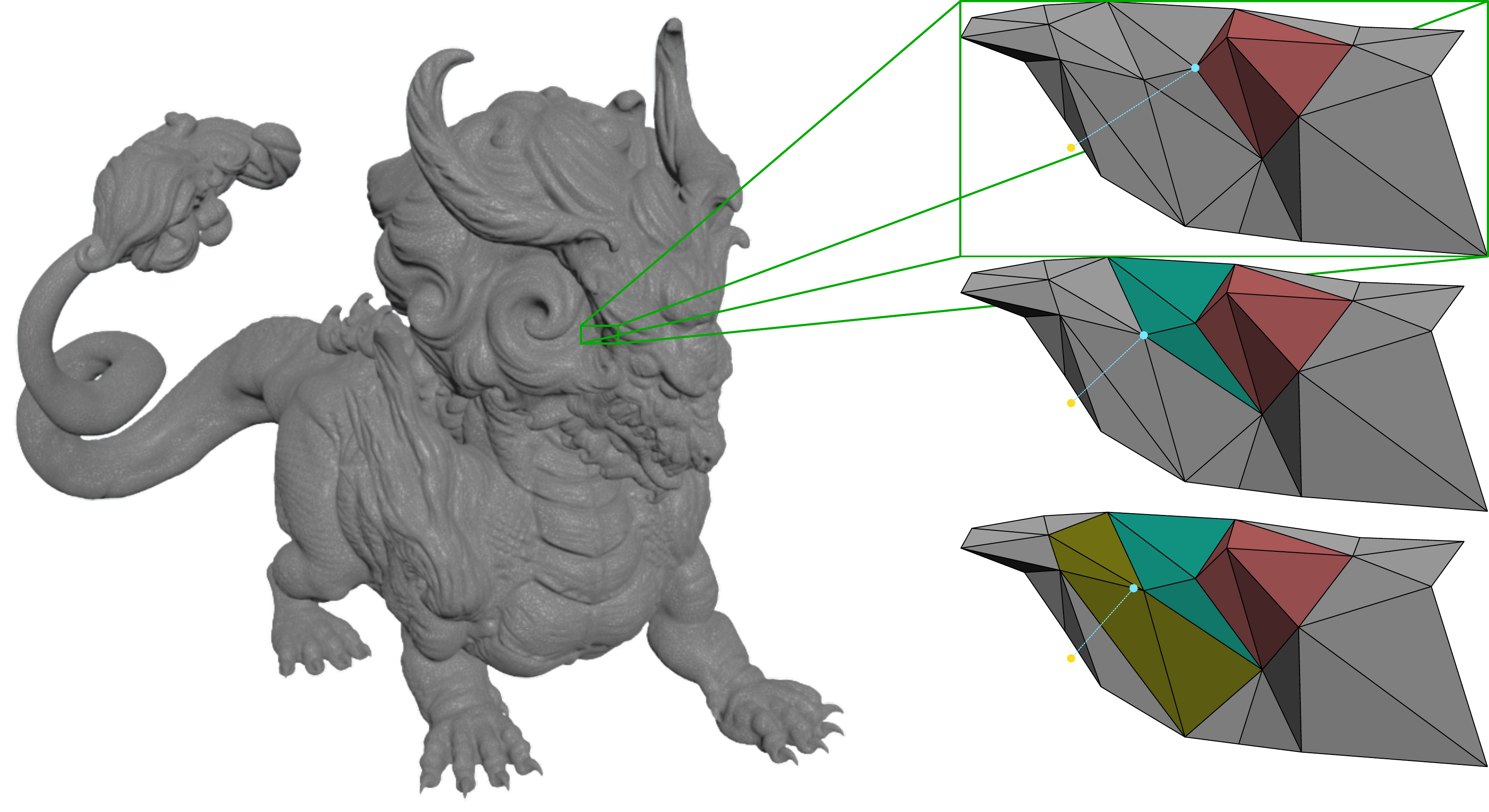}
    \caption{
        {\textbf{Patch expansion.}} The local patch $S_{i^V}$ corresponding to the yellow vertex is shown. The initial patch is indicated in red, and the closest facet is a vertex of the red patch. We add the missing incident triangles (turquoise) and recompute the closest facet. This is again a vertex with incident triangles not in the patch, so we repeat the process (with new triangles in dark yellow). The closest feature is now on an edge, and we proceed to the edge criteria for signing.
    }
    \label{fig:patch_expansion}
\end{figure}

In regions of higher curvature where the preliminary signs of vertices in $\Vs_e$ cannot be adopted in $\Vs$, we use an eikonal strategy \cite{osher:2003:LSMDIS} to propagate positive signs from $\Surf$ in the direction of the surface normal and minus signs in the opposite direction. This is well defined in light of the assumed existence of the pre-image $\TSurf$ of $\Surf$ under $\fms$. Here, each vertex $\xx^V_i$ in the volumetric extension $\Vs$ is associated with some collection of precursor meshes $\Vs_{e_i}$ where $\xx^V_i$ was created in the merge of vertices in the $\Vs_{e_i}$. 
This defines a local patch $S_{i^V}$ of surface triangles $t^S_{e_i}$ in $\Surf$ associated with $\xx^V_i$. 
When propagating signs from $\Surf$ to $\xx^V_i$, only these triangles are considered.
It is important to only use this local surface patch since there may be triangles in $\Surf$ that are geometrically close to $\xx^V_i$ but topologically distant.
Note that this precludes the use of global point-in-polygon algorithms based on ray casting or winding numbers since those will not give correct results when $\Surf$ has self-intersection.
Instead we adopt the local point-in-polygon method of Horn and Taylor \shortcite{horn:1989:theorem}.
First, we compute the closest mesh facet (triangle, edge, or point) in $S_{i^V}$ to $\xx^V_i$.
The closest facet calculation is performed by first storing $S_{i^V}$ in a CGAL surface mesh and then using its class functions and the locate function from the Polygon Mesh Processing package \cite{botsch:2020:surface,loriot:2020:polygon}.
If the closest facet is an edge or a point, we add triangles from $\Surf$ that are incident to the vertices in the edge or the point respectively to the patch $S_{i^V}$ (if they are not already in it). 
If more triangles were added, we recompute the closest mesh facet. 
We illustrate this process in Figures~\ref{fig:sign_conflict} and \ref{fig:patch_expansion}.
If the closest facet is a triangle, we compute the sign depending on the side of the plane containing the triangle that the point lies on. 
If the closest faces is an edge or point we use the conditions from \cite{horn:1989:theorem}, which we summarize below:
\begin{itemize}
    \item If the closest facet is an edge, then the sign is $-1$ if the edge is concave (as determined by the normals of the incident faces) and $+1$ if it is convex.
    \item If the closest facet is a vertex, then there exists a discrimination plane with an empty half-space. Choosing any such plane, the sign is $-1$ if the edges defining the plane are concave and $+1$ if they are convex.
\end{itemize}
A discrimination plane is defined by two non-collinear incident edges and it has an empty half-space if all incident faces and edges lie on one side of the plane or on the plane itself.

\section{Interior Extension Region Creation}\label{section:regioncopy}
\begin{figure}[h]
    \includegraphics[draft=\mydraft,width=\columnwidth]{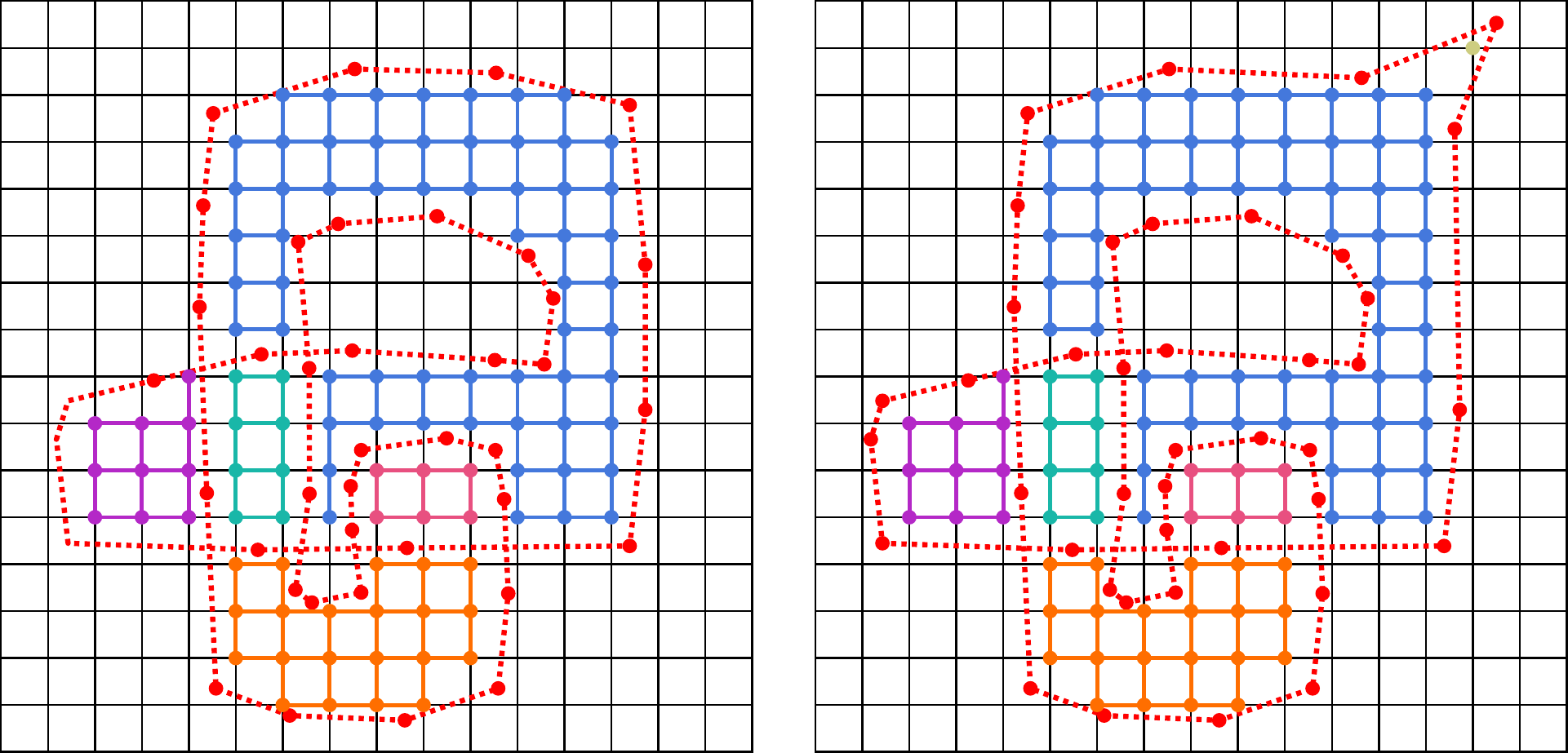}
    \caption{
        {\textbf{Region over-count.}} As the process of partitioning the grid only uses connectivity based on grid edges, it is possible for a contiguous region to be split into multiple regions. Shifting some of the vertices of $\mathcal{S}$ on the left results in the geometry on the right, which contains an additional region in the upper right corner since no edge connects this grid node to the larger blue region.
    }
    \label{fig:over_count}
\end{figure}

We grow the volumetric extension $\Vs$ on its interior boundary (defined by vertices with negative sign) to create the remainder of the volumetric mesh $\mathcal{V}$.
We determine where to grow the extension by examining connected components of the background grid defined by its intersections with $\Surf$.
We compute these components using depth-first search (as discussed in Section~\ref{subsection:merging}), where adjacency between nodes in the background grid is defined between edge neighbors not divided by $\Surf$.
We again use CGAL's intersection function from the 2D/3D Linear Kernel to determine whether or not an edge is divided.
This is a simplistic criterion which can lead to an over-count in the number of interior regions, as demonstrated in Figure \ref{fig:over_count}.
A more accurate criteria would use material connectivity determined from the intersection of the surface $\Surf$ with the relevant background grid cells, similar to the CSG operations in \cite{sifakis:2007:arbitrary}.
However, as noted in \cite{li:2018:immersion} these operations are extremely costly and our approach is robust to over-counting the number of interior regions since they are all merged together appropriately in the later stages of the algorithm.

\begin{figure}[h]
    \includegraphics[draft=\mydraft,width=\columnwidth]{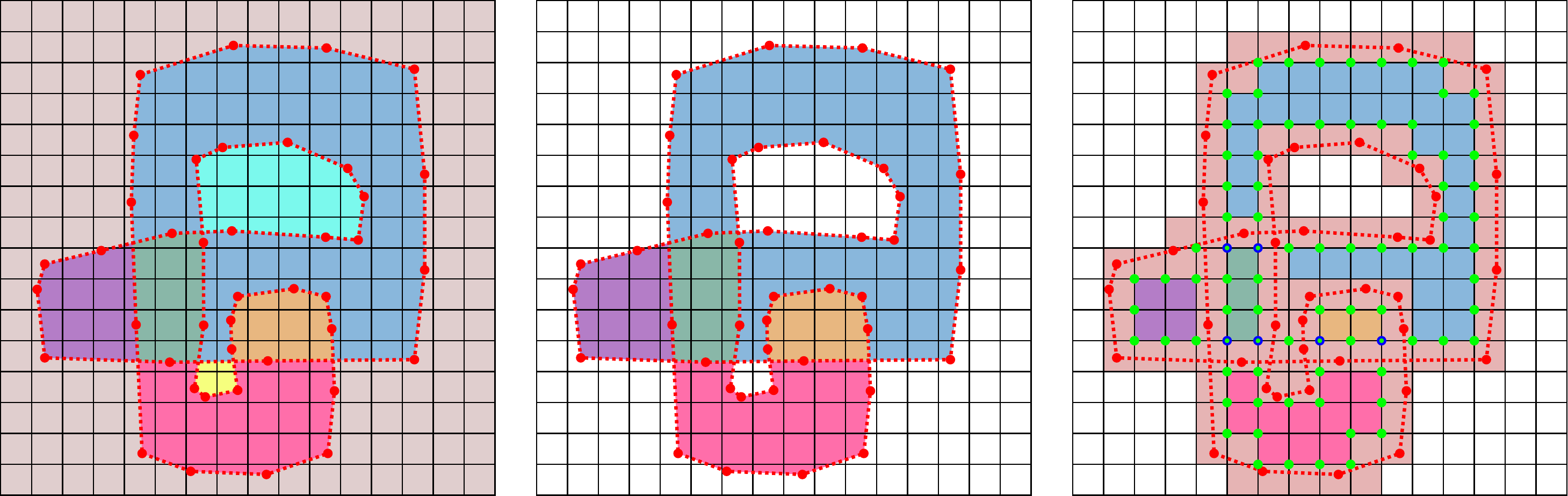}
    \caption{
        {\textbf{Connected regions.}} \textit{(Left)} The surface partitions the background grid into contiguous regions. \textit{(Middle)} The exterior regions are removed. \textit{(Right)} The volumetric extension $\mathcal{V}^S$ is shown, along with the negatively signed vertices in green. Multiple geometrically coincident vertices are indicated using blue circles with green centers.
    }
    \label{fig:regions}
\end{figure}

Each connected component of background grid nodes constitutes a contiguous region. Regions that have a grid node with at least one geometrically coincident vertex in $\xx^V$ with negative sign are defined to be interior. 
Exterior regions, those not containing a grid node with a geometrically coincident vertex in $\xx^V$ with negative sign, are discarded.
We create at least one hexahedron mesh $\mathcal{V}^{j^I,c}$ for each interior region $j^I$.
Multiple copies of interior meshes are created near self-intersecting portions of $\Surf$ since here they represent multiple overlapping portions of the volumetric domain. 
We illustrate this process in Figure~\ref{fig:regions}.
We note that as before, each hexahedron mesh $\mathcal{V}^{j^I,c}$ uses the common vertex array $\xx^V$.

\begin{figure}[h]
    \includegraphics[draft=\mydraft,width=\columnwidth]{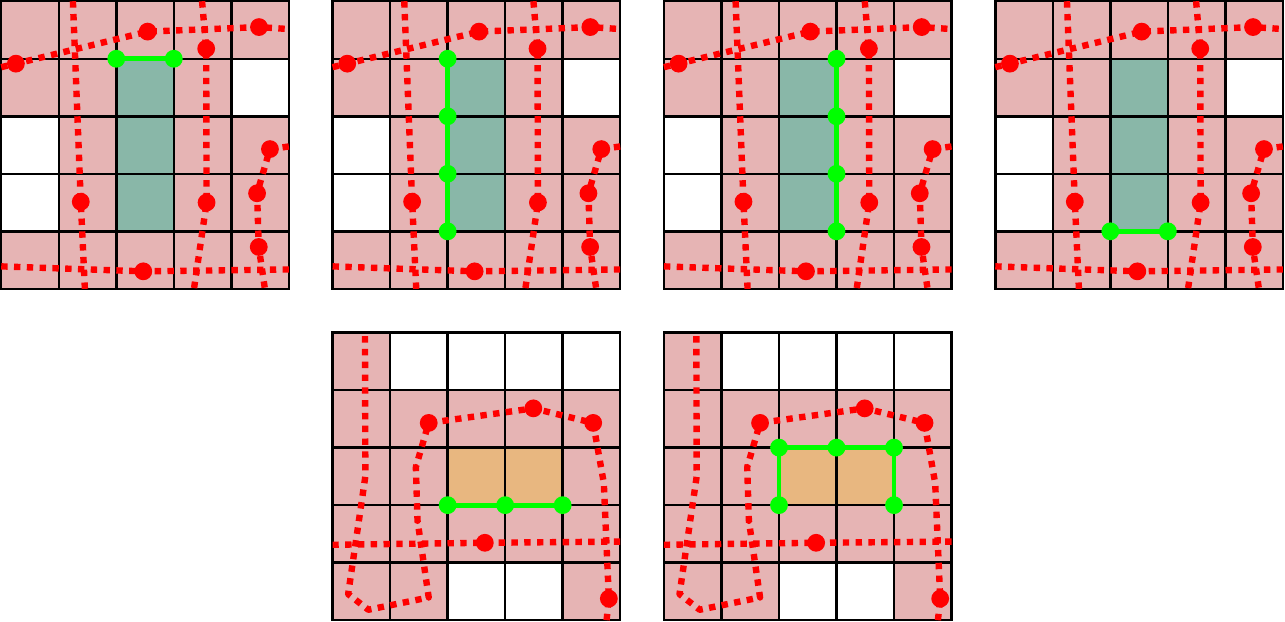}
    \caption{
        {\textbf{Copy counting.}} The two regions from Figure \ref{fig:regions} having multiple copies are shown. Each copy is displayed with its corresponding connected component of vertices with negative sign.
    }
    \label{fig:copies}
\end{figure}

We determine interior regions $j^I$ that require multiple copies as those with grid nodes that have more than one geometrically coincident vertex in $\xx^V$ with negative sign. 
For these regions, we create a copy $\mathcal{V}^{j^I,c}$ for each connected component $c$ of vertices in $\xx^V$ with negative sign that are geometrically coincident with a grid node in the region, as shown in Figure \ref{fig:copies}.
Adjacency between these vertices is defined if they are in a common hexahedron in the volumetric extension $\Vs$.
In general, this will be an over-count as multiple connected components may ultimately correspond to the same copy.
We note that this process is analogous to the cell creation portion of the method of Li and Barbi\v{c} \shortcite{li:2018:immersion}.
They show that in the case of simple immersions, the correct number of copies is equal to the winding number of the region.
We do not compute the winding number since our over-count is typically resolved during the merging process described in Section~\ref{section:merging_boundary}.
However, failure cases occur when the background uniform grid $\Gdx$ cannot resolve thin features or high-curvature in $\Surf$.
In these cases, an over-count that cannot be resolved in the later merging stages occurs.
The background grid must be refined to resolve these cases, however using a strategy similar to that of Wang et al. \shortcite{wang:2015:cutting} we use a topology-preserving coarsening strategy (see Section~\ref{section:coarsening}) after the algorithm has run to prevent excessively small element sizes and associated high element counts.
We also note again that unlike Li and Barbi\v{c} \shortcite{li:2018:immersion}, we cannot handle non-simple immersions.

\begin{figure}[h]
    \includegraphics[draft=\mydraft,width=\columnwidth]{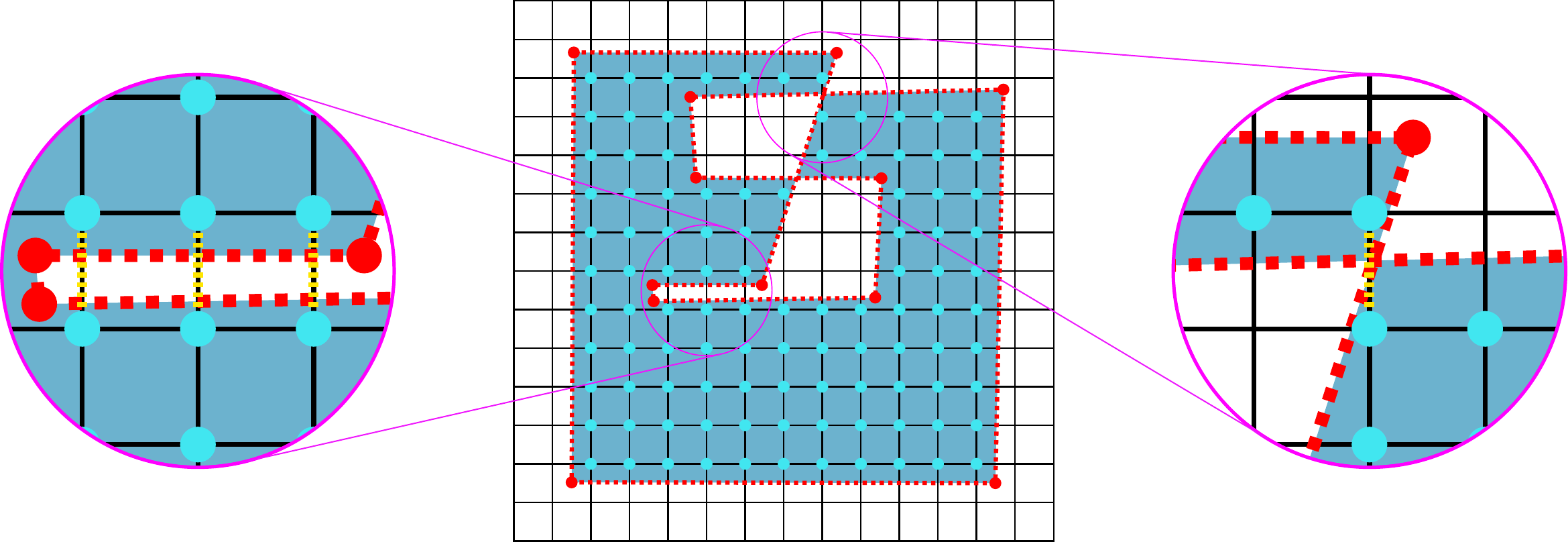}
    \caption{
        {\textbf{Edge cut criterion.}} Grid nodes $\xx_\ii$ of a region are shown, along with two examples showing that adjacent grid nodes may have their common edge cut by a triangle (cut edges are indicated by the dashed yellow lines). In this case, adjacencies are not built between the corresponding vertices in $\mathcal{V}_{\ii}^{j^I,0}$ to avoid unwanted sewing.
    }
    \label{fig:edge_cuts}
\end{figure}

As with $\Vs$, we construct the first copy of the hexahedron mesh for each interior region $\mathcal{V}^{j^I,0}$ from precursor hexahedron meshes $\mathcal{V}^{j^I,0}_\ii=(\xx^V,\mm^{V^{j^I,0}}_\ii)$. 
Here $\xx_\ii$ are the grid nodes in region $j^I$.
It should be noted that these are different than the vertices $\xx^V_i\in\xx^V$ and that $\ii=\left(i_0,i_1,i_2\right)$ is used to denote the grid multi-index associated with the node.
For each $\xx_\ii$, $\mm^{V{j^I,0}}_\ii$ consists of 8 hexahedra which are geometrically coincident with the 8 local background grid cells incident to $\xx_\ii$.
Copies of $\xx_\ii$ and the 26 background grid nodes surrounding $\xx_\ii$ (whether or not they are in region $j^I$) are introduced into $\xx^V$ to achieve this.
We again merge these precursors as described in Section~\ref{subsection:merging} where adjacencies between the vertices of $\xx^V$ are defined as follows.
For each pair of grid nodes $\xx_\ii$ and $\xx_\jj$ in region $j^I$, the geometrically coincident vertices in $\xx^V$ corresponding to the hexahedra of $\mathcal{V}^{j^I,0}_\ii$ and $\mathcal{V}^{j^I,0}_\jj$ are adjacent if $\xx_\ii$ and $\xx_\jj$ are connected by an edge in $\Gdx$ that is not cut by a triangle in $\Surf$.
This edge cut criteria prevents connection between geometrically close but topologically distant features, as illustrated in Figure \ref{fig:edge_cuts}.
We reemphasize that as described in Section~\ref{subsection:merging} the final $\mm^{V^{j^I,0}}$ is formed by concatenating all of the arrays $\mm^{V^{j^I,0}}_\ii$ (modified to account for merged vertex numbering) and removing any duplicated hexahedra. 
The remaining copies $\mathcal{V}^{j^I,c}$ are created by duplicating $\mm^{V^{j^I,0}}$ with new vertices distinct from those corresponding to $\mathcal{V}^{j^I,0}$ and any other copy.

\section{Interior Extension Region Merging}\label{section:merging_boundary}
Having created the interior extensions $\mathcal{V}^{j^I,c}$, the merging of these meshes with the volumetric extension $\Vs$ and with each other (to account for possible over-counting in their creation) is carried out in multiple steps.
We first merge hexahedra from $\Vs$ into $\mathcal{V}^{j^I,c}$ in a process described below.
We then determine which of the interior extensions should merge to each other, using hexahedra from $\Vs$ which merge into multiple $\mathcal{V}^{j^I,c}$ to generate a list of overlapping hexahedra between meshes of different regions and copies.
Next, we use these overlaps to determine which copies of the same region are duplicated and merge the duplicates together.
Finally, these overlapping hexahedra are used to define the adjacencies in the final merging process.
\begin{figure}[h]
    \includegraphics[draft=\mydraft,width=\columnwidth]{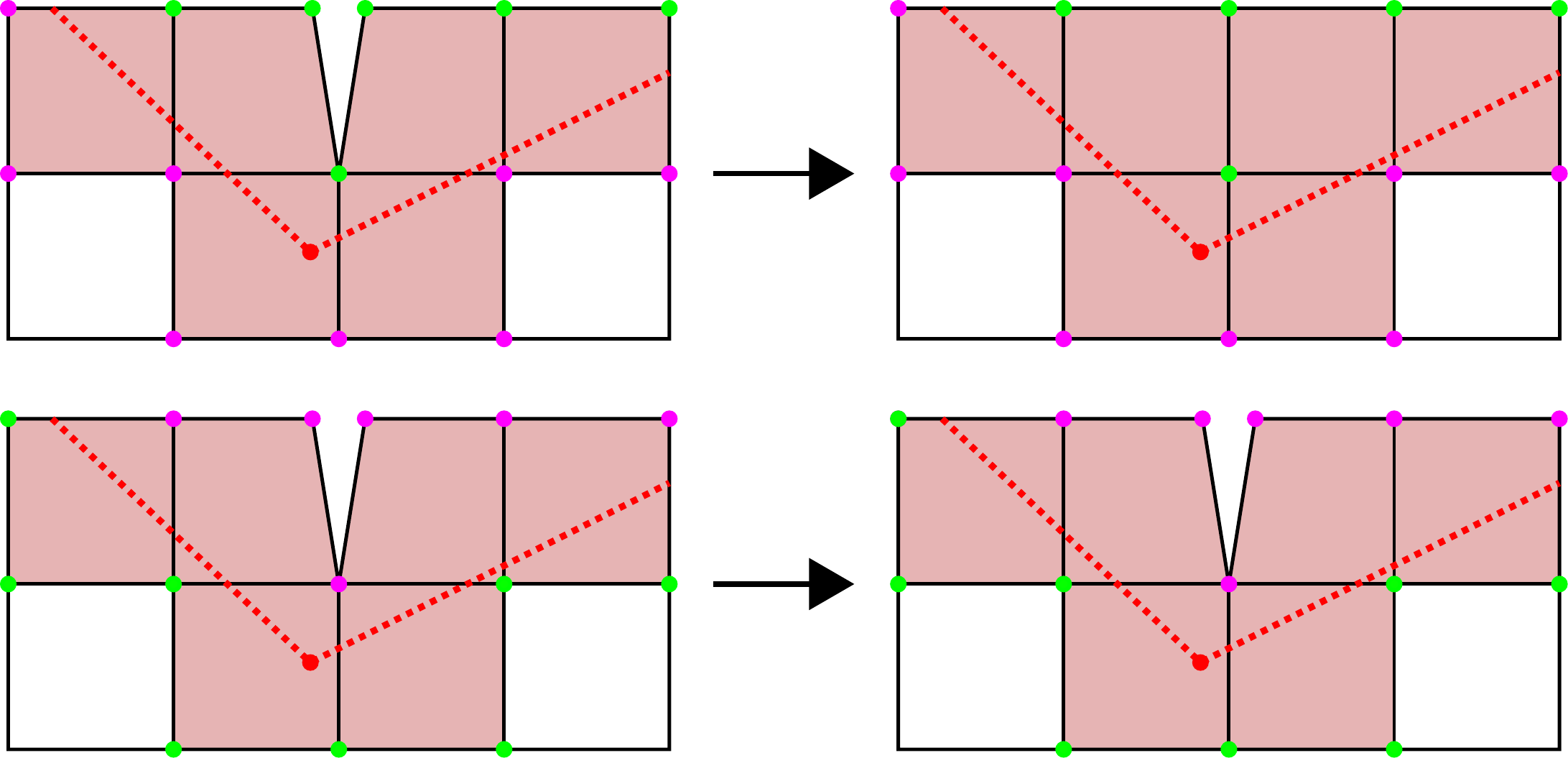}
    \caption{
        {\textbf{Preliminary merge.}} The construction of the volumetric extension $\mathcal{V}^S$ may result in geometrically coincident vertices which do not come from topologically distant parts of the mesh. Green vertices have negative signs, while purple vertices have positive sign. Above: The process in Section \ref{subsection:mergewithboundary} merges these vertices into a single vertex. Below: We do not merge coincident positive vertices, to avoid unnecessarily sewing the exterior.
    }
    \label{fig:node_merging}
\end{figure}

\subsection{Merge With Boundary}\label{subsection:mergewithboundary}
\begin{figure}[h]
    \includegraphics[draft=\mydraft,width=\columnwidth]{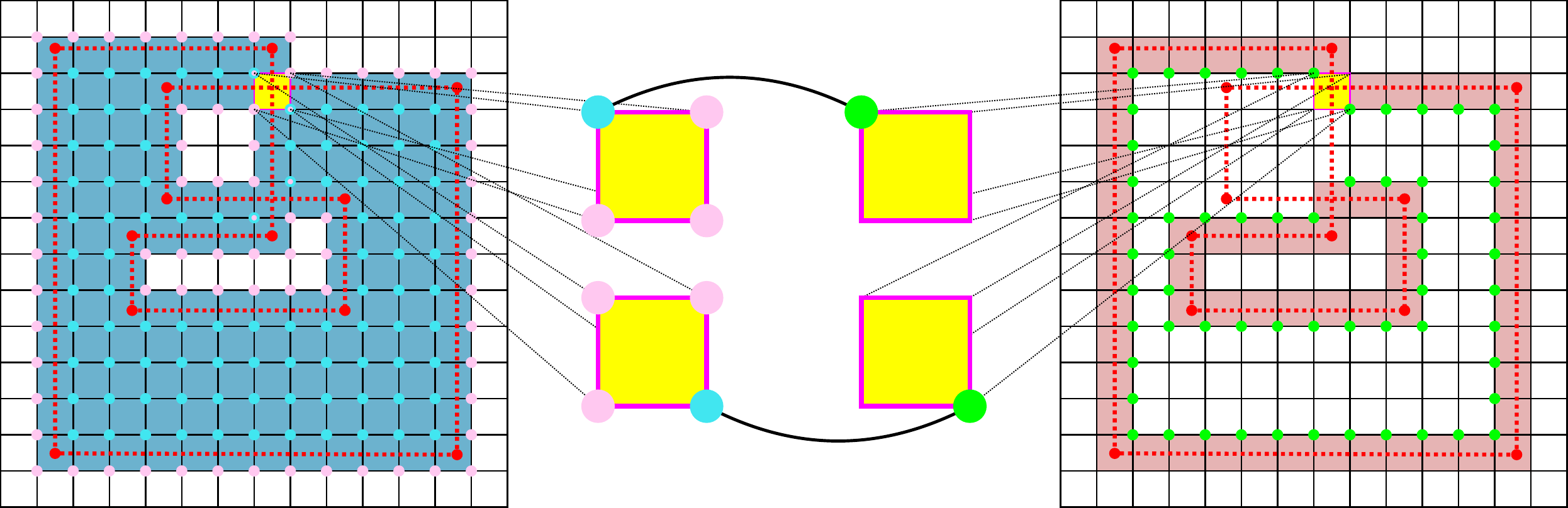}
    \caption{
        {\textbf{Vertex adjacency.}} The merge process between vertices of $\mathcal{V}^{j^I,c}$ and $\mathcal{C}_c^{j^I}$. For the cell highlighted in yellow, there are 2 hexahedra from $\mathcal{V}^{j^I,c}$ and therefore 4 pairs of geometrically coincident vertices. The two negatively signed vertices (in green) from $\mathcal{C}_c^{j^I}$ are matched to the vertices which came from an interior connected component (marked in cyan) and not the ones which did not (marked in pink).
    }
    \label{fig:adjacency_def}
\end{figure}

Recall from Section \ref{section:regioncopy} that in regions with more than one copy, we create a copy $\mathcal{V}^{j^I,c}$ for each connected component $c$ of vertices in $\xx^V$ located in region $j^I$ with negative sign.
We use $\mathcal{C}^{j^I}_c$ to denote the collection of these nodes in the connected component $c$.
For regions with only one copy, $\mathcal{C}^{j^I}_0$ instead denotes the collection of all vertices in $\xx^V$ located in region $j^I$ with negative sign, as we do not generate connected components in this case.
Not that for these single copy regions, the vertices of $\mathcal{C}^{j^I}_0$ need not be connected (see the geometry of Figure \ref{fig:adjacency_def}, where the vertices $\mathcal{C}^{j^I}_0$ are composed of two connected components on the outer and inner boundaries).
We merge vertices of $\mathcal{V}^{j^I,c}$ with vertices in $\mathcal{C}^{j^I}_c$ using the merge described in Section \ref{subsection:merging}.
Before this merge, we first perform a preliminary merge of vertices in $\mathcal{C}^{j^I}_c$ which are geometrically coincident.
Here, two vertices of $\xx^V$ are adjacent if they are geometrically coincident and both in $\mathcal{C}^{j^I}_c$.
The effect of this preliminary merge is to close unwanted interior voids without `sewing' the exterior and without merging topologically distant vertices of $\mathcal{V}^S$, as shown in Figure \ref{fig:node_merging}.
The merge between the vertices of $\mathcal{V}^{j^I,c}$ and $\mathcal{C}^{j^I}_c$ is then defined by the following adjacency.
Vertices of $\mathcal{V}^{j^I,c}$ and $\mathcal{C}^{j^I}_c$ are adjacent if they are geometrically coincident and the vertex of $\mathcal{V}^{j^I,c}$ was created from an interior connected component of vertices in the $\mathcal{V}^{j^I,0}_\ii$ that gave rise to $\mathcal{V}^{j^I,c}$ via the merge described in Section~\ref{section:regioncopy}.
Here, an interior connected component is one that contains the center vertex (as opposed to one of the surrounding 26 vertices) introduced in the creation of $\mathcal{V}^{j^I,0}_\jj$ for some grid node $\xx_\jj$ in the region $j^I$.
This requirement effectively means that vertices of $\mathcal{C}^{j^I}_c$ should only merge to the those vertices of $\mathcal{V}^{j^I,c}$ which are actually interior to the region, and not the vertices which are overlapping from a topologically far part of $\mathcal{V}^{j^I,c}$. We illustrate this in Figure \ref{fig:adjacency_def}. Note that after this merge has been performed, we update the indices in $\mathcal{C}^{j^I}_c$ accordingly as this set will be used in latter steps of the merging procedure.

\begin{figure}[h]
    \includegraphics[draft=\mydraft,width=\columnwidth]{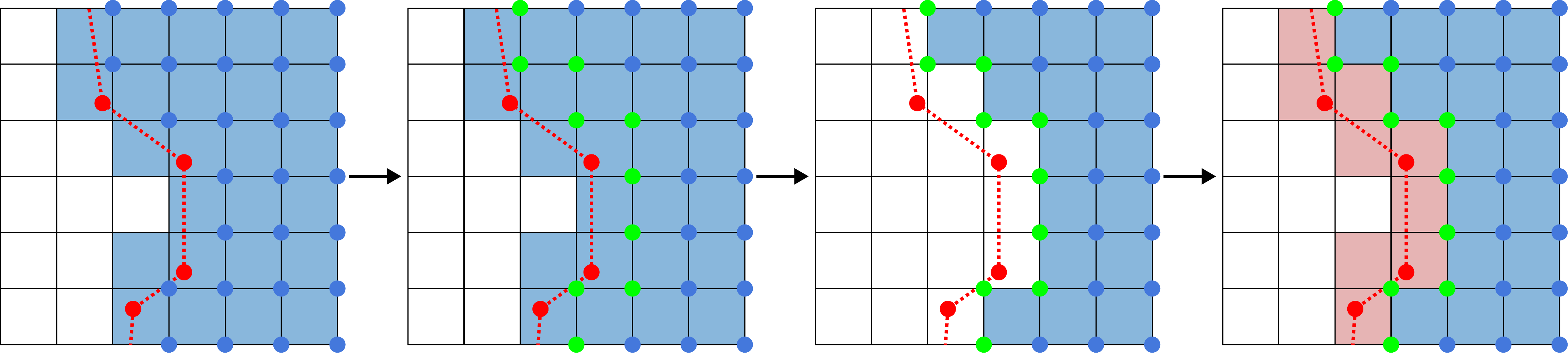}
    \caption{
        {\textbf{Merge with boundary.}} We illustrate the process of Section \ref{subsection:mergewithboundary} following the preliminary merge of negatively signed vertices. First, specific vertices of $\mathcal{V}^{j^I,c}$ are merged with vertices of $\mathcal{C}^{j^I}_c$. Next, hexahedra to be replaced are removed from the $\mathcal{V}^{j^I,c}$. Finally, copies of hexahedra from $\mathcal{V}^S$ are added to this mesh.
    }
    \label{fig:boundary_merge}
\end{figure}

We next use a strategy different to that in Section~\ref{subsection:merging} for merging hexahedral elements in $\Vs$ to their geometrically coincident counterparts in$\mathcal{V}^{j^I,c}$.
This modified merging strategy is designed to prefer the structure of $\Vs$ over that in $\mathcal{V}^{j^I,c}$.
For instance, if two hexahedra of $\Vs$ are geometrically coincident but share only vertices on one face, then they will still have this connectivity after merging to $\mathcal{V}^{j^I,c}$.
We merge the hexahedra in $\Vs$ incident to the vertices in $\mathcal{C}^{j^I}_c$ to their geometrically coincident counterparts in $\mathcal{V}^{j^I,c}$.
Specifically, for each vertex $\xx_i^V$ with $i\in \mathcal{C}^{j^I}_c$ and $k^{\Vs}_i\in\mathcal{I}^{\Vs}_i$, the hexahedron $\lfloor \frac{k^{\Vs}_i}{8}\rfloor$ is marked for merging.
We denote the collection of hexahedra in $\Vs$ marked to be merged with their counterparts in copy $c$ of region $j^I$ as $\mathcal{I}^{j^I,c}_H$.
Note that it is possible that some hexahedra of $\Vs$ are not included in any such collection.
To perform this modified merging procedure, we first remove hexahedra from $\mm^{V^{j^I,c}}$ that are geometrically coincident with a hexahedron from $\mathcal{I}^{j^I,c}_H$ and incident to a vertex in $\mathcal{C}^{j^I}_c$.
Note that a hexahedron in $\mm^{V^{j^I,c}}$ can only be incident to a node in $\mathcal{C}^{j^I}_c$ after the merge described in the previous paragraph has been completed.
Next, copies of the hexahedra in $\mathcal{I}^{j^I,c}_H$ are added to $\mm^{V^{j^I,c}}$.
The process following the preliminary merge is outlined in Figure \ref{fig:boundary_merge}.

\subsection{Overlap Lists}
We next merge differing regions $\mathcal{V}^{j^I_0,c}$ along their appropriately defined common boundaries.
The boundary region between any two region copy meshes $\mathcal{V}^{j^I_0,c_0}$ and $\mathcal{V}^{j^I_1,c_1}$ is grown from seeds which we define by hexahedra in the respective meshes that are equal and in $\Vs$.
For example, suppose that $\mathcal{V}^{j^I_0,c_0}$ and $\mathcal{V}^{j^I_1,c_1}$ contain such a hexahedron. In this case there are hexahedra with indices $h^{V^{j^I_0,c_0}}_{e_0}, h^{V^{j^I_1,c_1}}_{f_0}\in\mathbb{N}$ sharing the same vertices as a hexahedron in $\Vs$ with index $h^{V^S}_{g_0}\in\mathbb{N}$ such that 
\begin{align}
    m^{V^{j^I_0,c_0}}_{8h^{V^{j^I_0,c_0}}_{e_0} + i^e} = m^{V^{j^I_1,c_1}}_{8h^{V^{j^I_1,c_1}}_{f_0} + i^e}= m^{V^{S}}_{8h^{V^{S}}_{g_0} + i^e}, \  i^e\in\left\{0,1,\hdots,7\right\}.\label{eq:hex_copy}
\end{align}
When these hexahedra exist in two region copies ${j^I_0,c_0}$ and ${j^I_1,c_1}$ we use the notation $\qq=({j^I_0,c_0},{j^I_1,c_1})$ to denote a pair of region copies with common boundary (that which will eventually merge).
We define $\ss^\qq_0=(h^{V^{j^I_0,c_0}}_{e_0},h^{V^{j^I_1,c_1}}_{f_0})$ as a seed between the pair of region copies.
Furthermore, we use
$\mb{p}^\qq=[\ss^\qq_0,\ldots,\ss^\qq_{N^\qq_s-1}]$ to denote the collection of all such seeds between $j^I_0,c_0$ and $j^I_1,c_1$ with $N^\qq_s$ being the number of seeds.
This collection, which we call an overlap list, is grown into the complete overlapping common boundary between $j^I_0,c_0$ and $j^I_1,c_1$.

\begin{figure}[h]
    \includegraphics[draft=\mydraft,width=\columnwidth]{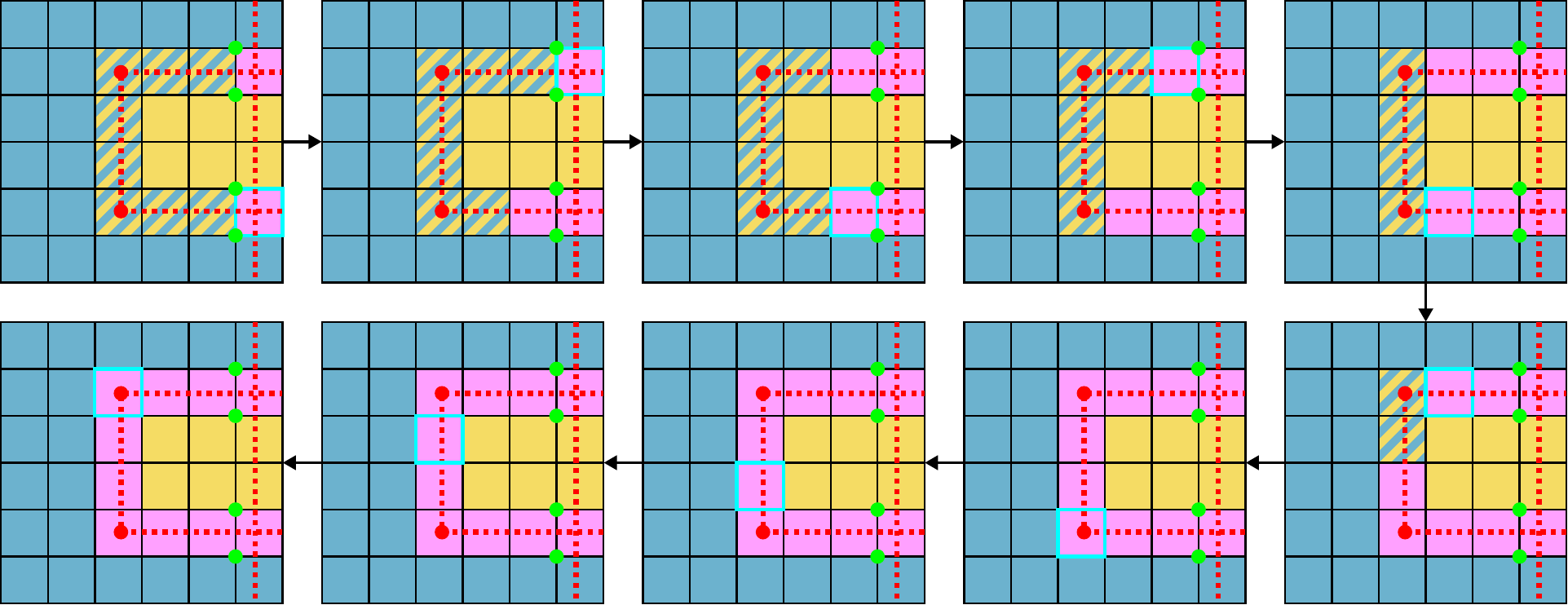}
    \caption{
        {\textbf{Overlap lists.}} A closeup of the overlap region from the geometry of Figure \ref{fig:adjacency_def} is shown here. At the upper left, the seeds for the overlap between the two copies are shown in purple, as well as the incident negative vertices (green) to the seeds from each copy. At each step, the current seed is marked with a cyan border. New geometrically coincident neighbors of the seed hexahedra are then added in the next step. When all seeds have been traversed, the process stops.
    }
    \label{fig:overlap_list}
\end{figure}

We expand the initial seed collections $\mb{p}^\qq$ by first marking background grid cells geometrically coincident with hexahedra in the seeds as being visited.
Then, starting with the seed $\ss^\qq_0$, we compute the neighbor hexahedra of each hexahedron in the seed (the neighbors of a hexahedron are those which share a common vertex).
Geometrically coincident neighbors of the two hexahedra in the seed are added to $\mb{p}^\qq$ if the background grid cell to which they are geometrically coincident is unvisited.
We then mark the cell as visited, and continue until every seed has been processed in this way.
At the end of this expansion, $\mb{p}^\qq$ is a list of overlapping hexahedra that will be used to sew the regions together.
We illustrated this process in Figure~\ref{fig:overlap_list}.

\subsection{Deduplication}
\begin{figure}[h]
    \includegraphics[draft=\mydraft,width=\columnwidth]{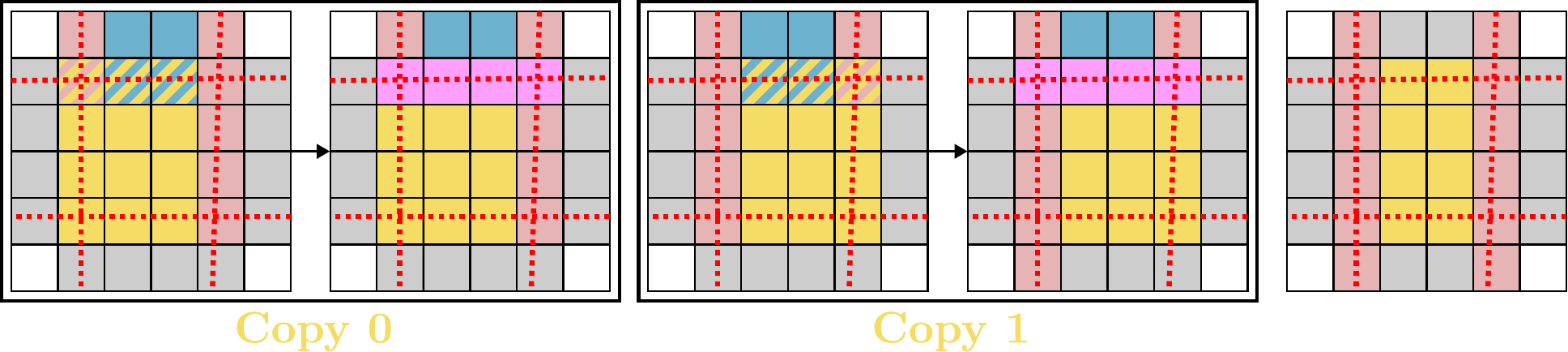}
    \caption{
        {\textbf{Deduplication.}} We show two of the four copies of the central region (yellow), corresponding to the right and left segments of $\mathcal{V}^S$. Each of copies 0 and 1 create an overlap list with the upper region (blue). The overlap list for copy 0 creates a pair between a non-boundary yellow hexahedron and a boundary hexahedron from the blue region. This boundary hexahedron is in a pair with a boundary hexahedron of copy 1, allowing us to deduce that copies 0 and 1 of the yellow region are duplicates. We then repeat the boundary merge process to create a deduplicated copy with complete boundary information.
    }
    \label{fig:deduplication}
\end{figure}

As mentioned in Section \ref{section:regioncopy}, the number of copies is generally an over count.
We use the overlap lists $\pp^\qq$ to deduce which copies $c$ of a region $j^I$ are redundant.
For each hexahedron $h^S_e$ in $\Vs$, we create a list of hexahedra from geometrically coincident counterparts in interior region copies.
This list is formed by considering each pair $\qq$: if either hexahedron in a seed of $\mb{p}^\qq$ is a copy of $h^S_e$ (i.e. it uses the same vertices in $\xx^S$ as in Equation~\eqref{eq:hex_copy}), both hexahedra in the seed are added to the list associated with $h^S_e$.
Note that while the hexahedron pairs of the initial seeds in $\mb{p}^\qq$ are both copies of hexahedra from $\mathcal{V}^S$ in accordance with Equation~\eqref{eq:hex_copy}, subsequent seeds added during the overlap process may have both, one, or neither hexahedra equal to copies of hexahedra from $\mathcal{V}^S$.
Should any list for any hexahedron $h^S_e$ in $\mathcal{V}^S$ contain hexahedra from multiple copies $c_0$ and $c_1$ of the same region $j^I$, copies $c_0$ and $c_1$ are considered to be redundant duplicates of each other.
Redundant copies are merged using the process of Section \ref{subsection:mergewithboundary}. This process is shown in Figure \ref{fig:deduplication}.

For each region, we compute connected components of its copies using duplication as the notion of adjacency.
For each connected component of copies, we take the copy with the smallest index $c_i$ as the representative copy.
However, this copy's mesh only has the vertices of the component $c_i$.
Likewise, only copies of the hexes in $\mathcal{I}^{c_i}_H$ are in $\mathcal{V}^{j^I,c_i}$.
We remedy this by repeating the merge with boundary process of Section \ref{subsection:mergewithboundary} on updated data.
Specifically, we replace the connected component $c_i$ of vertices with the union of all components $c_j$ for copies in the connected component of copies.
We then form an updated collection of incident hexahedra $\mathcal{I}^{c_i}_H$ before repeating the boundary merge process.
Finally, we update the overlap lists.
Any overlap list corresponding to a duplicated copy is recreated using the minimum representative in place of the original copy to account for updated hexahedron ordering.
Redundant overlap lists resulting from this update are then discarded.

\subsection{Final Merge}
We now merge the vertices of $\xx^V$ using the pattern of Section \ref{subsection:merging} with adjacencies defined by the overlap lists.
For each seed $\ss$ in an overlap list, the geometrically coincident nodes of the two hexahedra in $\ss$ are considered adjacent.
We then create the final mesh $\mathcal{V}$ by combining all of the arrays $\mm^{V^{j^I,c}}$ from copies which are either the minimum representative, or not duplicated.
Recall from Section \ref{subsection:merging} that some hexahedra of $\Vs$ are not copied into any copy's mesh.
We add all such hexahedra to $\mathcal{V}$ to guarantee that $\Vs$ is contained in this final mesh, completing the interior extension region merging process.

\section{Coarsening}\label{section:coarsening}
\begin{figure}[h]
    \includegraphics[draft=\mydraft,width=\columnwidth]{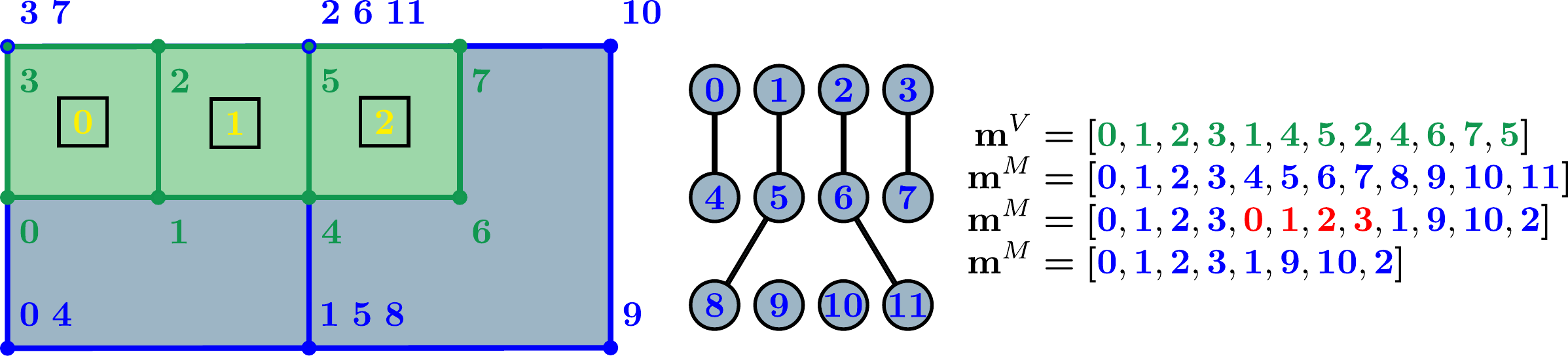}
    \caption{
        {\textbf{Coarsening.}} An example of fine mesh connections. Hexahedra 0 and 1 are totally connected, while hexahedra 1 and 2 are connected by a face. After merging the vertices of the coarse mesh (blue), the duplicated hexahedron (indicated in red) is removed.
    }
    \label{fig:coarsening}
\end{figure}

Our method requires high-resolution (small $\Delta x$) background grids for high-curvature/detailed surfaces. 
We provide a topology-aware coarsening strategy to provide user control over the final volumetric mesh resolution/element counts.
After the hexhedron mesh $\mathcal{V}$ is created, we coarsen the underlying grid by doubling $\Delta x$. 
We then create a maximal coarse mesh $\mathcal{M}$ based on the fine mesh $\mathcal{V}$. 
For each index $m_j^V$ in $\mathcal{V}$, we define the initial connectivity for $\mathcal{M}$ as $m_j^M=j$.
We then bin the center of each fine hexahedron $h^M\in\mathbb{N}^{N^M_e}$ into the coarsened grid and keep track of its multi-dimensional grid index $\ii^{h^M}$.
We initialize the position array $\xx^M$ for $\mathcal{M}$ from the coarse grid cell corners of cell $\ii^{h^M}$.
Specifically, for each hexahedron in $h^M$ in $\mathcal{M}$ we define
$\xx^M_{8h^M+i^e}=\xx^{2\Delta x}_{\ii^{h^M}} + \oo_{i^e}$ where $\oo_{i^e}$ is an offset from the coarse cell center $\xx^{2\Delta x}_{\ii^{h^M}}$ to the eight respective corners of the coarse grid cell $\ii^{h^M}$.
To build the final coarsened mesh, we merge portions of the maximal coarse mesh using Section~\ref{subsection:merging} where adjacencies are defined from a hexahedron-wise notion of connectivity.
Two maximal coarse hexahedra $h^M_0$ and $h^M_1$ are connected if their corresponding fine hexahedra $h^V_0=h^M_0$ and $h^V_1=h^M_1$ share a face $\ff_i^V=\left[f_{i0}^V,f_{i1}^V,f_{i2}^V,f_{i3}^V\right]\in\mathbb{N}^4$ in $\mathcal{V}$. 
We define two types of connection: totally connected and partially connected.
Maximal coarse hexahedra are totally connected if they have the same coarse grid index $\ii^{h^M_0}=\ii^{h^M_1}$ and their corresponding fine hexahedra $h^V_0$ and $h^V_1$ are not geometrically coincident. 
Maximal coarse hexahedra are partially connected if they are connected but are not totally connected.
We define vertex adjacency from our notions of hexahedron connectivity.
If two hexahedra $h^{M}_{0}$ and $h^{M}_{1}$ in the maximal coarse mesh are totally connected, then their eight respective geometrically coincident vertices are defined to be adjacent, i.e. vertex
$m^{M}_{8h^{M}_{0} + i^e}$ is adjacent to vertex $m^{M}_{8h^{M}_{1} + i^e}$, $0\leq i^e<8$. 
If they are partially connected, then their corresponding fine hexahedra $h_0^V, h_1^V$ share a face $\ff_i^V = \left[f_{i0}^V,f_{i1}^V,f_{i2}^V,f_{i3}^V\right]$. 
We then identify an analogous face in each of $h_0^V$ and $h_1^V$ which we define in terms of the indices $k^V_{0\alpha},k^V_{1\alpha}$, $\alpha\in\left\{0,1,2,3\right\}$.
Only the vertices corresponding to the analogous face are defined to be adjacent
\begin{align}
    m^{M}_{8h^{M}_{0} + k^V_{0\alpha}} = m^{M}_{8h^{M}_{1} + k^V_{1\alpha}}, \  \alpha \in \left\{0,1,2,3\right\}.
\end{align}
There are two cases that define the analogous face. 
First, if the fine hexahedron counterparts $h_0^V, h_1^V$ are geometrically coincident, then the analogous face is the one on the analogous side of the coarse hexahedron. If they are not geometrically coincident, then the analogous face is the one geometrically coincident with the fine face defined from $\ff_i^V$.
The general coarsening procedure is illustrated in Figure \ref{fig:coarsening}.

\section{Hexahedron Mesh To Tetrahedron Mesh Conversion}\label{section:HextoTet}
\begin{figure}[h]
    \includegraphics[draft=\mydraft,width=\columnwidth]{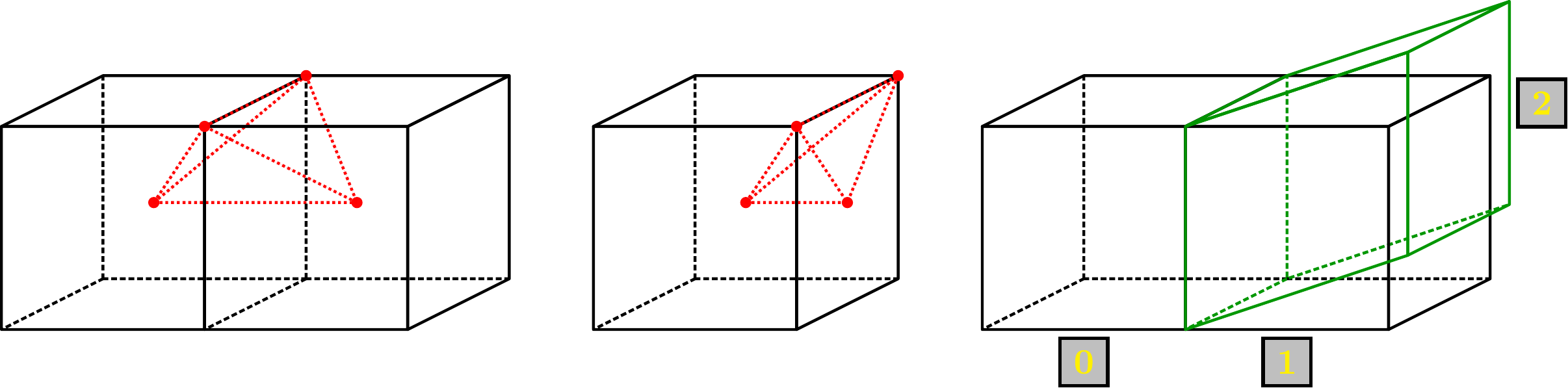}
    \caption{
        {\textbf{Hexahedra tetrahedralization}}. \textit{(Left)} a standard interior face in $\mathcal{V}$. The centers of the two incident hexahedra are combined with two face vertices to form the tetrahedra (red). \textit{(Middle)} a standard boundary face uses a face center instead of the missing incident hexahedron center. \textit{(Right)} a non-standard interior face is shown. The right-most incident hexahedra are geometrically coincident. We form hexahedra pairs/faces (0,1), (0,2) and treat them respectively as standard interior, as in the left-most image.
    }
    \label{fig:hextotet}
\end{figure}

We design a topologically-aware BCC-based approach for the creation of a tetrahedron mesh $\mathcal{T}$ from the hexahedron mesh $\mathcal{V}$.
We initialize the particle array for the tetrahedron mesh $\xx^T$ to be the same as $\xx^V$, but we add a new vertex in the center of each hexahedron and each boundary face.
Tetrahedra are computed from the faces in the mesh $\mathcal{V}$.
Normally a face in $\mathcal{V}$ would have one (boundary face) or two (interior face) incident hexahedra.
However, since $\mathcal{V}$ is comprised of many geometrically coincident hexahedra there are more cases.
We classify them as: standard boundary face (one incident hexahedraon), standard interior face (two non-geometrically coincident incident hexahedra), non-standard interior (more than two incident hexahedra, some geometrically coincident and some not geometrically coincident) and non-standard boundary (more than one incident hexahderon, all geometrically coincident).
Each face contributes four tetrahedra to $\mathcal{T}$ in the case of standard boundary and standard interior faces.
The tetrahedra consist of two vertices from the face and the cell centers on either side of the face in the case of standard interior faces.
In the case of standard boundary faces, the face center is used in place of the second hexahedron center.
For non-standard interior faces, we take all pairs of non-geometrically coincident incident hexahedra and add tetrahedra as if their common face was a standard interior face.
For non-standard boundary faces, tetrahedra are added for each incident hexahedron as if it were incident to a standard boundary face.
We illustrate this procedure in Figure \ref{fig:hextotet}.

\section{Examples}\label{section:examples}

We consider a variety of examples in both two and three dimensions.  To illustrate the capabilities of the final mesh connectivites, we treat the objects as deformable solids and run a finite element (FEM) simulation \cite{sifakis:2012:course}.  Performance statistics for the 3D examples are presented in Table \ref{tbl:perf}.  All experiments were run on a workstation with a single Intel\textsuperscript{\textregistered} Core\texttrademark{} i9-10980XE CPU at 3.00GHz.

\subsection{2D Examples}

\subsubsection{Single Overlap}

\begin{figure}
    \centering
    \begin{subfigure}[b]{0.24\columnwidth}
        \includegraphics[draft=\mydraft,width=\linewidth,trim={300px 0 250px 0},clip]{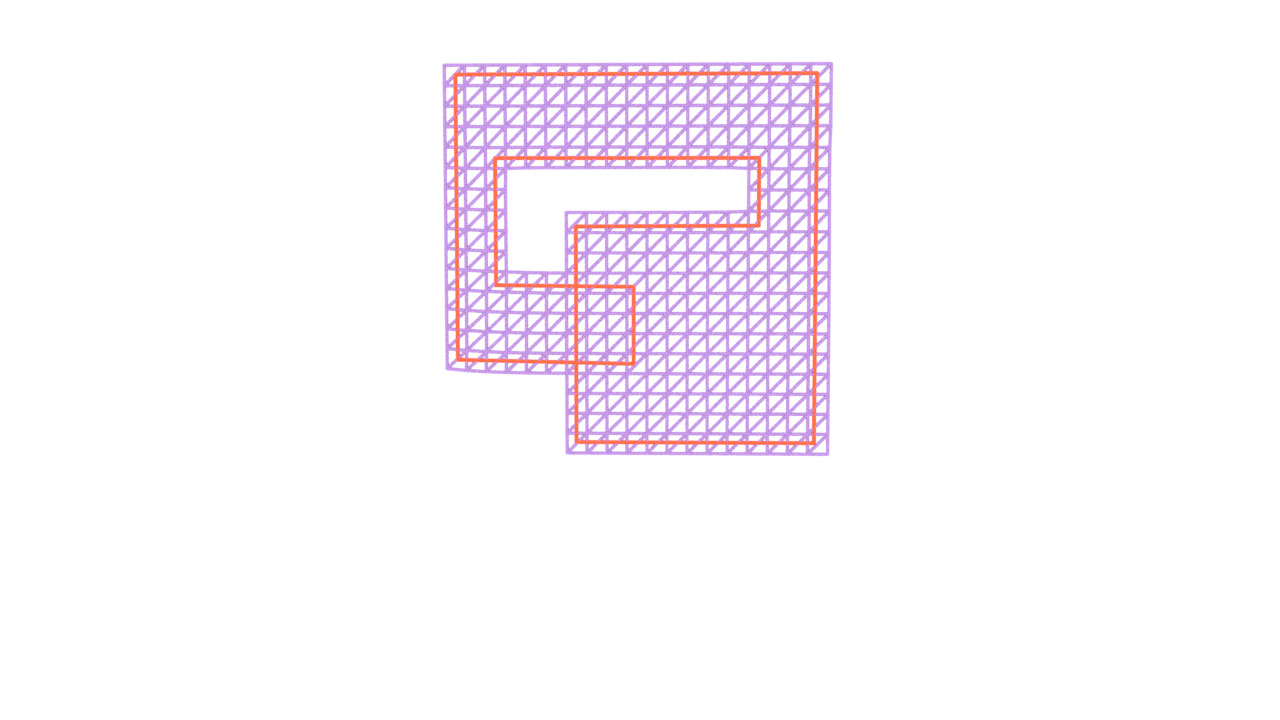}
        \caption{Frame 0}
    \end{subfigure}
    \begin{subfigure}[b]{0.24\columnwidth}
        \includegraphics[draft=\mydraft,width=\linewidth,trim={300px 0 250px 0},clip]{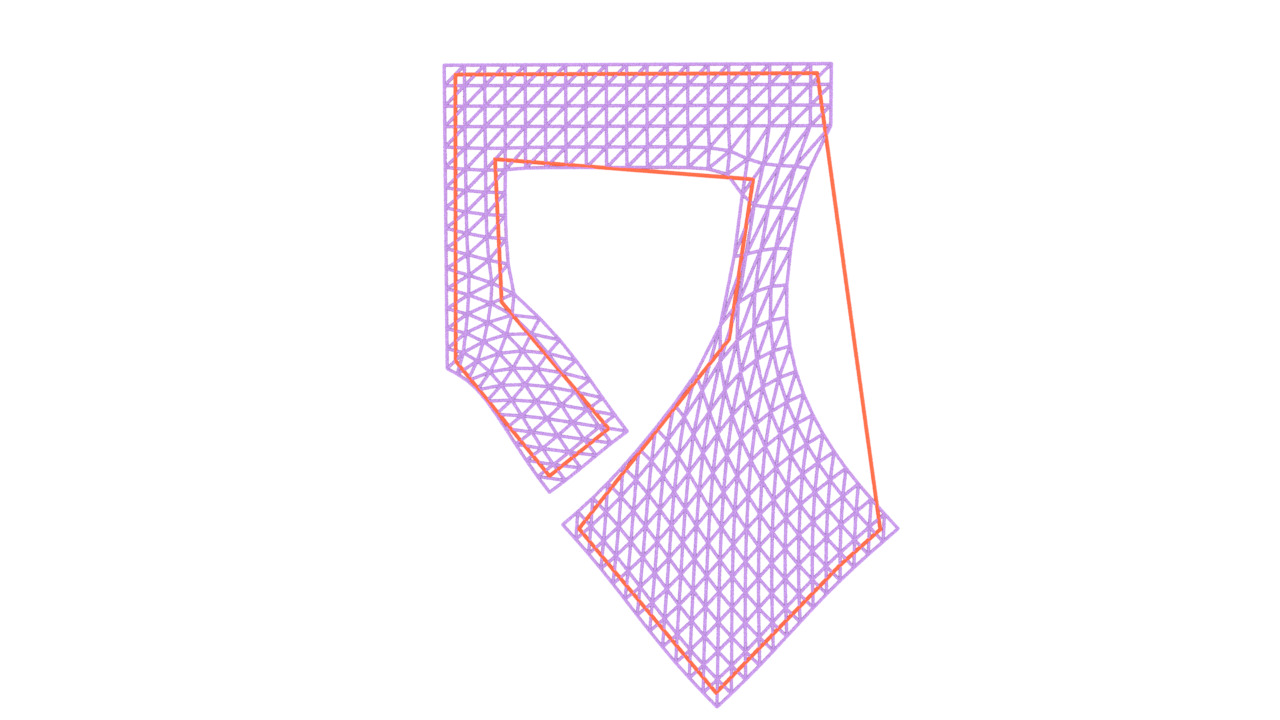}
        \caption{Frame 11}
    \end{subfigure}
    \begin{subfigure}[b]{0.24\columnwidth}
        \includegraphics[draft=\mydraft,width=\linewidth,trim={300px 0 250px 0},clip]{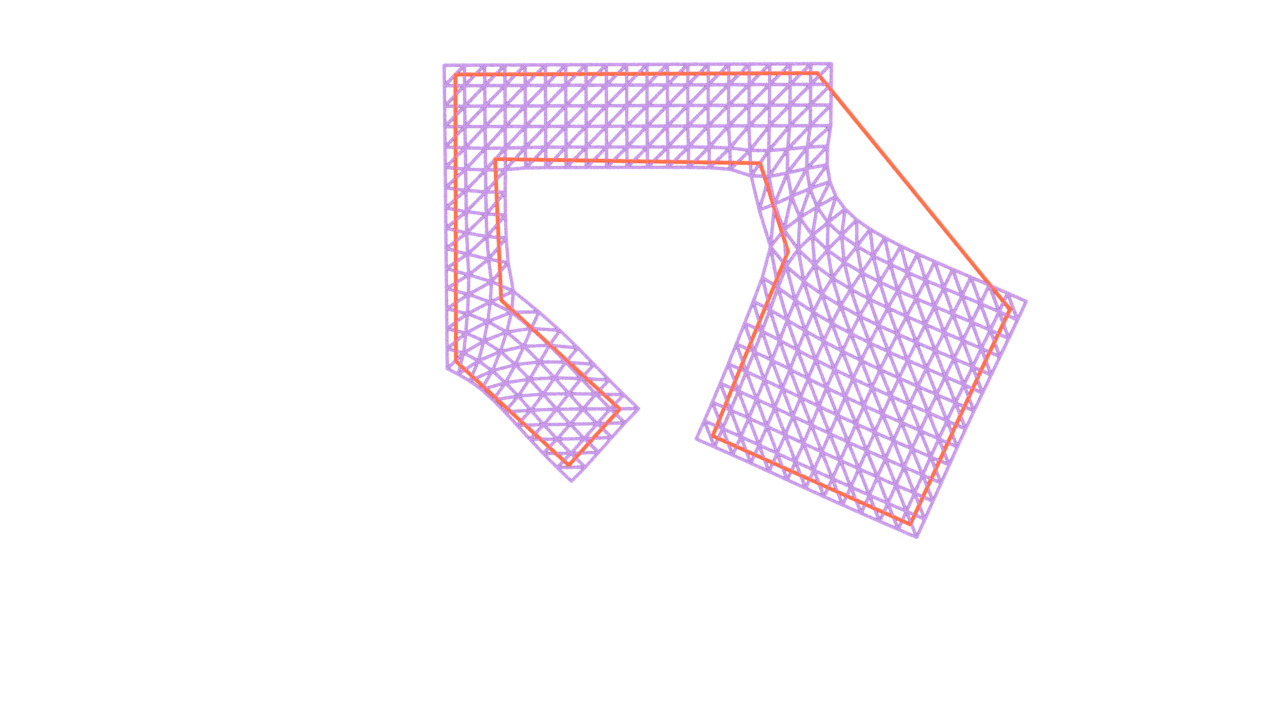}
        \caption{Frame 27}
    \end{subfigure}
    \begin{subfigure}[b]{0.24\columnwidth}
        \includegraphics[draft=\mydraft,width=\linewidth,trim={300px 0 250px 0},clip]{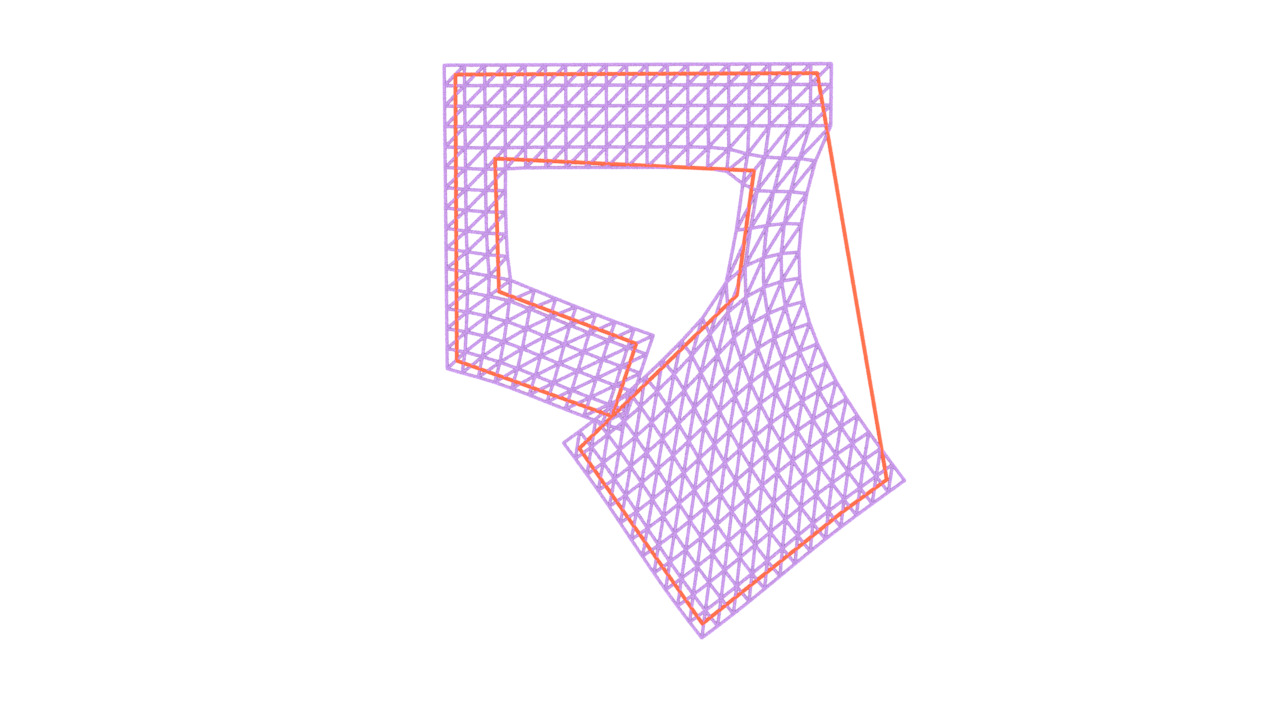}
        \caption{Frame 60}
    \end{subfigure}
    \caption{A self-intersecting shape is suspended from a ceiling.  The geometry deforms under gravity, and both sides freely move regardless of the initial overlap.}
    \label{fig:single-overlap}
\end{figure}

Figure \ref{fig:single-overlap} shows a deformable FEM simulation using a volumetric mesh produced by our algorithm.  As evidenced by the geometry's ability to separate and freely move, our algorithm produces a mesh that properly resolves the single self-intersection present in the initial configuration.

\subsubsection{Ribbon}

\begin{figure}
    \centering
    \begin{subfigure}[b]{0.24\columnwidth}
        \includegraphics[draft=\mydraft,width=\linewidth,trim={400px 0 400px 0},clip]{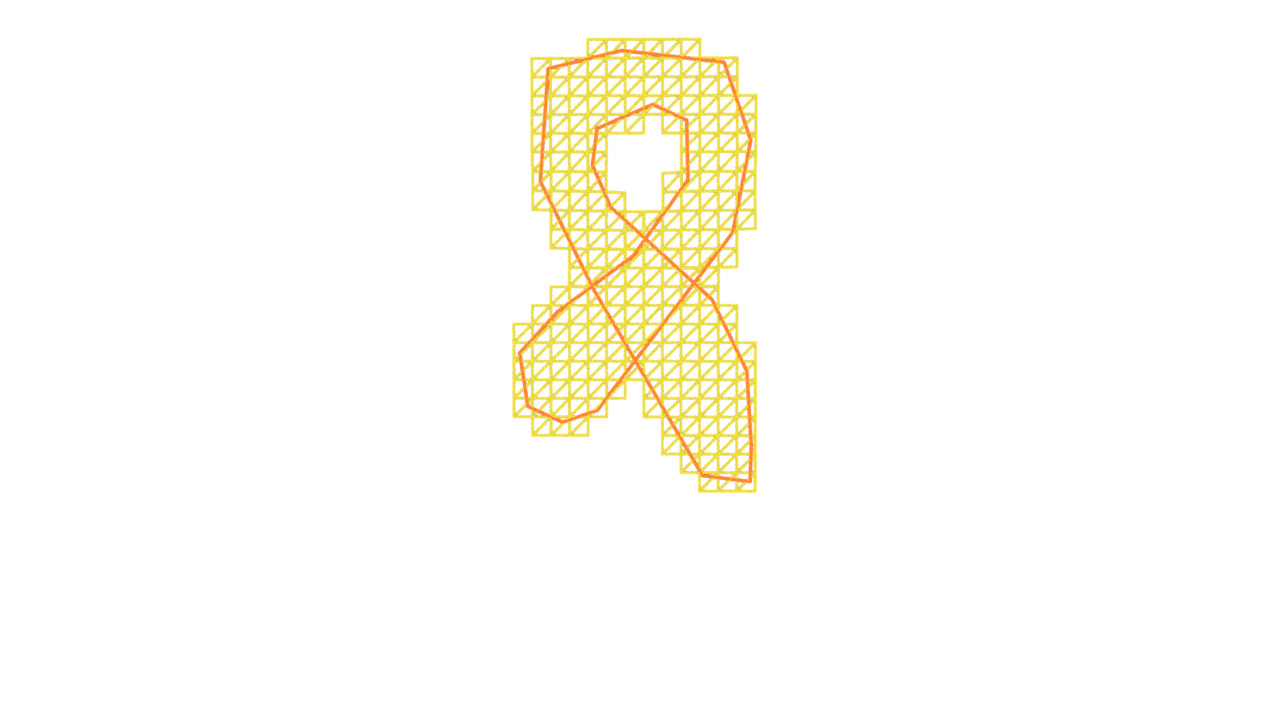}
        \caption{Frame 0}
    \end{subfigure}
    \begin{subfigure}[b]{0.24\columnwidth}
        \includegraphics[draft=\mydraft,width=\linewidth,trim={400px 0 400px 0},clip]{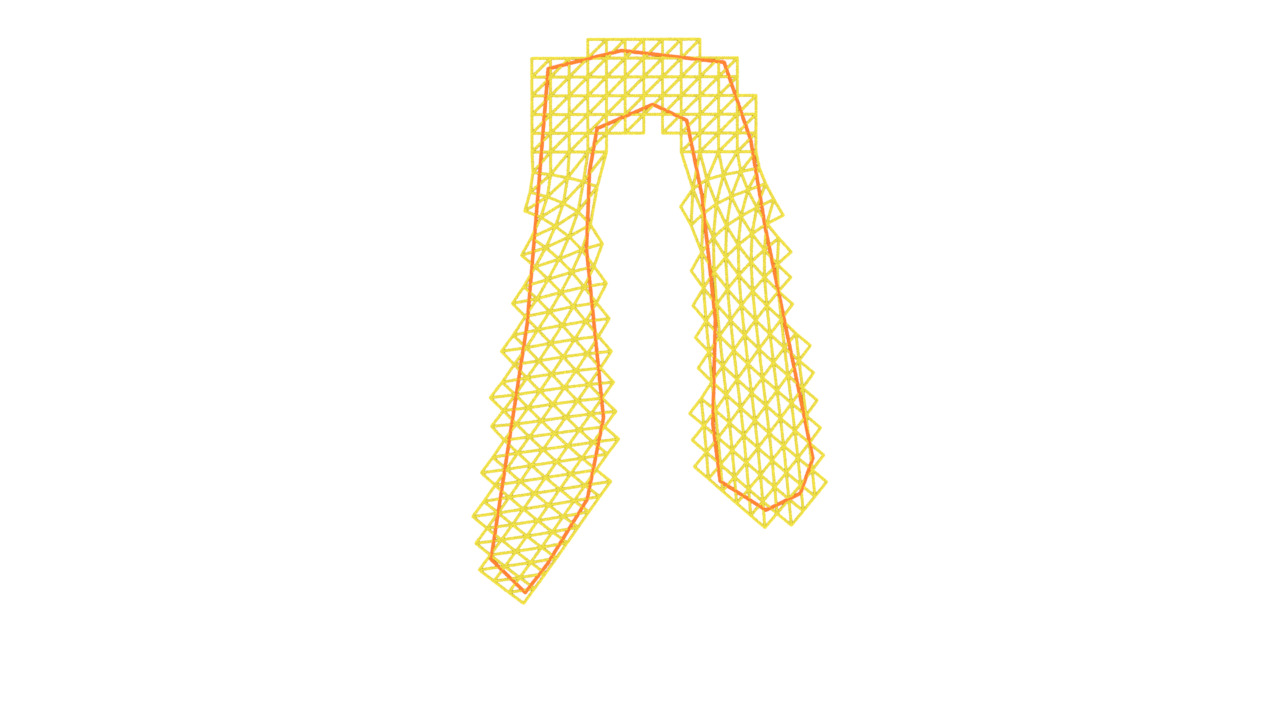}
        \caption{Frame 14}
    \end{subfigure}
    \begin{subfigure}[b]{0.24\columnwidth}
        \includegraphics[draft=\mydraft,width=\linewidth,trim={400px 0 400px 0},clip]{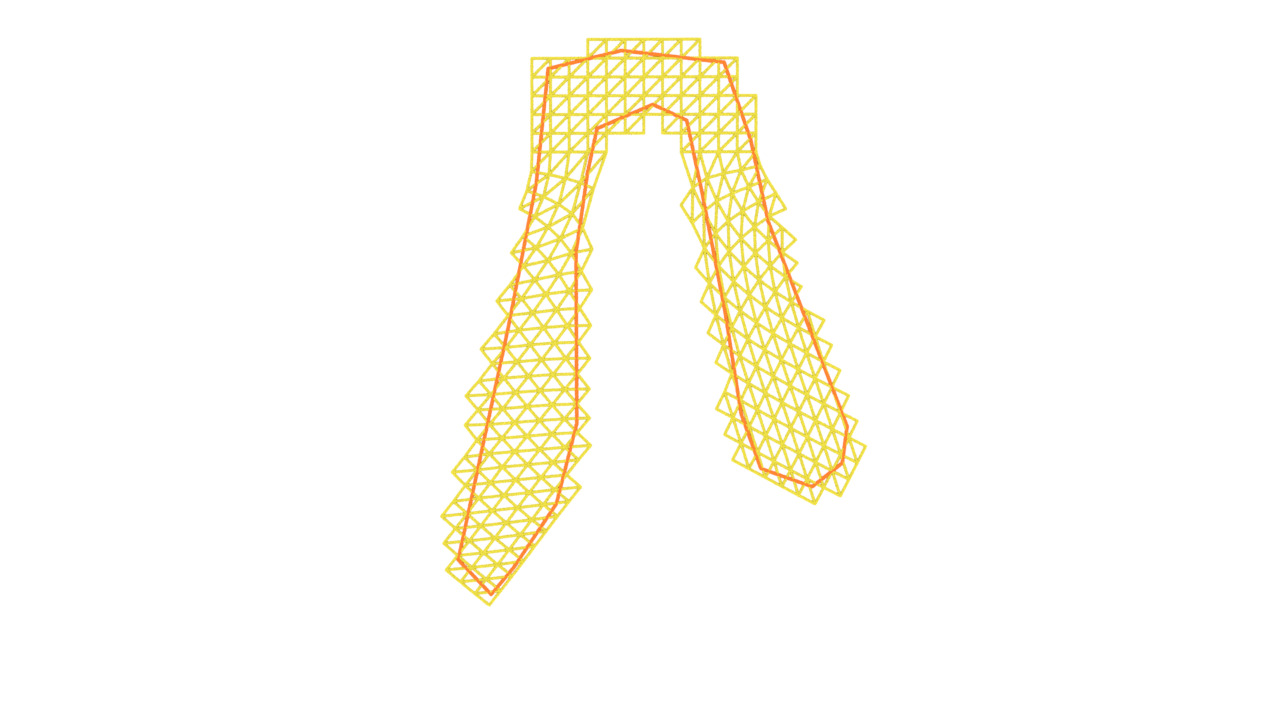}
        \caption{Frame 59}
    \end{subfigure}
    \begin{subfigure}[b]{0.24\columnwidth}
        \includegraphics[draft=\mydraft,width=\linewidth,trim={400px 0 400px 0},clip]{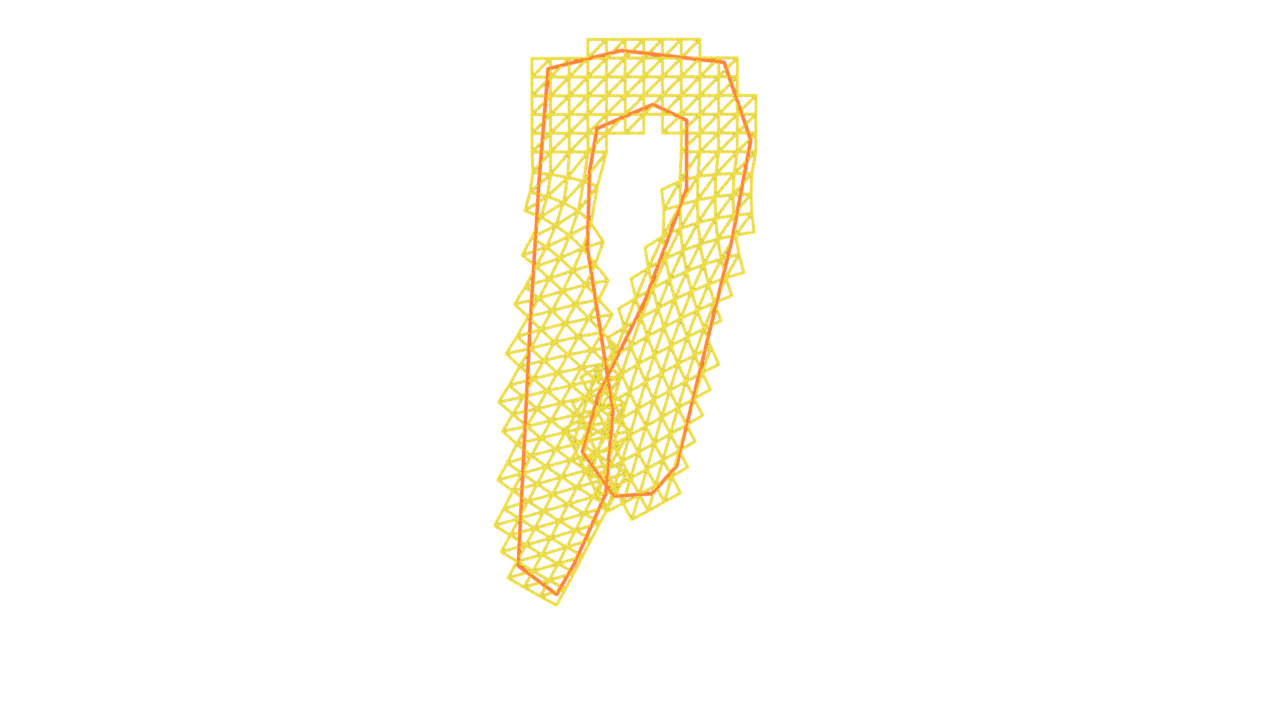}
        \caption{Frame 74}
    \end{subfigure}
    \caption{A ribbon with a more complicated initial self-intersection is also treated properly by our method.}
    \label{fig:ribbon}
\end{figure}

Our algorithm can also handle more complex self-intersections.  In Figure \ref{fig:ribbon}, one end of a ribbon shape passes through the other, partitioning the surface into several components.  These intersections are successfully resolved, and the mesh is allowed to move as in the previous example.

\subsubsection{Face}

\begin{figure}
    \centering
    \begin{subfigure}[b]{0.24\columnwidth}
        \includegraphics[draft=\mydraft,width=\linewidth,trim={400px 150px 400px 0},clip]{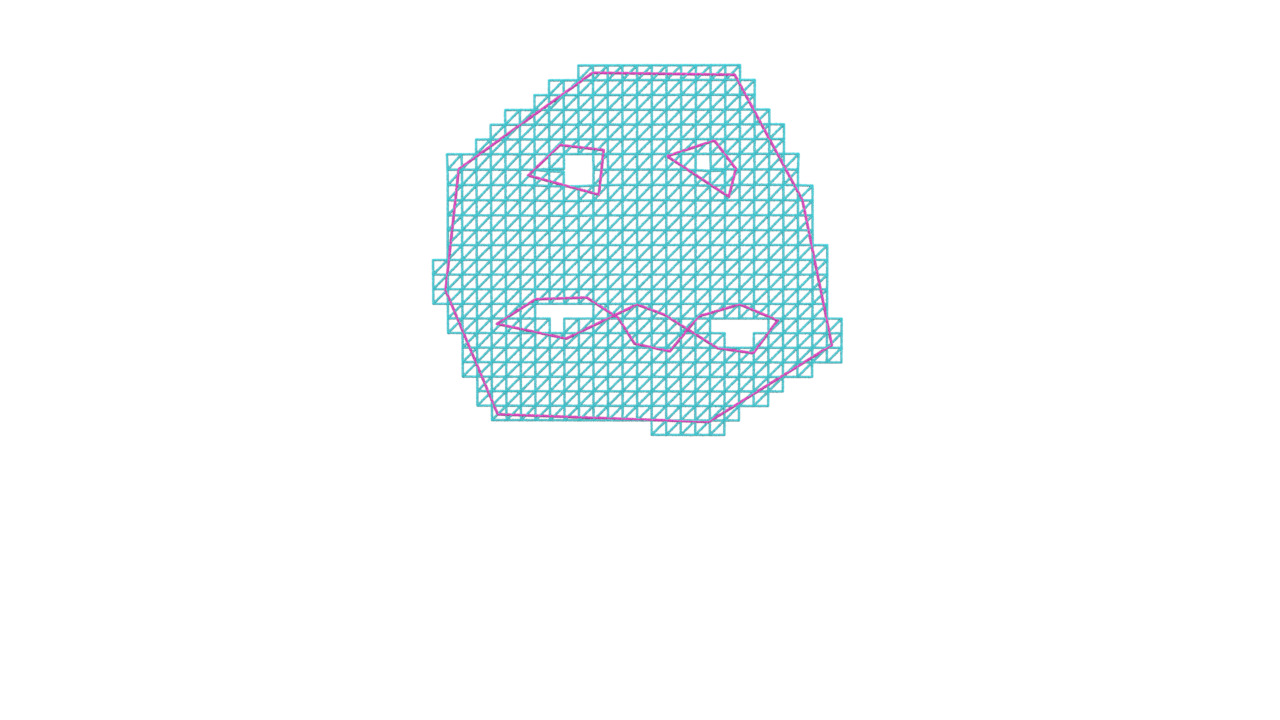}
        \caption{Frame 0}
    \end{subfigure}
    \begin{subfigure}[b]{0.24\columnwidth}
        \includegraphics[draft=\mydraft,width=\linewidth,trim={400px 150px 400px 0},clip]{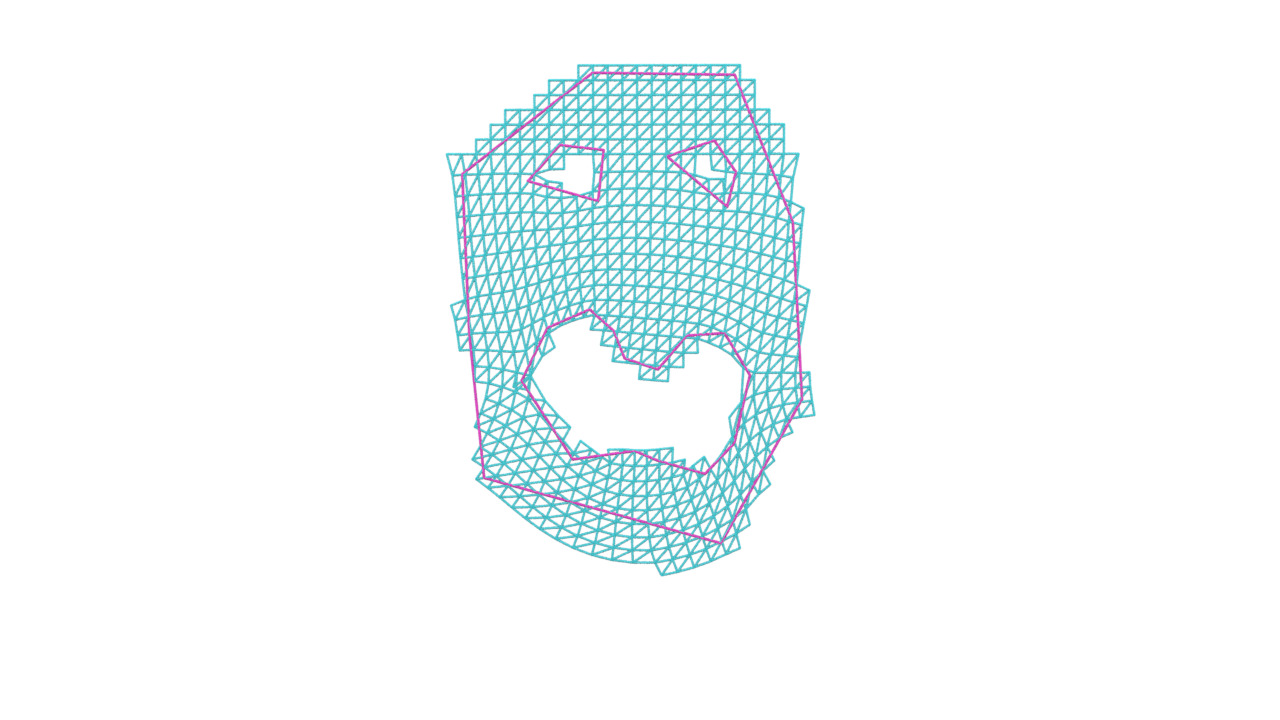}
        \caption{Frame 8}
    \end{subfigure}
    \begin{subfigure}[b]{0.24\columnwidth}
        \includegraphics[draft=\mydraft,width=\linewidth,trim={400px 150px 400px 0},clip]{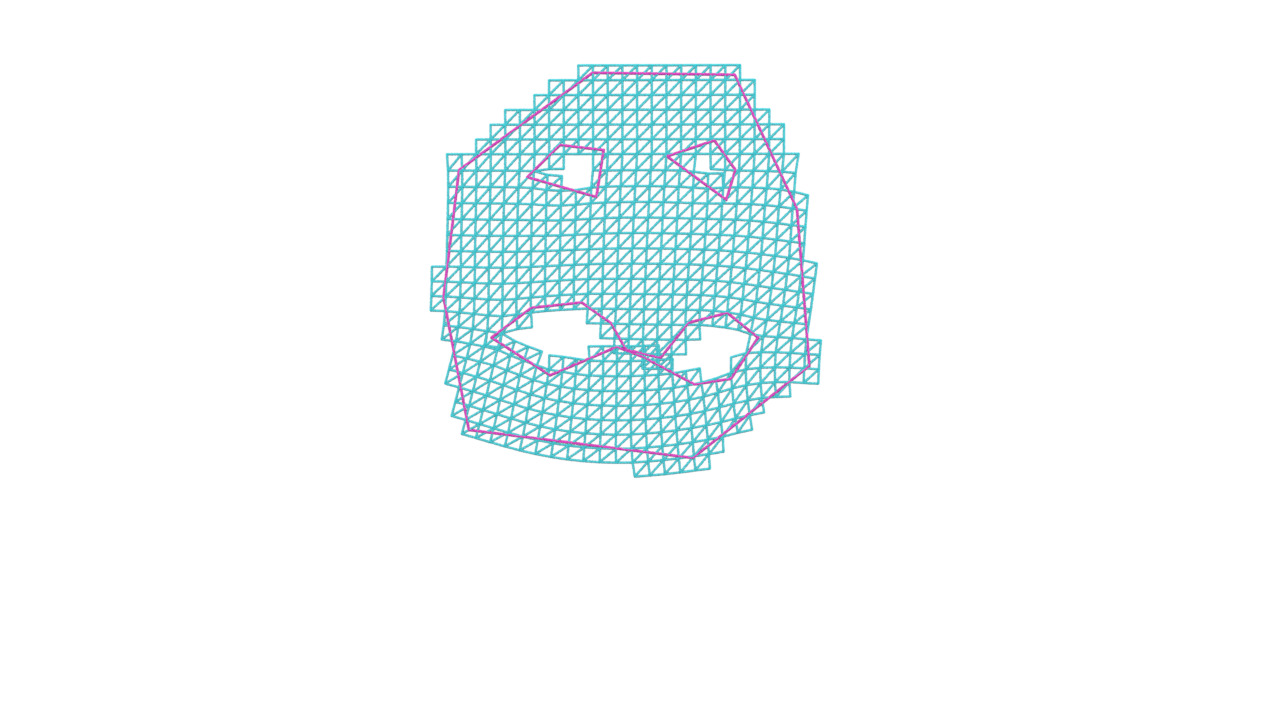}
        \caption{Frame 21}
    \end{subfigure}
    \begin{subfigure}[b]{0.24\columnwidth}
        \includegraphics[draft=\mydraft,width=\linewidth,trim={400px 150px 400px 0},clip]{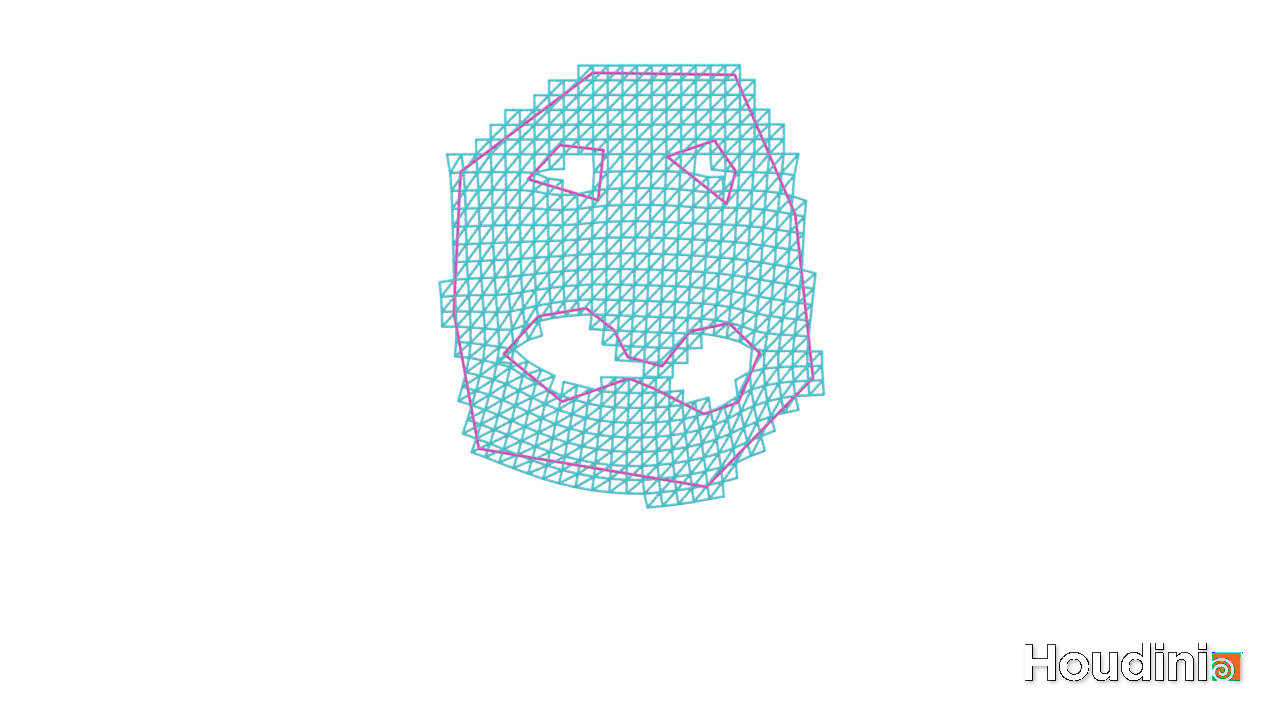}
        \caption{Frame 92}
    \end{subfigure}
    \caption{A face with multiple boundary components and initially self-intersecting lips is successfully animated.}
    \label{fig:face}
\end{figure}

Figure \ref{fig:face} demonstrates a similar scenario.  In this case, the lips of the face geometry initially overlap; and, as an added challenge, the boundary of the input geometry consists of multiple disconnected components.  Our method successfully treats cases like these by design.

\subsection{3D Examples}

\begin{table}[tbp]
    \caption{Performance of generating volumetric meshes using our algorithm for various 3D examples.  All times are in seconds and represent the total runtime of the algorithm.}
    \resizebox{\columnwidth}{!}{%
        \begin{tabular}{@{}l | ll | ll@{}}
            \toprule
            Example & Grid dim. & $\Delta x$ & \# Hex & Time (s) \\
            \midrule
            Two Boxes & 66$\times$64$\times$86 & 0.00955671 & 256368 & 2.80219 \\
            Simple Overlap & 194$\times$64$\times$194 & 0.00328125 & 1606296 & 24.0179 \\
            Double M\"obius & 294$\times$288$\times$64 & 0.0347391 & 903653 & 33.6324 \\
            Twin Bunnies & 162$\times$166$\times$128 & 0.0203027 & 1525821 & 31.1815 \\
            Dragon & 512$\times$690$\times$520 & 0.0708709 & 20110457 & 303.301 \\
            Fancy Ball & 130$\times$132$\times$128 & 2.82671 & 515400 & 25.8388 \\
            Head & 512$\times$830$\times$718 & 0.000501962 & 62444819 & 839.951 \\
            Sacht & 52$\times$104$\times$42 & 4.26331 & 112682 & 9.64888 \\
            \bottomrule
        \end{tabular}
    }
    \label{tbl:perf}
\end{table}

\subsubsection{Two Boxes \& Simple Overlap}

\begin{figure}
    \centering
    \begin{subfigure}[b]{0.49\columnwidth}
        \includegraphics[draft=\mydraft,width=\linewidth]{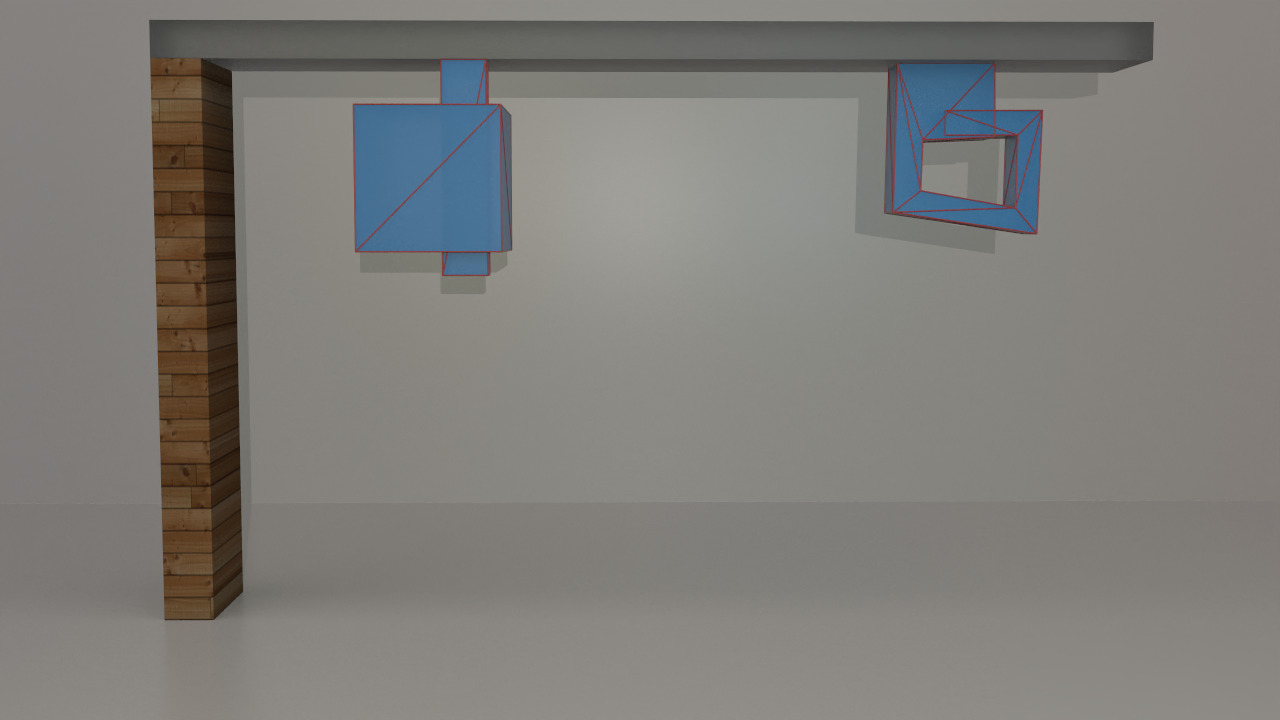}
        \caption{Frame 4}
    \end{subfigure}
    \begin{subfigure}[b]{0.49\columnwidth}
        \includegraphics[draft=\mydraft,width=\linewidth]{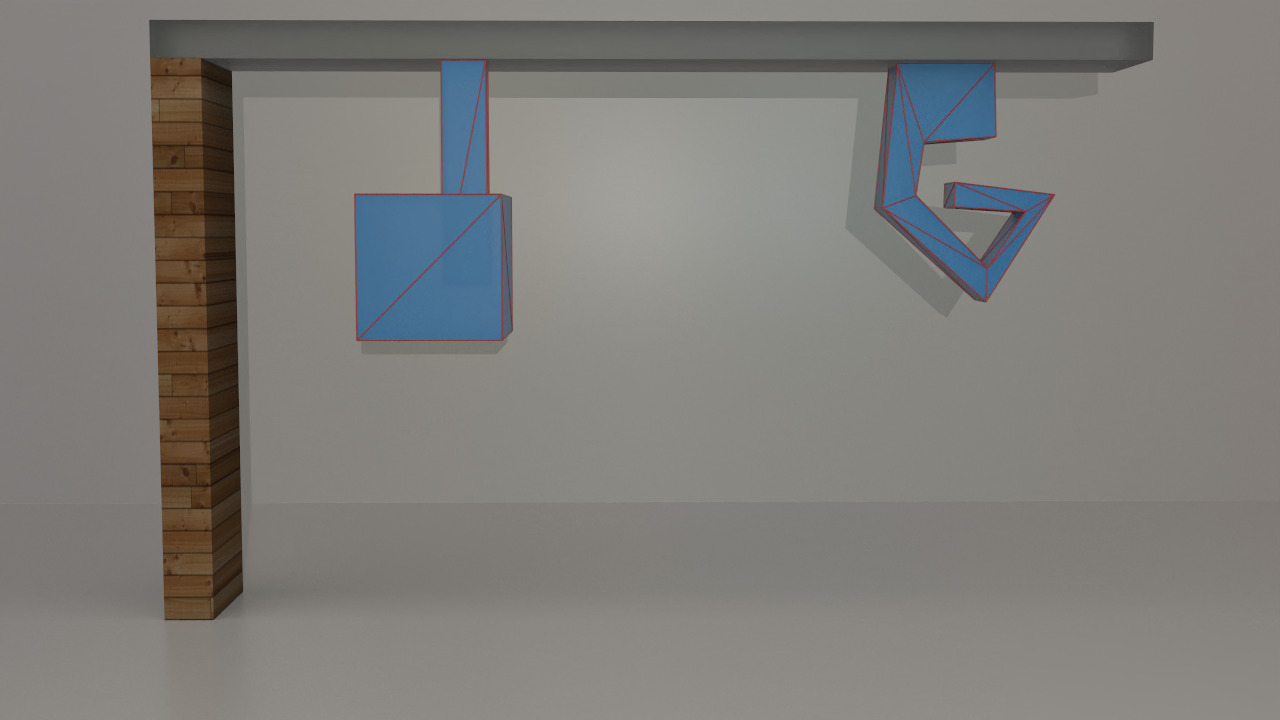}
        \caption{Frame 9}
    \end{subfigure}
    \hfill
    \begin{subfigure}[b]{0.49\columnwidth}
        \includegraphics[draft=\mydraft,width=\linewidth]{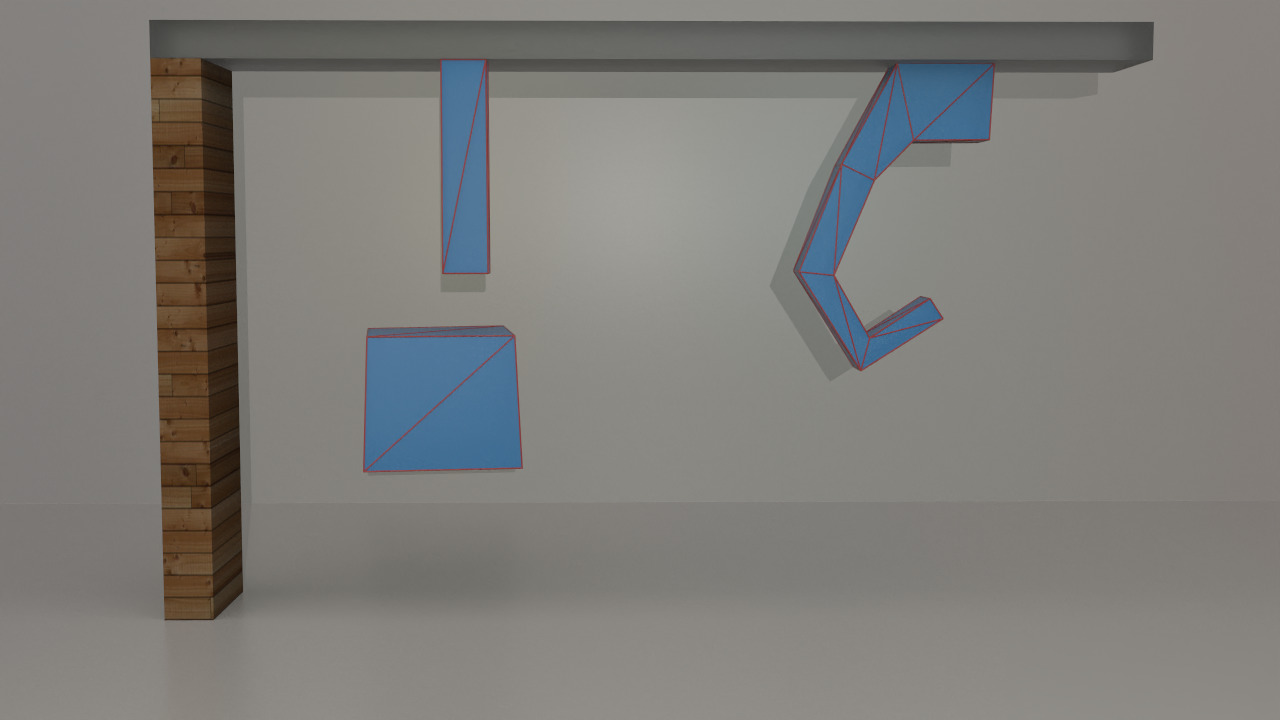}
        \caption{Frame 33}
    \end{subfigure}
    \begin{subfigure}[b]{0.49\columnwidth}
        \includegraphics[draft=\mydraft,width=\linewidth]{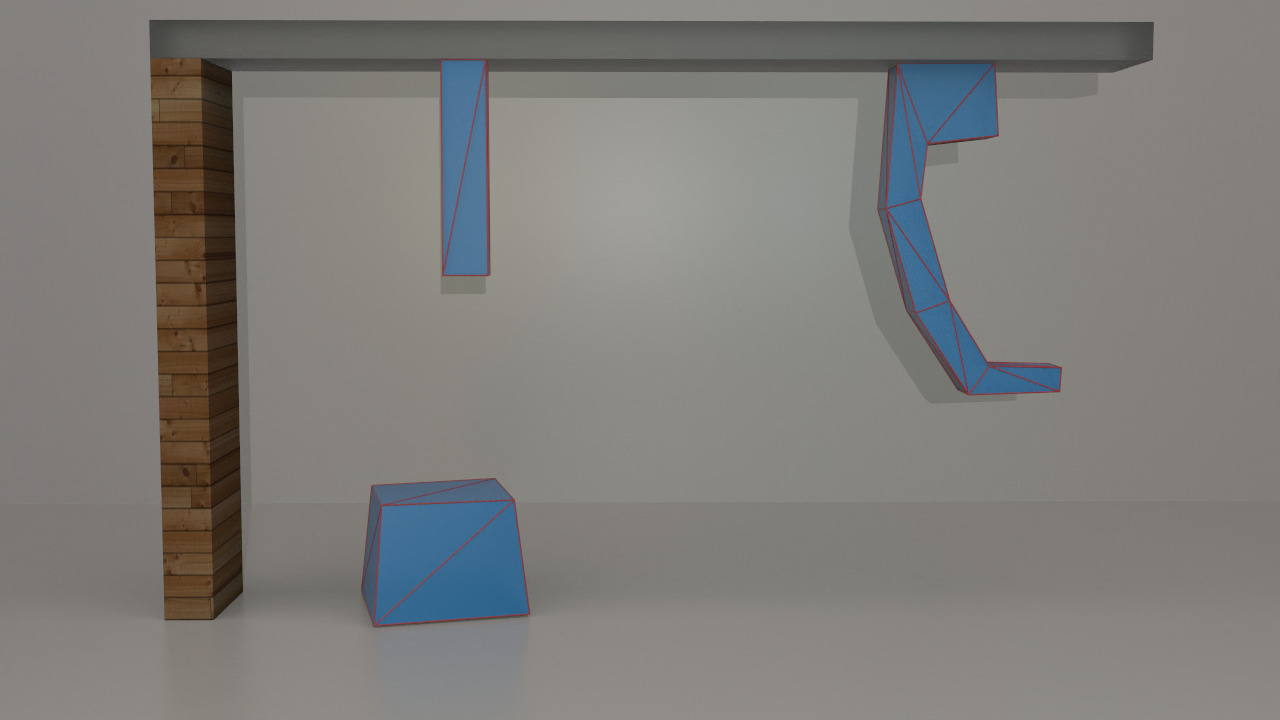}
        \caption{Frame 48}
    \end{subfigure}
    \hfill
    \caption{Simple self-intersecting 3D geometries are able to separate and unfurl with our algorithm.}
    \label{fig:simple-overlap}
\end{figure}

We begin our 3D examples by demonstrating that our algorithm is able to quickly generate consistent meshes for simple self-intersecting geometries.  In Figure \ref{fig:simple-overlap}, basic hand-made geometries are allowed to separate and unfurl from their initial self-intersecting states.  The two boxes in the left-hand side of each subfigure were meshed using a background grid resolution of $66\times64\times86$ cells and $\Delta x = .00955671$, taking $2.80219\text{s}$ to generate the resulting 256,368 hexahedra in the output mesh.  The simple overlapping shape in the right-hand side of each subfigure was meshed using a grid with $194\times64\times194$ cells and $\Delta x = .00328125$, resulting in 1,606,296 hexahedra in the output mesh.

\subsubsection{Double M\"obius}

\begin{figure}
    \centering
    \begin{subfigure}[b]{\columnwidth}
        \includegraphics[draft=\mydraft,width=\linewidth,trim={230px 580px 260px 10px},clip]{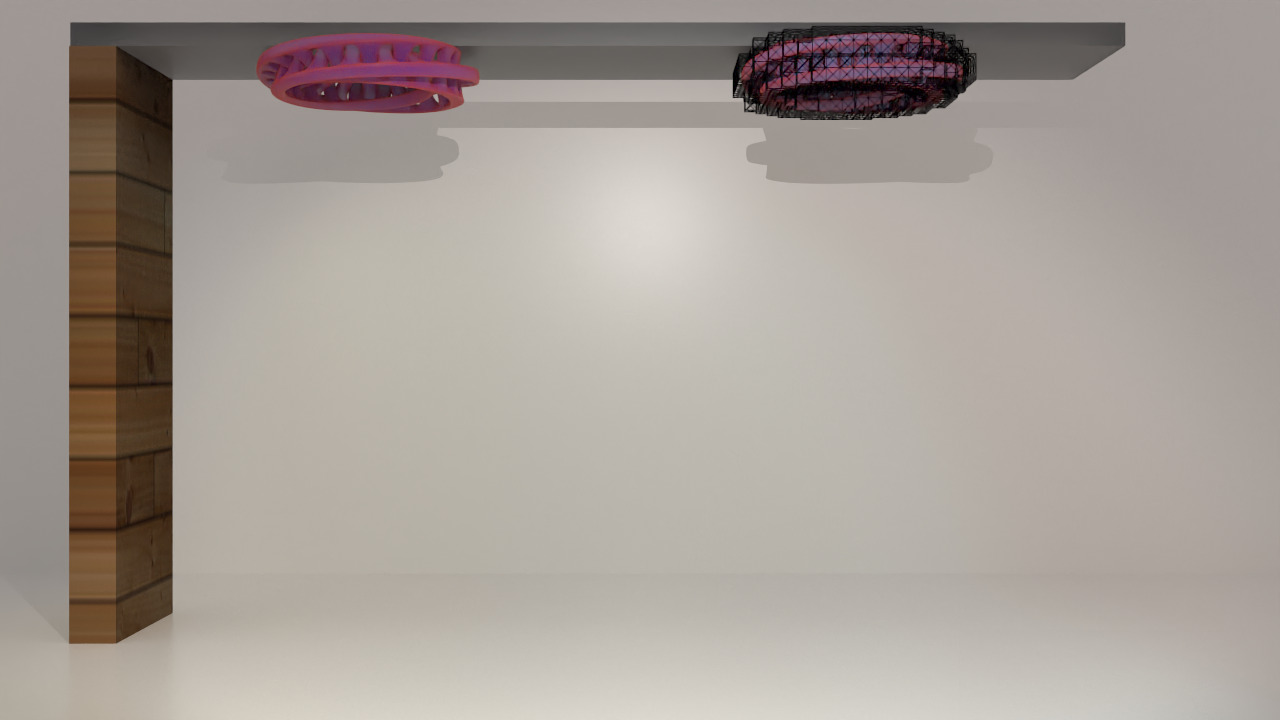}
        \caption{Frame 0}
    \end{subfigure}
    \hfill
    
    \begin{subfigure}[b]{0.49\columnwidth}
        \includegraphics[draft=\mydraft,width=\linewidth]{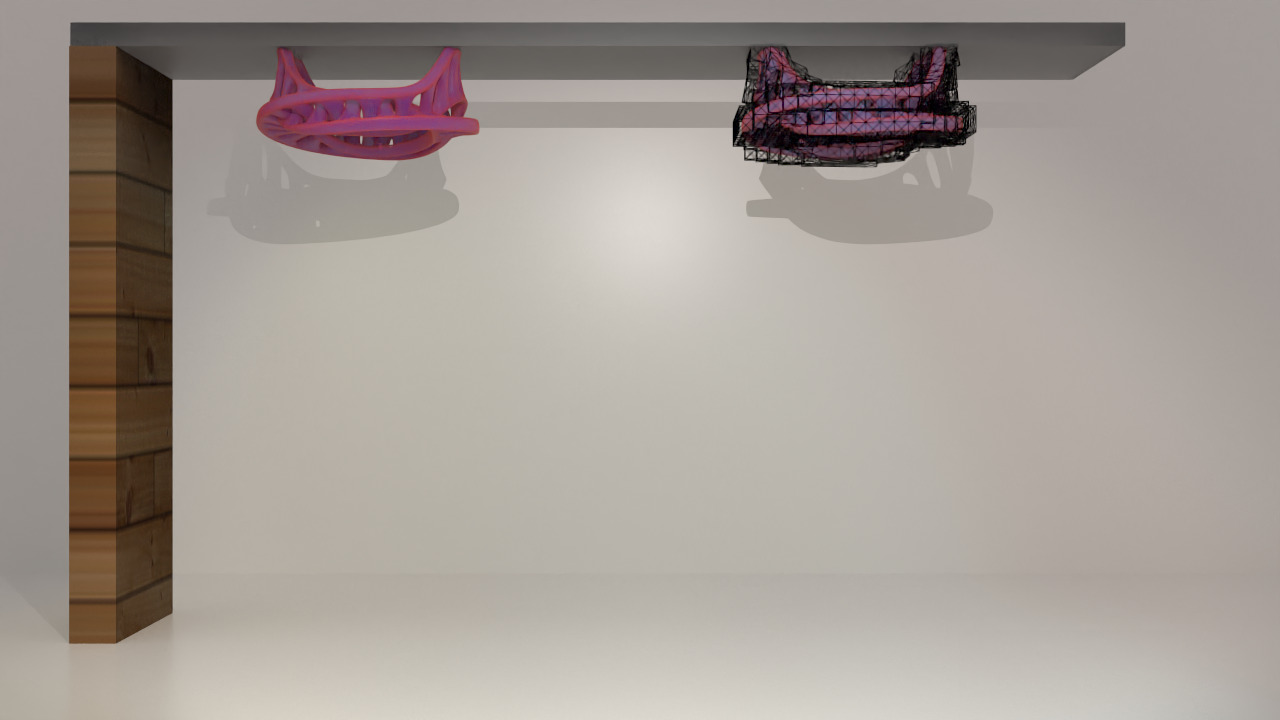}
        \caption{Frame 20}
    \end{subfigure}
    \begin{subfigure}[b]{0.49\columnwidth}
        \includegraphics[draft=\mydraft,width=\linewidth]{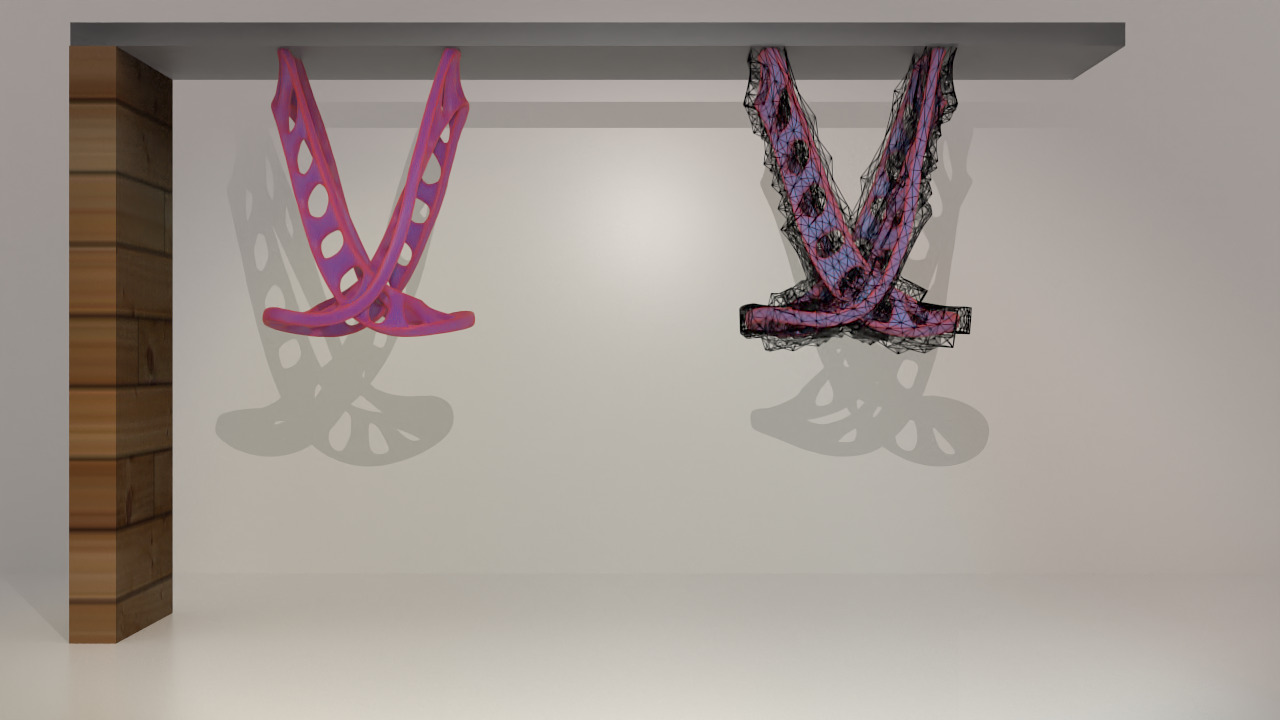}
        \caption{Frame 44}
    \end{subfigure}
    \hfill
    \begin{subfigure}[b]{0.49\columnwidth}
        \includegraphics[draft=\mydraft,width=\linewidth]{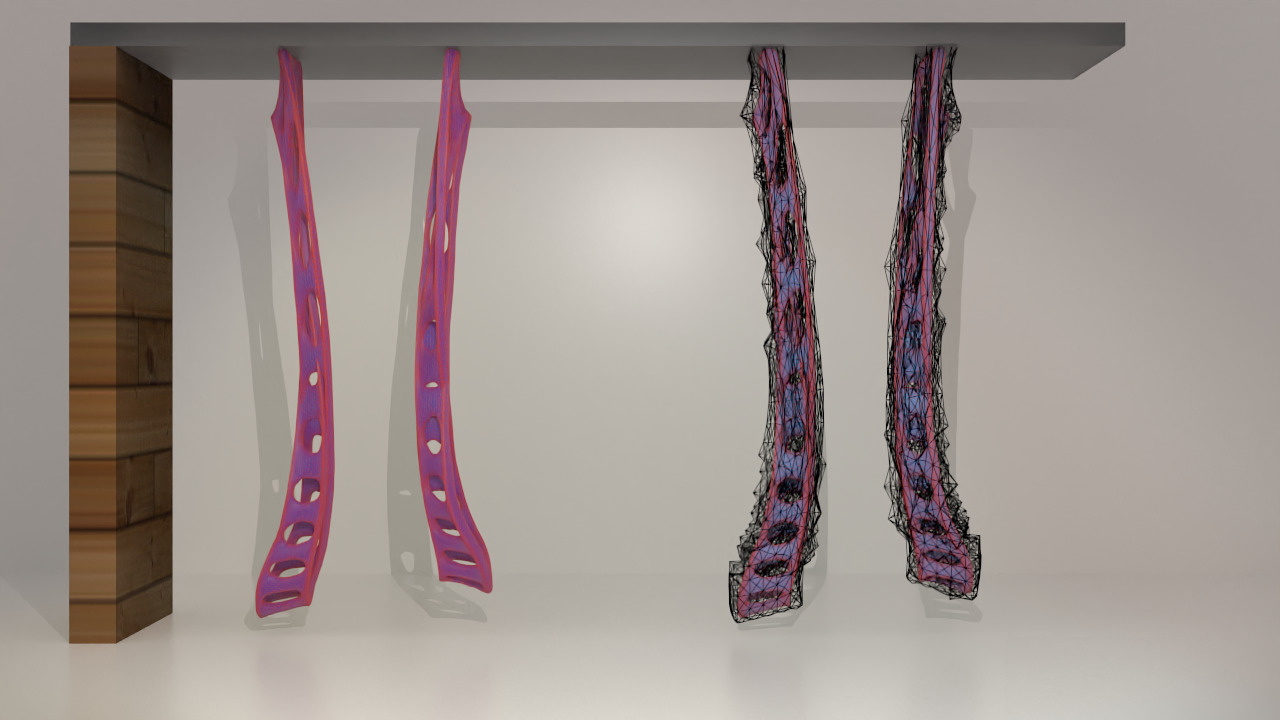}
        \caption{Frame 78}
    \end{subfigure}
    \begin{subfigure}[b]{0.49\columnwidth}
        \includegraphics[draft=\mydraft,width=\linewidth]{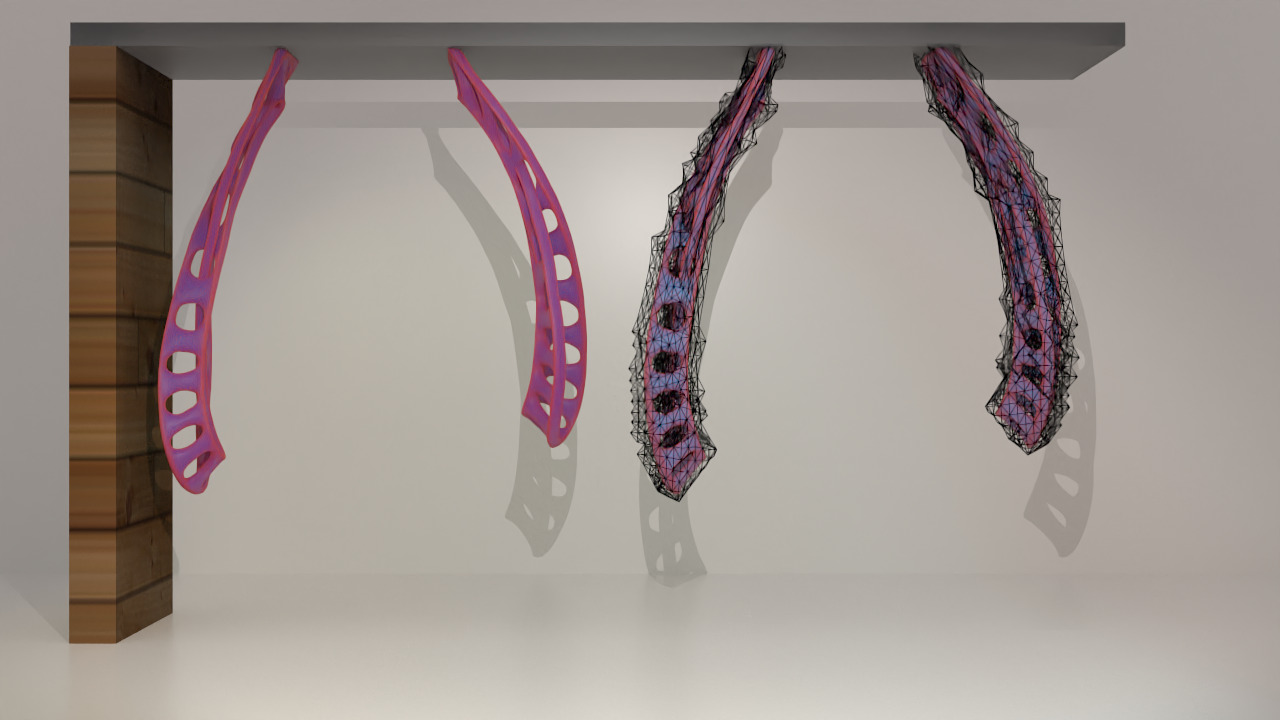}
        \caption{Frame 110}
    \end{subfigure}
    \hfill
    \caption{Two intersecting M\"obius-strip-like geometries (pink) naturally fall and separate under our method.  The associated hexahedron meshes are shown in the right half of each frame.}
    \label{fig:double-mobius}
\end{figure}

Figure \ref{fig:double-mobius} shows two M\"obius-strip-like geometries\footnote{``\href{https://www.thingiverse.com/thing:126984}{Mobius Bangle}'' by \href{https://www.thingiverse.com/creative_hacker}{Creative\_Hacker} is licensed under \href{https://creativecommons.org/licenses/by/4.0/}{CC BY 4.0}.} falling and separating under the effects of gravity, despite substantial intersections at the start of the simulation.  This example was run using a background grid with $294\times288\times64$ cells and a $\Delta x$ of $0.0347391$.  The resulting hexahedron mesh has 903,653 elements.  Generating the volumetric mesh using our algorithm takes $33.6324\text{s}$.

\begin{figure}
    \centering
    \begin{subfigure}[b]{\columnwidth}
        \includegraphics[draft=\mydraft,width=\linewidth,trim={130px 590px 310px 30px},clip]{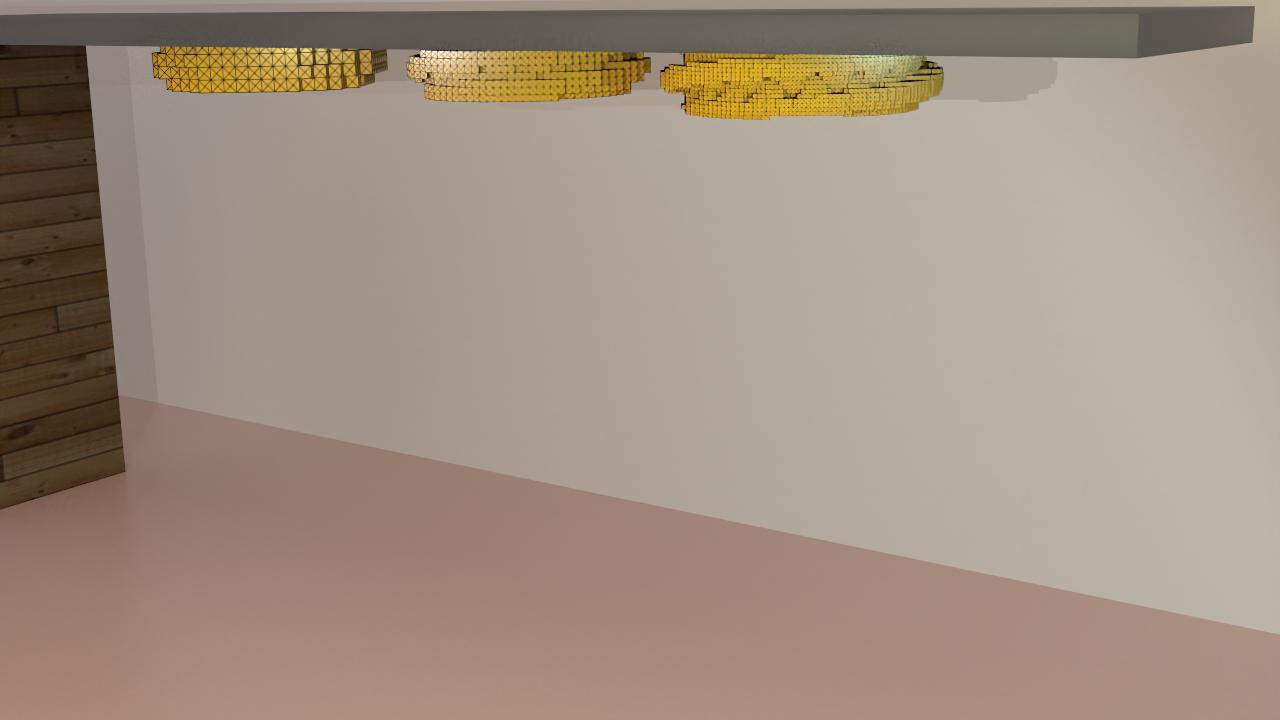}
        \caption{Frame 0}
    \end{subfigure}
    \hfill
    
    \begin{subfigure}[b]{0.49\columnwidth}
        \includegraphics[draft=\mydraft,width=\linewidth]{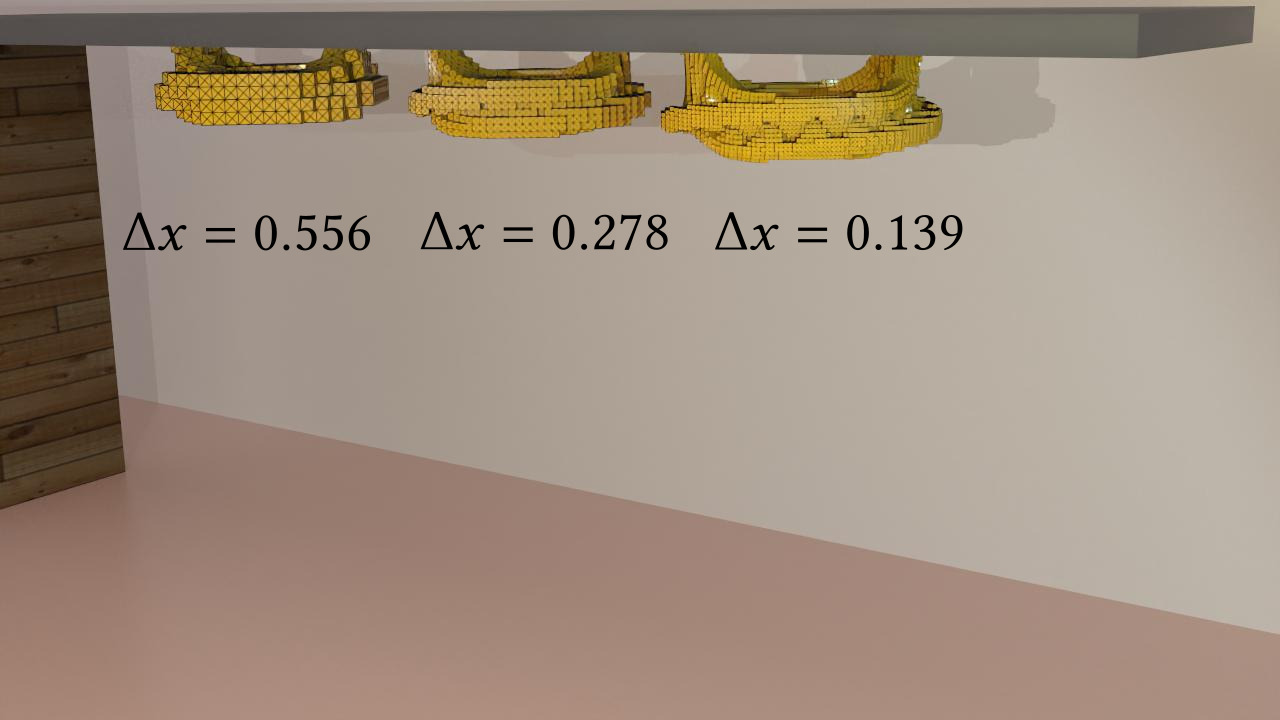}
        \caption{Frame 16}
    \end{subfigure}
    \begin{subfigure}[b]{0.49\columnwidth}
        \includegraphics[draft=\mydraft,width=\linewidth]{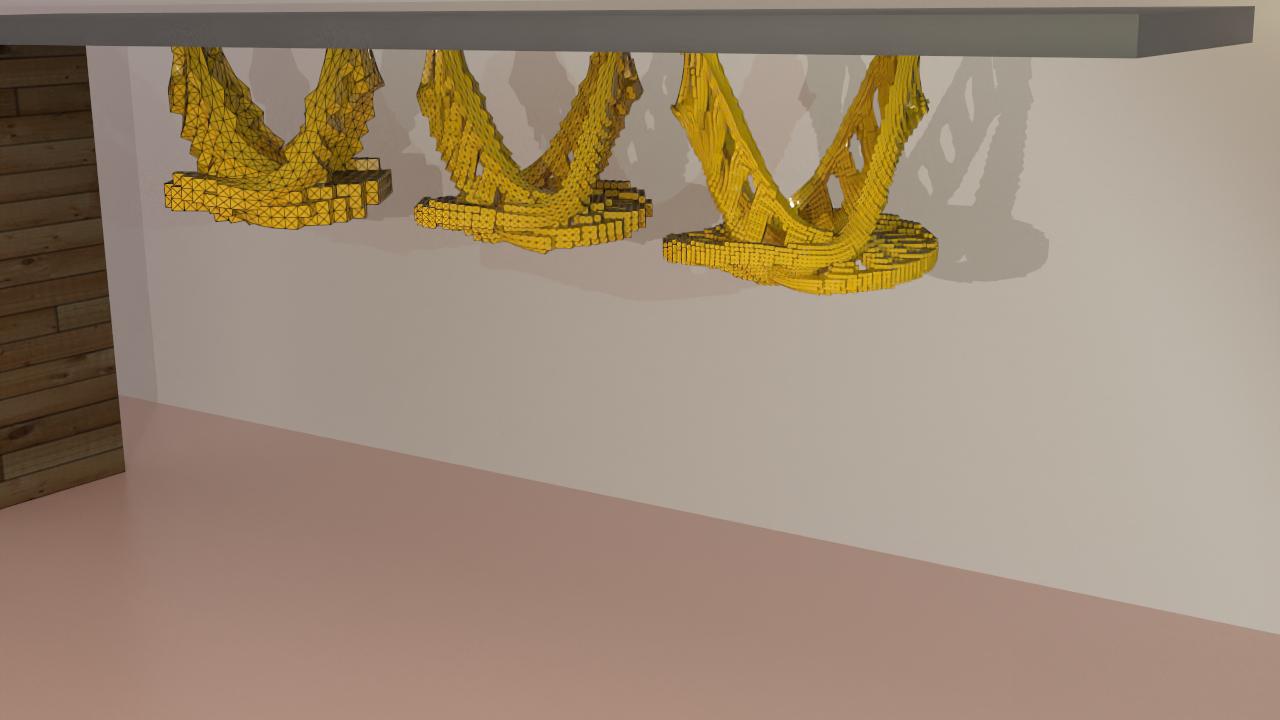}
        \caption{Frame 33}
    \end{subfigure}
    \hfill
    \begin{subfigure}[b]{0.49\columnwidth}
        \includegraphics[draft=\mydraft,width=\linewidth]{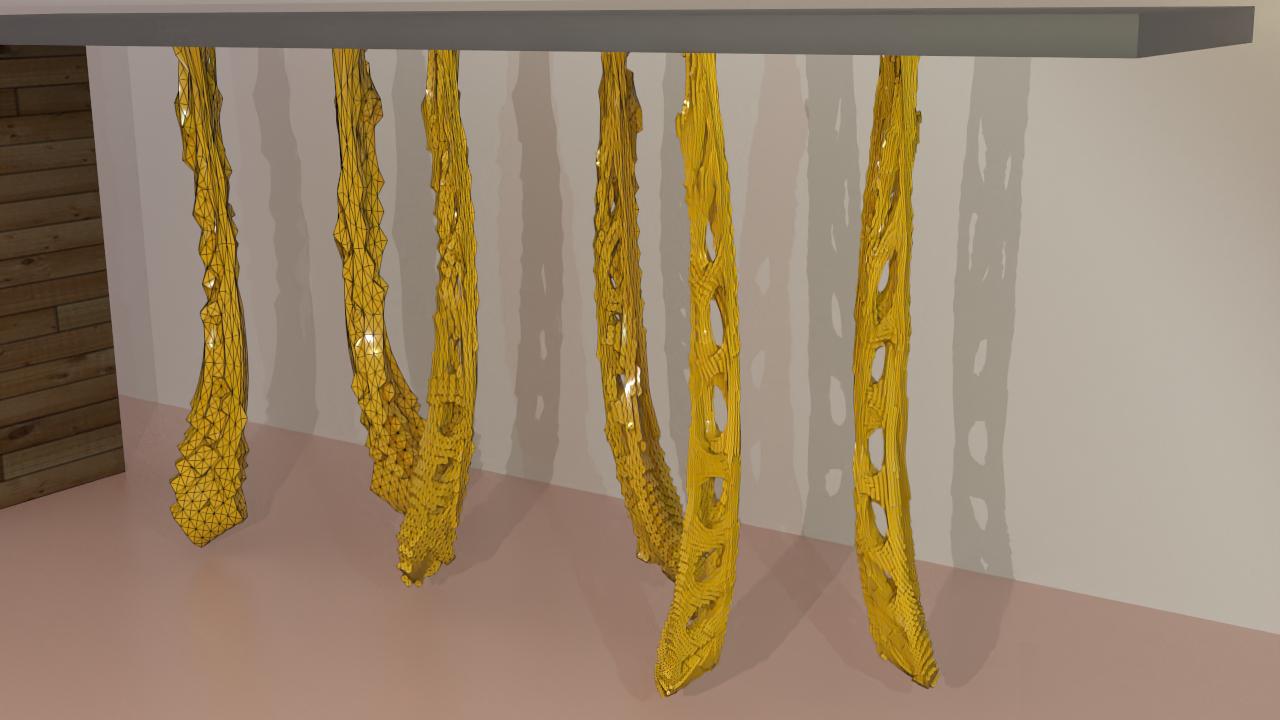}
        \caption{Frame 84}
    \end{subfigure}
    \begin{subfigure}[b]{0.49\columnwidth}
        \includegraphics[draft=\mydraft,width=\linewidth]{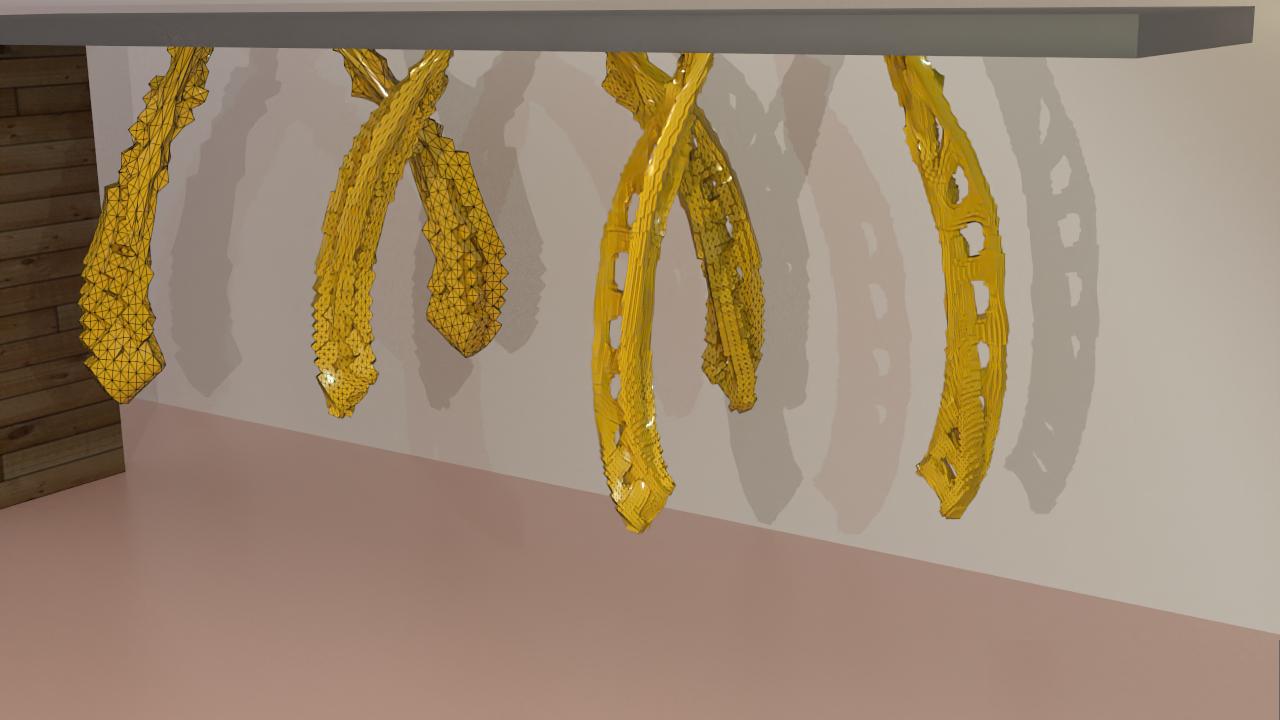}
        \caption{Frame 115}
    \end{subfigure}
    \hfill
    \caption{Running the example shown in Figure \ref{fig:double-mobius} at different spatial resolutions.  In each frame, from left to right, the background grids have $\Delta x = 0.556$, $0.278$, and $0.139$.}
    \label{fig:double-mobius-resolution}
\end{figure}

We also consider repeating this example at multiple spatial resolutions in order to demonstrate the effect of resolution on the quality of meshing results (see Figure \ref{fig:double-mobius-resolution}).  The coarsest grid (corresponding to the leftmost meshes in each subfigure) is $21\times19\times5$ with $\Delta x = 0.556$.  An intermediate grid resolution of $39\times37\times9$ cells with $\Delta x = 0.278$ corresponds to the middle meshes in each subfigure.  The rightmost meshes in each subfigure come from using a grid with $75\times73\times17$ cells with $\Delta x = 0.139$.  Proper separation is achieved at all three of these tested resolutions, and in particular, our algorithm performs quite well on this example even at extremely low spatial resolution.

\subsubsection{Twin Bunnies}

Another standard example is the Stanford bunny.  Figure \ref{fig:twin-bunnies} demonstrates that two almost completely overlapping bunny meshes can naturally separate under our method.  No issues are encountered as different segments of the bunnies pass through one another.  This example uses a grid resolution of $162\times166\times128$ cells with $\Delta x = 0.0203027$, resulting in a mesh with 1,525,821 hexahedra.

\subsubsection{Dragon}

\begin{figure}
    \centering
    \begin{subfigure}[b]{0.49\columnwidth}
        \includegraphics[draft=\mydraft,width=\linewidth,trim={250px 50px 250px 50px},clip]{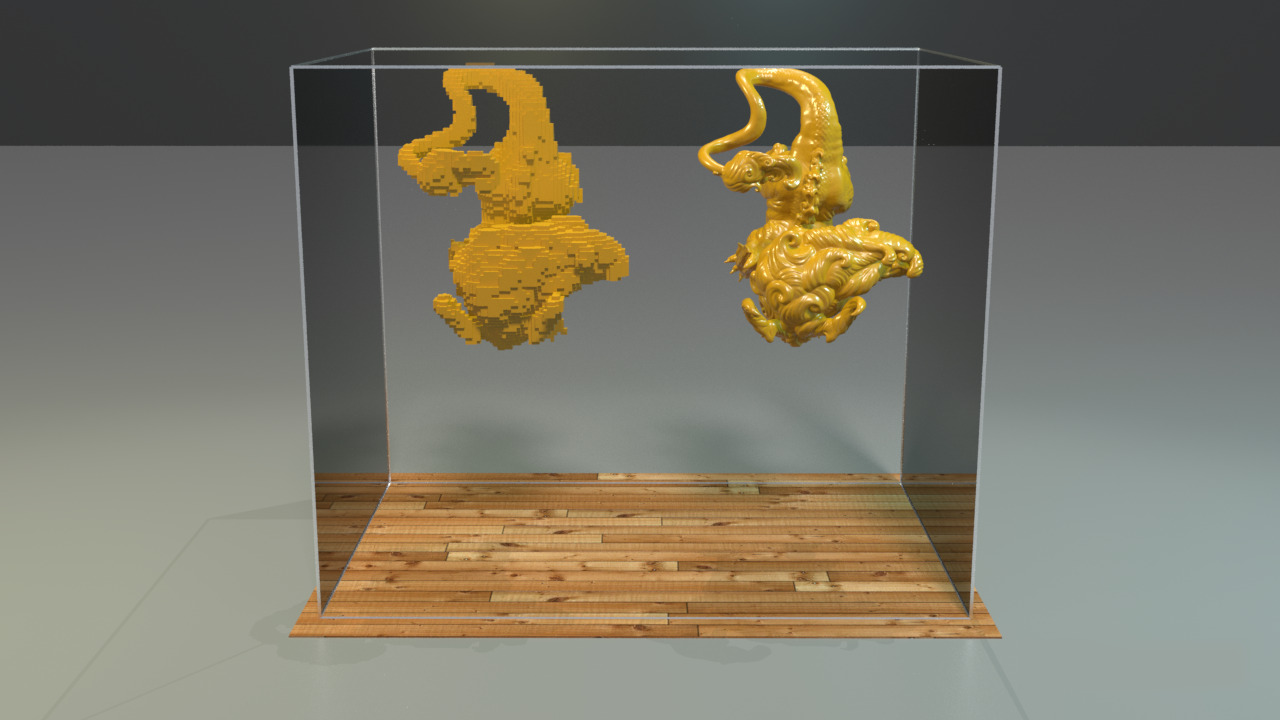}
        \caption{Frame 0}
    \end{subfigure}
    \begin{subfigure}[b]{0.49\columnwidth}
        \includegraphics[draft=\mydraft,width=\linewidth,trim={250px 50px 250px 50px},clip]{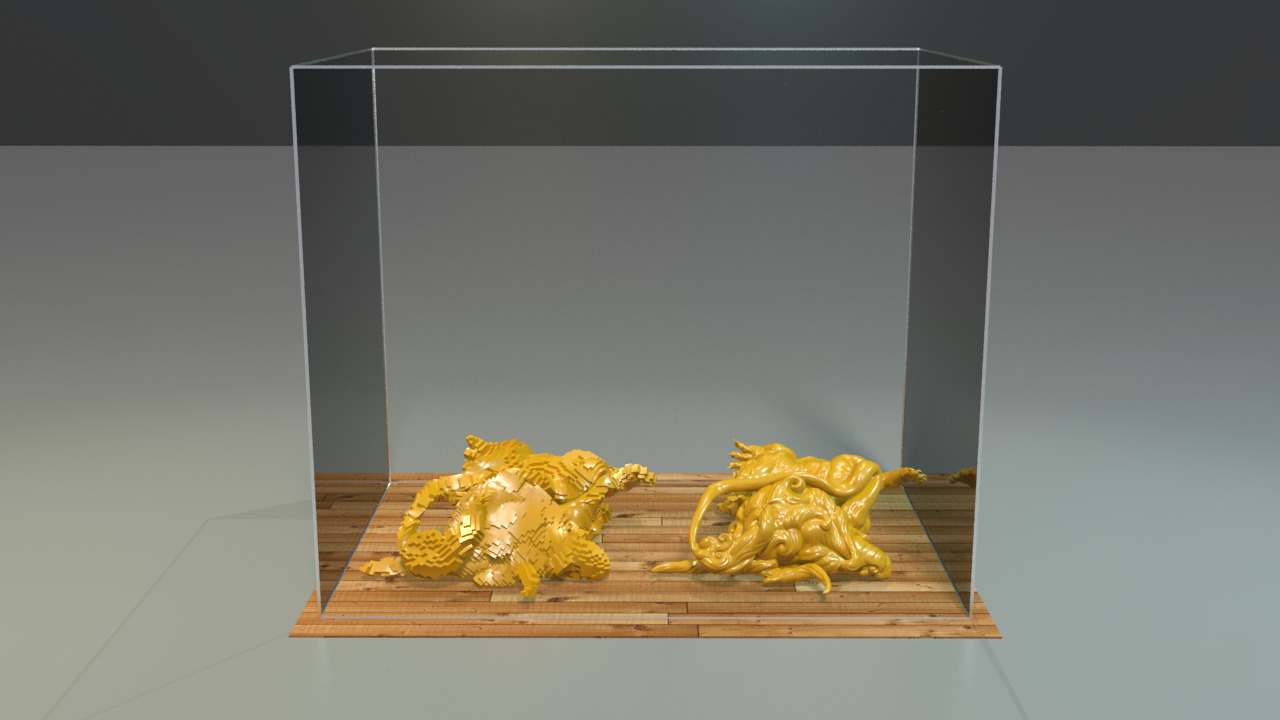}
        \caption{Frame 100}
    \end{subfigure}
    \hfill
    \begin{subfigure}[b]{0.49\columnwidth}
        \includegraphics[draft=\mydraft,width=\linewidth,trim={250px 50px 250px 50px},clip]{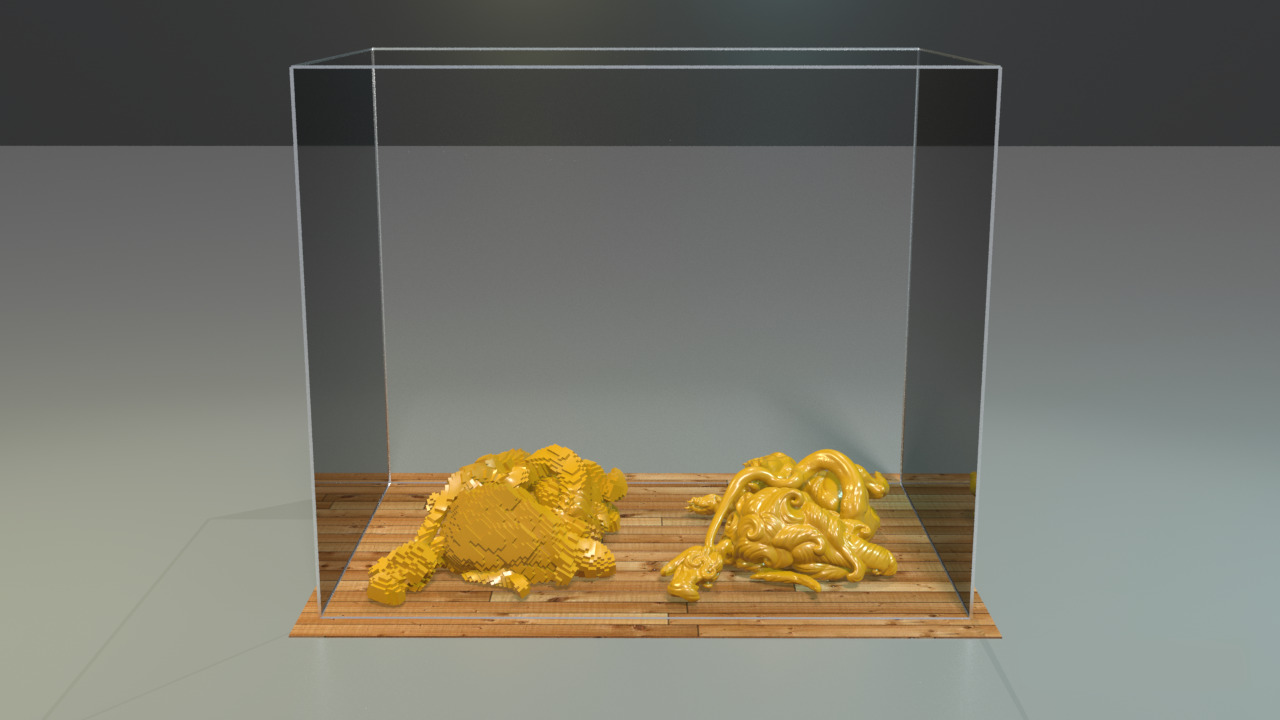}
        \caption{Frame 200}
    \end{subfigure}
    \begin{subfigure}[b]{0.49\columnwidth}
        \includegraphics[draft=\mydraft,width=\linewidth,trim={250px 50px 250px 50px},clip]{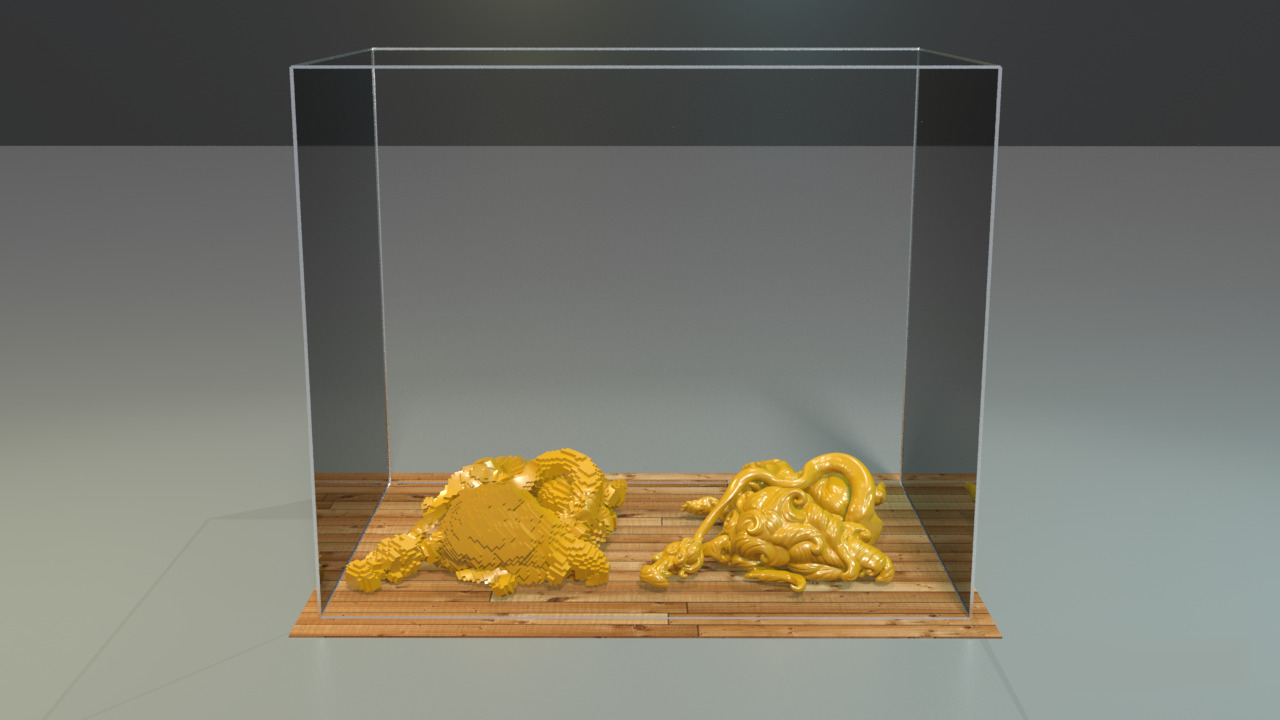}
        \caption{Frame 300}
    \end{subfigure}
    \hfill
    \caption{A complex mesh of a dragon is allowed to fall under gravity.  The left-hand side of each subfigure shows the deforming mesh we generate, and each right-hand side shows the corresponding surface mesh.}
    \label{fig:dragon}
\end{figure}

The most complicated geometry we test our method on is the dragon\footnote{\href{https://www.cgtrader.com/free-3d-print-models/art/sculptures/asian-dragon-86daa8ef-302b-465a-b164-b3c76817c877}{``Asian Dragon''} by \href{https://www.cgtrader.com/lalo-bravo}{Lalo-Bravo}.} shown in Figure \ref{fig:dragon} (and also shown in Figure \ref{fig:patch_expansion}).  Adequate resolution is required in order to resolve all the fine-scale features of this mesh; accordingly, we use a grid resolution of $512\times690\times520$ cells with $\Delta x = 0.0708709$.  Our final mesh, generated in five minutes, contains just over 20 million hexahedra.

\subsubsection{Fancy Ball}

\begin{figure}
    \centering
    \begin{subfigure}[b]{0.49\columnwidth}
        \includegraphics[draft=\mydraft,width=\linewidth,trim={170px 50px 170px 0},clip]{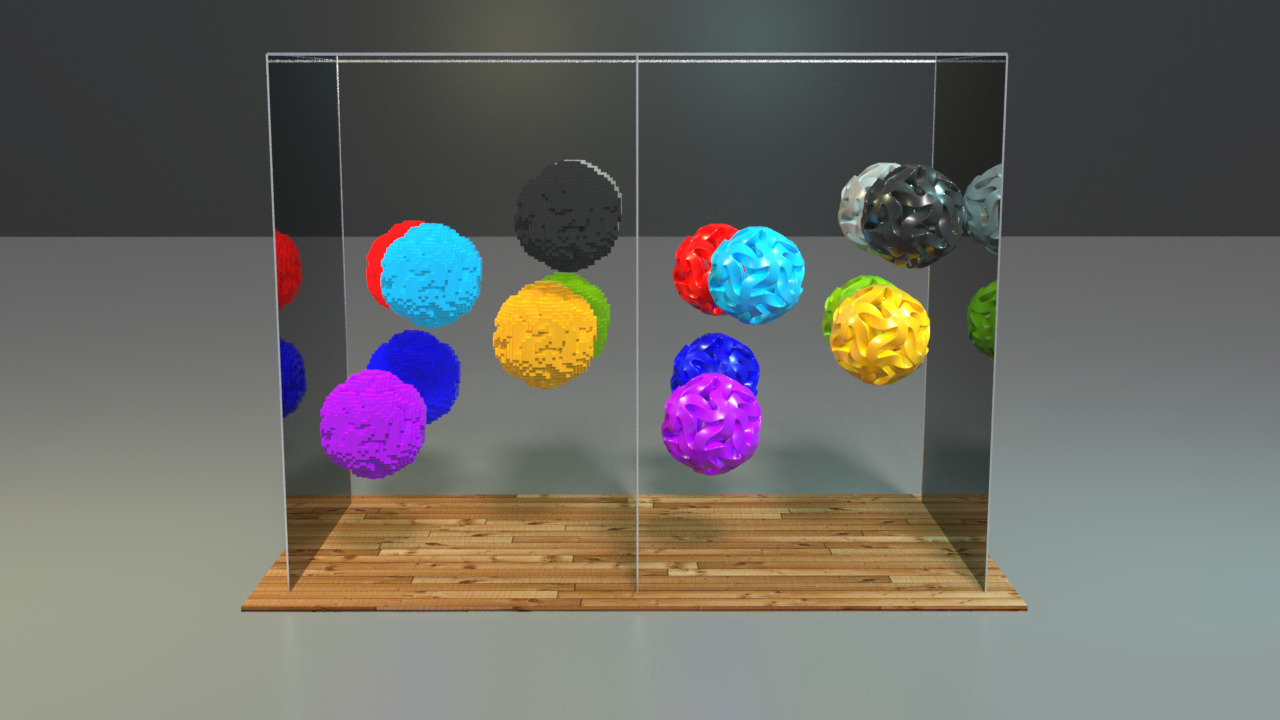}
        \caption{Frame 20}
    \end{subfigure}
    \begin{subfigure}[b]{0.49\columnwidth}
        \includegraphics[draft=\mydraft,width=\linewidth,trim={170px 50px 170px 0},clip]{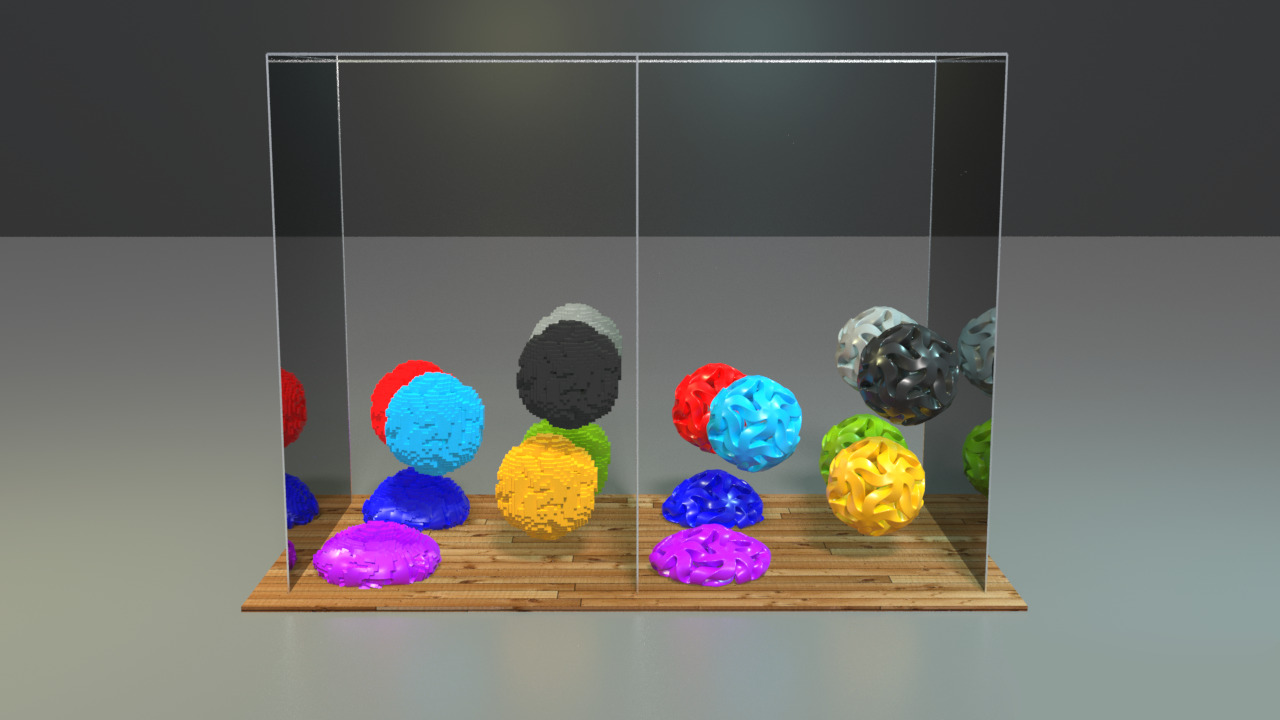}
        \caption{Frame 35}
    \end{subfigure}
    \hfill
    \begin{subfigure}[b]{0.49\columnwidth}
        \includegraphics[draft=\mydraft,width=\linewidth,trim={170px 50px 170px 0},clip]{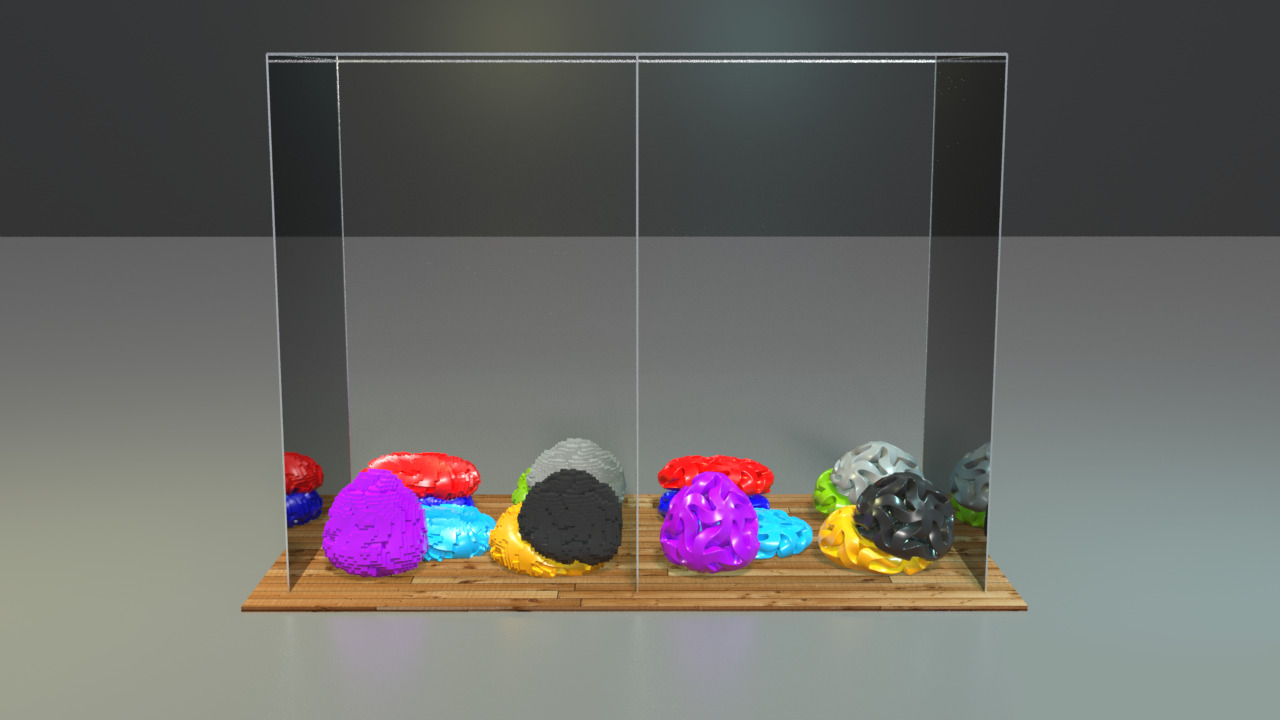}
        \caption{Frame 45}
    \end{subfigure}
    \begin{subfigure}[b]{0.49\columnwidth}
        \includegraphics[draft=\mydraft,width=\linewidth,trim={170px 50px 170px 0},clip]{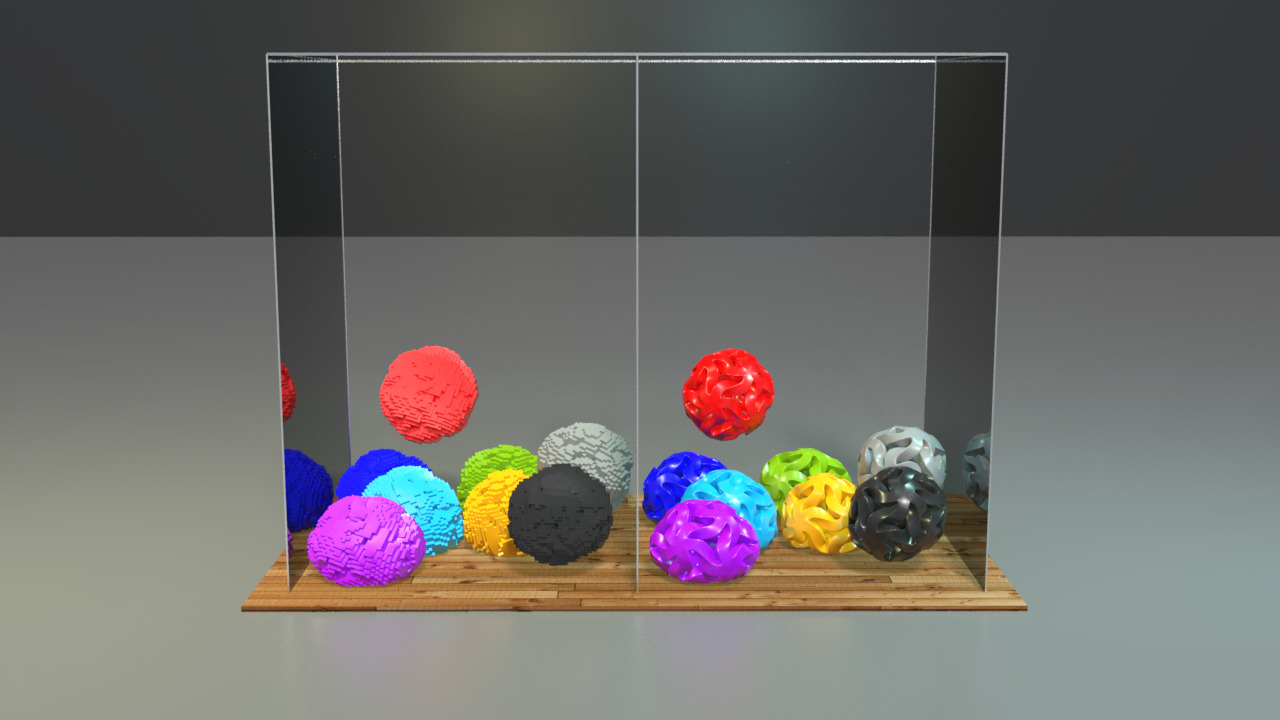}
        \caption{Frame 80}
    \end{subfigure}
    \hfill
    \caption{Several ball-like geometries with intricate slices and holes are successfully meshed with our algorithm and then deform and collide under an FEM simulation.}
    \label{fig:fancy-ball}
\end{figure}

Figure \ref{fig:fancy-ball} shows another interesting case where several ball-like geometries\footnote{\href{https://www.turbosquid.com/3d-models/object-abstract-3d-model-1707700}{``Abstract object''} by \href{https://www.turbosquid.com/Search/Artists/sonic-art}{sonic art}.} deform and collide after being meshed with our algorithm.  Each ball has a number of thin cuts and fine-scale features, which our algorithm is able to resolve using a grid with $130\times132\times128$ cells and $\Delta x = 2.82671$.  The 515,400 resulting hexahedra are generated in $25.8388\text{s}$.

\subsubsection{Head}

Modeling of the human body often gives rise to self-intersection. This is particularly common in the faces, where lip geometries often self-intersect.  To that end, we consider a real-world head geometry in Figure \ref{fig:head-3d}. Note that the lips separate effectively. This example results in a volumetric mesh with over 62 million elements, using a background grid resolution of $512\times830\times718$ cells and $\Delta x = 0.000501962$.  Generating the hexahedron mesh takes $839.951\text{s}$. 

\subsubsection{Collection}

\begin{figure}
    \centering
    \begin{subfigure}[b]{0.49\columnwidth}
        \includegraphics[draft=\mydraft,width=\linewidth,trim={0 0 350px 0},clip]{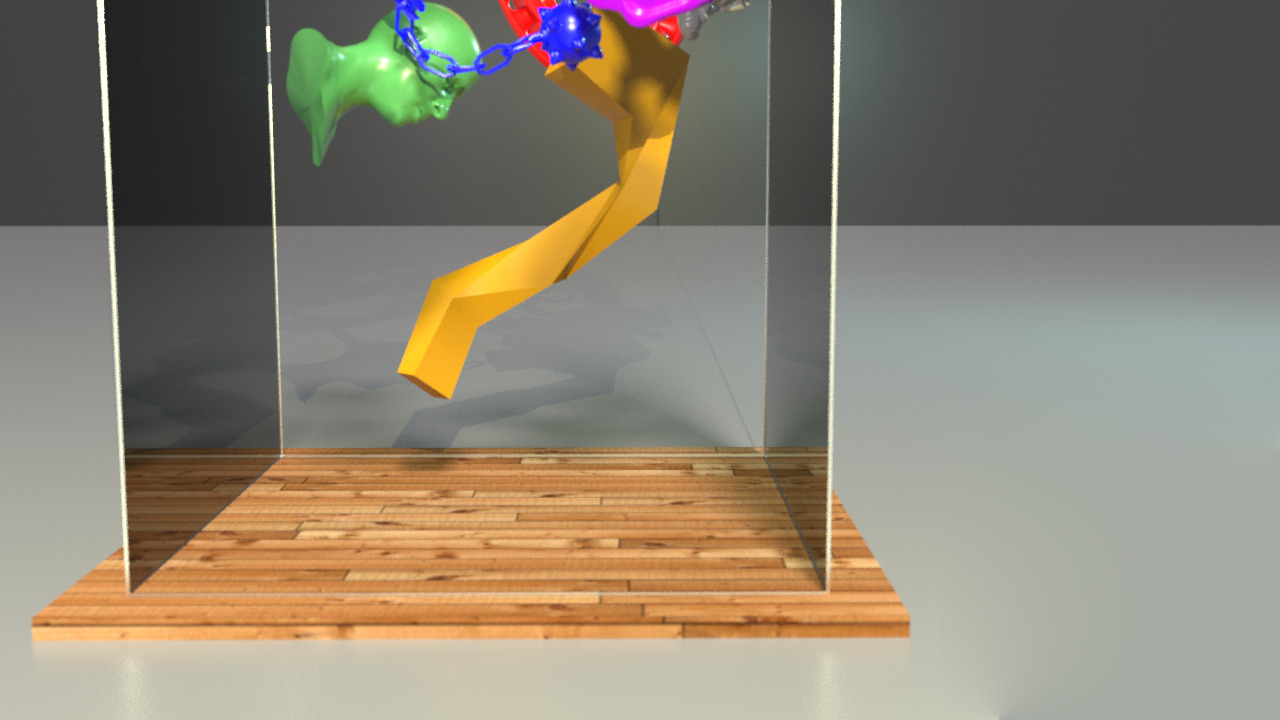}
        \caption{Frame 60}
    \end{subfigure}
    \begin{subfigure}[b]{0.49\columnwidth}
        \includegraphics[draft=\mydraft,width=\linewidth,trim={0 0 350px 0},clip]{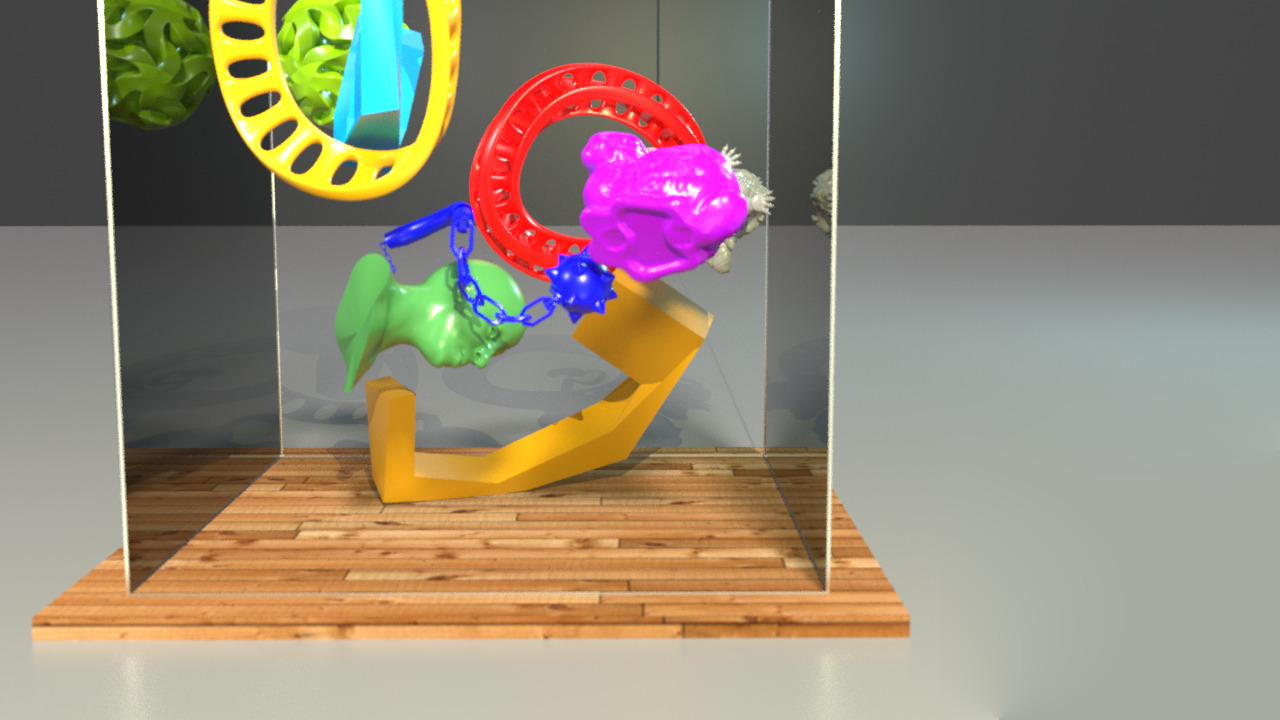}
        \caption{Frame 80}
    \end{subfigure}
    \hfill
    \begin{subfigure}[b]{0.49\columnwidth}
        \includegraphics[draft=\mydraft,width=\linewidth,trim={0 0 350px 0},clip]{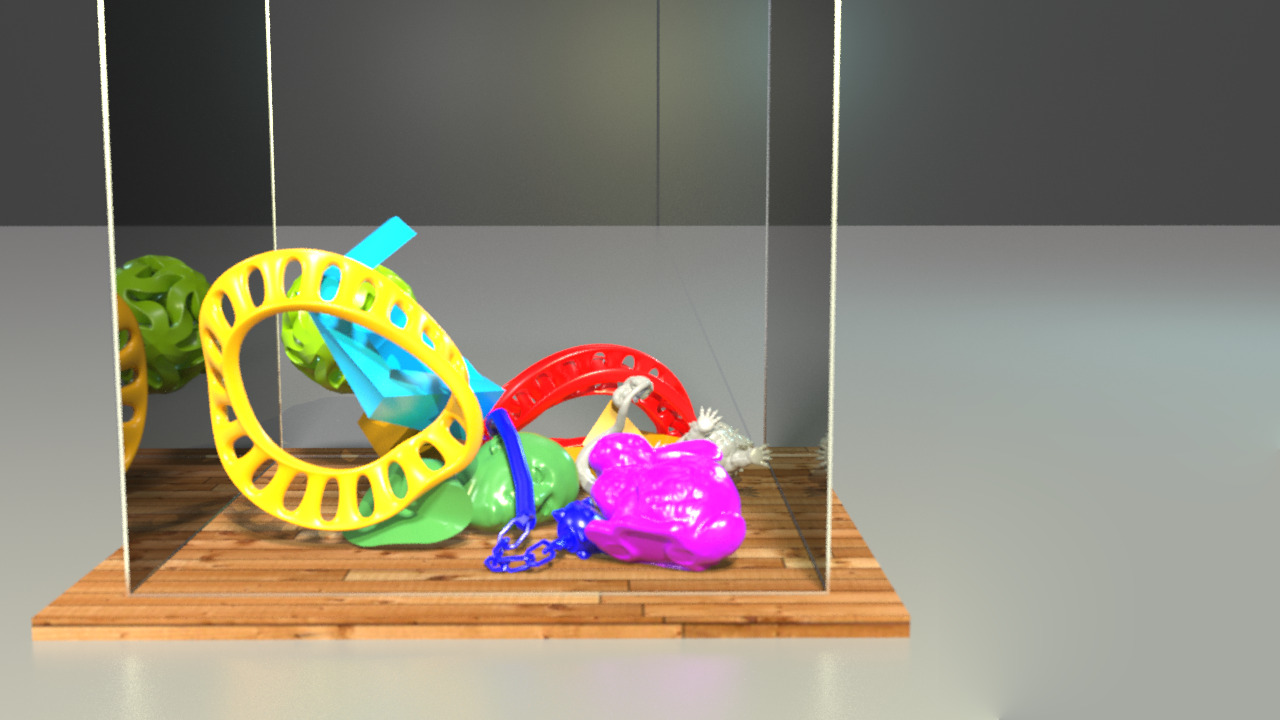}
        \caption{Frame 100}
    \end{subfigure}
    \begin{subfigure}[b]{0.49\columnwidth}
        \includegraphics[draft=\mydraft,width=\linewidth,trim={0 0 350px 0},clip]{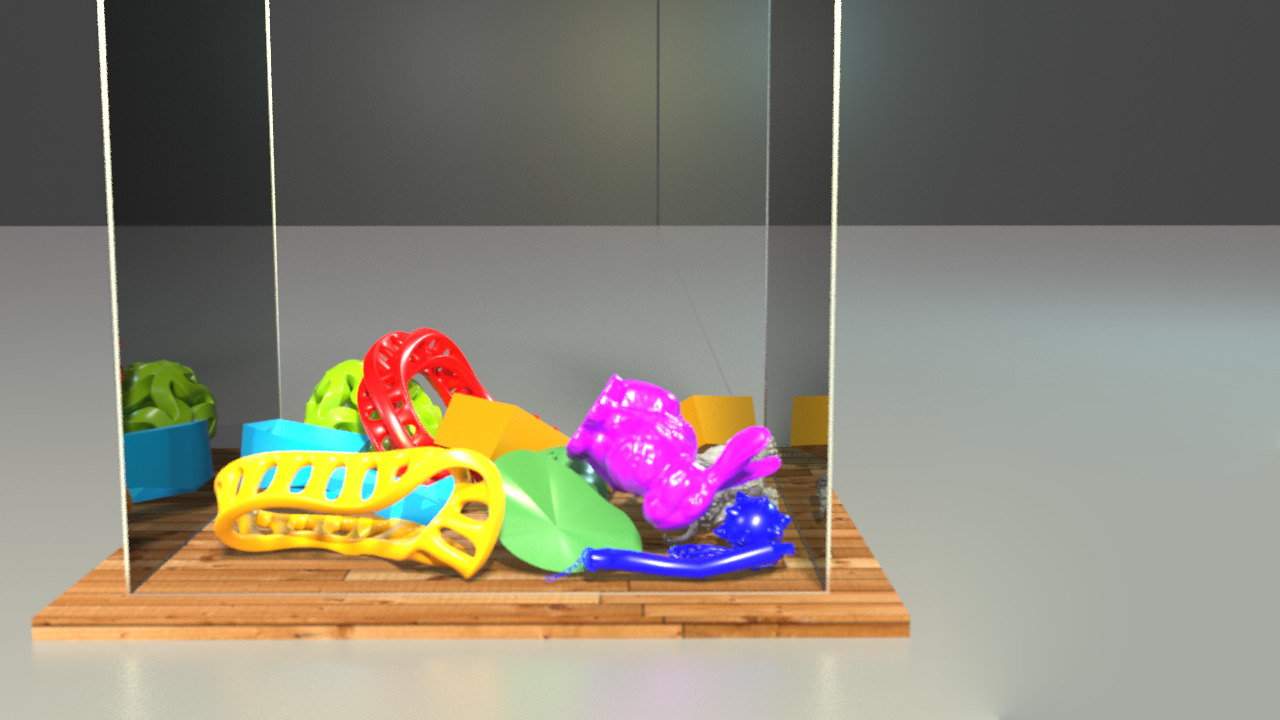}
        \caption{Frame 200}
    \end{subfigure}
    \hfill
    \caption{We simulated dropping our 3D examples into a box with a FEM sim.}
    \label{fig:collection}
\end{figure}

Various objects from 3D examples are dropped in a tank in Figure \ref{fig:collection}.  The objects naturally deform and collide without meshing or simulation issues.

\subsubsection{Sacht et al.\ Mesh}

\begin{figure}
    \centering
    \begin{subfigure}[b]{0.49\columnwidth}
        \includegraphics[draft=\mydraft,width=\linewidth]{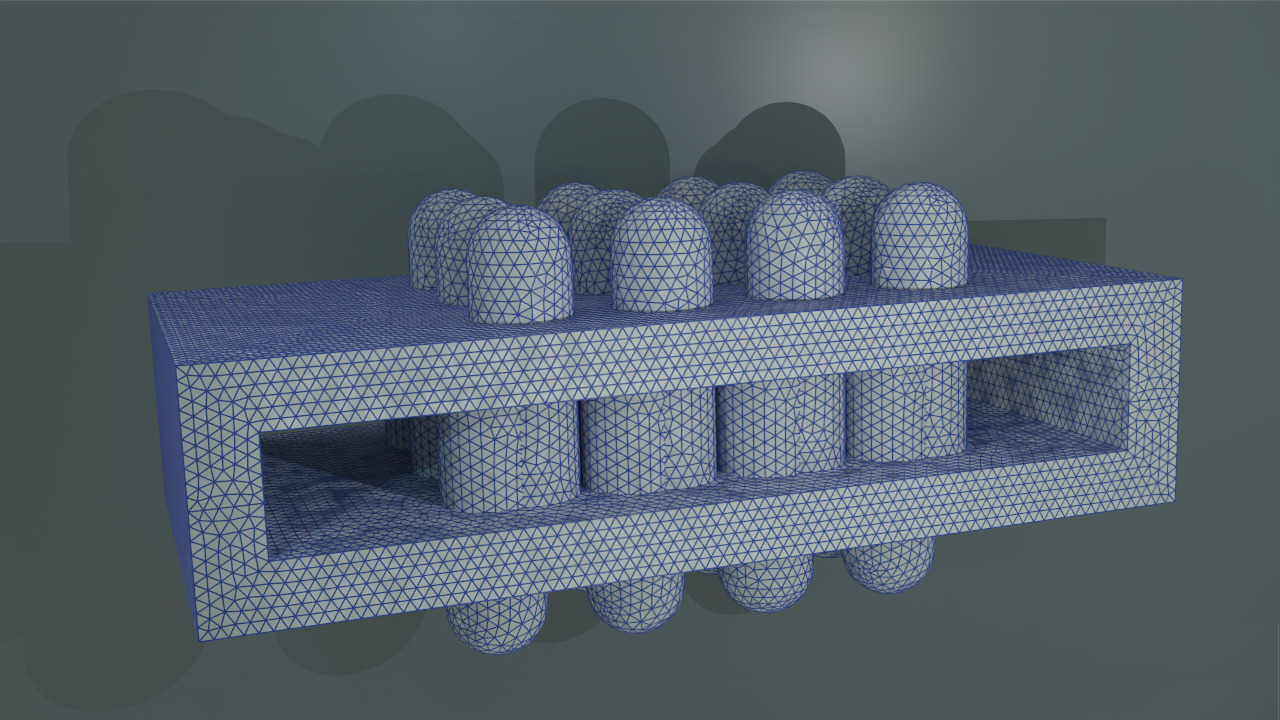}
        \caption{Initial State}
    \end{subfigure}
    \begin{subfigure}[b]{0.49\columnwidth}
        \includegraphics[draft=\mydraft,width=\linewidth]{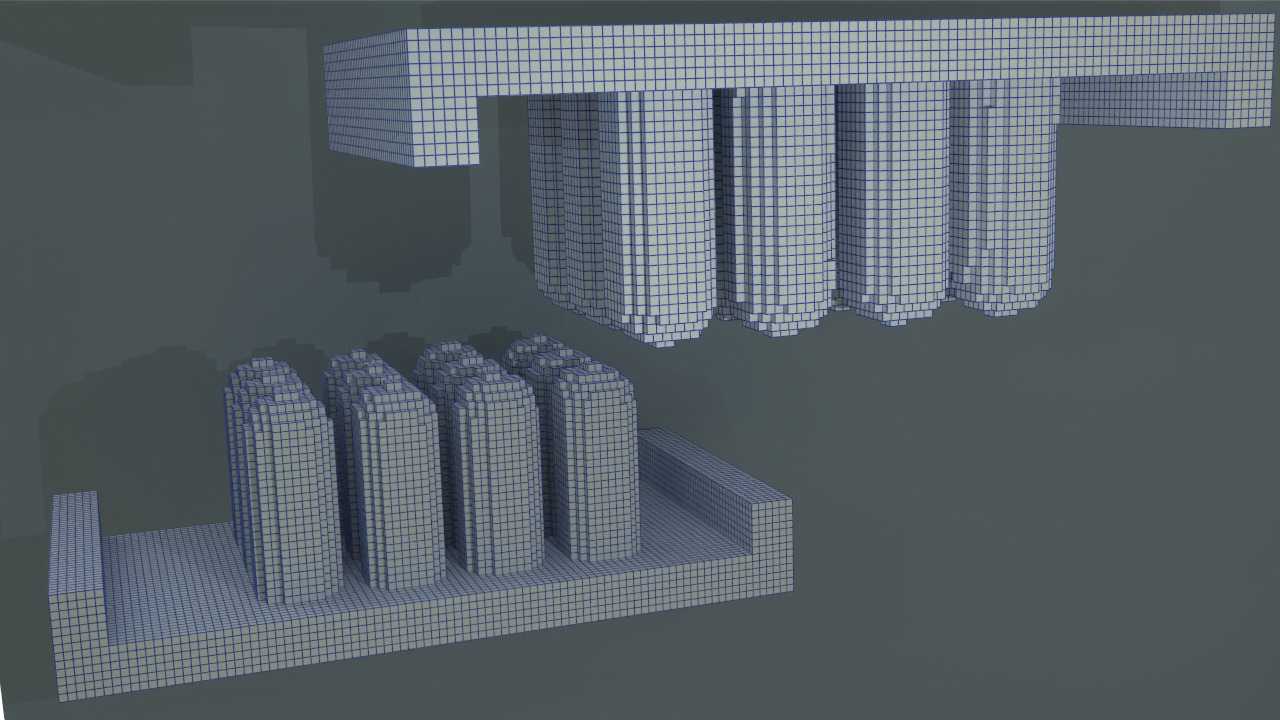}
        \caption{Separation}
    \end{subfigure}
    \caption{Our method can successfully separate the torus and bristle geometry proposed in \cite{sacht:2013:consistent}.}
    \label{fig:sacht-comparison}
\end{figure}

Finally, we demonstrate that our method, like that of Li and Barbi\v{c} \shortcite{li:2018:immersion}, can successfully separate the geometry shown in Figure \ref{fig:sacht-comparison} that is not supported by the method of Sacht et al. \shortcite{sacht:2013:consistent}.  In \cite{sacht:2013:consistent}, the bristles in this geometry get locked by the surrounding torus.  However, both our method and \cite{li:2018:immersion} properly resolve all self-intersections.  Of note, for a similar number of output mesh elements (112,682 vs.\ 112,554), our method runs noticeably faster than that of Li and Barbi\v{c} \shortcite{li:2018:immersion} ($9.65\text{s}$ vs.\ $22.5\text{s}$).

\section{Discussion and Limitations}\label{section:discussion}
Our method has various limitations, most of which are attributed to our reduced use of exact/adaptive precision arithmetic.
The most prominent limitations of our approach are in the types of input surface mesh $\mathcal{S}$ that we support.
Fine-scale features, e.g., thin parallel sheets, can cause negatively signed vertices to be located in regions of the grid corresponding to an incorrect region.
This may result in exterior regions erroneously generating copies, or interior regions creating extra copies which will not be correctly merged or deduplicated.
In these pathological cases, the output mesh will have undesirable extraneous collections of hexahedra.
We resolve these issues by refining the background grid, but very fine features may require refinement to an unreasonable resolution.
However, our coarsening approach is designed to mitigate this.
Even using added resolution and subsequent coarsening, our methodological simplifications prevent us from handling certain classes of cases that Li and Barbi\v{c} \shortcite{li:2018:immersion} can handle, e.g.,\ we cannot resolve non-simple immersions.
It would be interesting to investigate whether our minimal-exact-arithmetic approach could be extended to handle non-simple immersions as well.
Other future work includes improvements to the algorithm to handle known pathological cases without the need for refinement and subsequent coarsening, as well as improved detection mechanisms for such cases.\\

Lastly, Figure~\ref{fig:cases} illustrates an interesting case which neither our approach, that of Li and Barbi\v{c} \shortcite{li:2018:immersion} nor that of Sacht et al. \shortcite{sacht:2013:consistent} can handle.
In this case, which is common near e.g. elbows and even shoulders in an upper torso, a portion of the domain overlaps in such a way that $\fms$ must have negative Jacobian determinant in some regions. 
Our approach returns a mesh for this case, but it does not properly copy the overlap region and one of the two copies that would be required is rejected.
I.e. our approach does not give a result consistent with creating a mesh in ${\TSurf}^V$ and pushing it forward under $\fms$.
In Li and Barbi\v{c} \shortcite{li:2018:immersion}, this is noted as a case for which an immersion does not exist and 
Sacht et al. \shortcite{sacht:2013:consistent} explicitly require the Jacobian determinant of $\fms$ to be non-negative.
However, this is a commonly occurring case which would be beneficial to resolve.

%
%
%
%

\bibliographystyle{ACM-Reference-Format}
\bibliography{references}


%% file: paper.bbl

\begin{thebibliography}{56}


\ifx \showCODEN    \undefined \def \showCODEN     #1{\unskip}     \fi
\ifx \showDOI      \undefined \def \showDOI       #1{#1}\fi
\ifx \showISBNx    \undefined \def \showISBNx     #1{\unskip}     \fi
\ifx \showISBNxiii \undefined \def \showISBNxiii  #1{\unskip}     \fi
\ifx \showISSN     \undefined \def \showISSN      #1{\unskip}     \fi
\ifx \showLCCN     \undefined \def \showLCCN      #1{\unskip}     \fi
\ifx \shownote     \undefined \def \shownote      #1{#1}          \fi
\ifx \showarticletitle \undefined \def \showarticletitle #1{#1}   \fi
\ifx \showURL      \undefined \def \showURL       {\relax}        \fi
\providecommand\bibfield[2]{#2}
\providecommand\bibinfo[2]{#2}
\providecommand\natexlab[1]{#1}
\providecommand\showeprint[2][]{arXiv:#2}

\bibitem[\protect\citeauthoryear{Angelidis, Cani, Wyvill, and King}{Angelidis
  et~al\mbox{.}}{2006}]%
        {angelidis:2006:swirling}
\bibfield{author}{\bibinfo{person}{A. Angelidis}, \bibinfo{person}{M.-P. Cani},
  \bibinfo{person}{G. Wyvill}, {and} \bibinfo{person}{S. King}.}
  \bibinfo{year}{2006}\natexlab{}.
\newblock \showarticletitle{Swirling-sweepers: Constant-volume modeling}.
\newblock \bibinfo{journal}{\emph{Graph. Models}} \bibinfo{volume}{68},
  \bibinfo{number}{4} (\bibinfo{year}{2006}), \bibinfo{pages}{324--332}.
\newblock


\bibitem[\protect\citeauthoryear{Attene}{Attene}{2010}]%
        {attene:2010:lightweight}
\bibfield{author}{\bibinfo{person}{M. Attene}.}
  \bibinfo{year}{2010}\natexlab{}.
\newblock \showarticletitle{A lightweight approach to repairing digitized
  polygon meshes}.
\newblock \bibinfo{journal}{\emph{The visual computer}} \bibinfo{volume}{26},
  \bibinfo{number}{11} (\bibinfo{year}{2010}), \bibinfo{pages}{1393--1406}.
\newblock


\bibitem[\protect\citeauthoryear{Barill, Dickson, Schmidt, Levin, and
  Jacobson}{Barill et~al\mbox{.}}{2018}]%
        {barill:2018:fast}
\bibfield{author}{\bibinfo{person}{G. Barill}, \bibinfo{person}{N. Dickson},
  \bibinfo{person}{R. Schmidt}, \bibinfo{person}{D. Levin}, {and}
  \bibinfo{person}{A. Jacobson}.} \bibinfo{year}{2018}\natexlab{}.
\newblock \showarticletitle{Fast winding numbers for soups and clouds}.
\newblock \bibinfo{journal}{\emph{ACM Trans. Graph.}} \bibinfo{volume}{37},
  \bibinfo{number}{4} (\bibinfo{year}{2018}), \bibinfo{pages}{1--12}.
\newblock


\bibitem[\protect\citeauthoryear{Belytschko and Black}{Belytschko and
  Black}{1999}]%
        {belytschko:1999:elastic}
\bibfield{author}{\bibinfo{person}{T. Belytschko} {and} \bibinfo{person}{T.
  Black}.} \bibinfo{year}{1999}\natexlab{}.
\newblock \showarticletitle{Elastic crack growth in finite elements with
  minimal remeshing}.
\newblock \bibinfo{journal}{\emph{Int. J. Num. Meth. Engr.}}
  \bibinfo{volume}{45}, \bibinfo{number}{5} (\bibinfo{year}{1999}),
  \bibinfo{pages}{601--620}.
\newblock


\bibitem[\protect\citeauthoryear{B{\'e}zier}{B{\'e}zier}{1970}]%
        {bezier:1970:numerical}
\bibfield{author}{\bibinfo{person}{P. B{\'e}zier}.}
  \bibinfo{year}{1970}\natexlab{}.
\newblock \showarticletitle{Numerical control: mathematics and applications}.
\newblock  (\bibinfo{year}{1970}).
\newblock


\bibitem[\protect\citeauthoryear{Blank}{Blank}{1967}]%
        {blank:1967:extending}
\bibfield{author}{\bibinfo{person}{S. Blank}.} \bibinfo{year}{1967}\natexlab{}.
\newblock \emph{\bibinfo{title}{Extending immersions of the circle}}.
\newblock \bibinfo{thesistype}{Ph.D. Dissertation}. \bibinfo{school}{Brandeis
  University}, \bibinfo{address}{Waltham, Mass.}
\newblock


\bibitem[\protect\citeauthoryear{Botsch, Sieger, Moeller, and Fabri}{Botsch
  et~al\mbox{.}}{2020}]%
        {botsch:2020:surface}
\bibfield{author}{\bibinfo{person}{M. Botsch}, \bibinfo{person}{D. Sieger},
  \bibinfo{person}{P. Moeller}, {and} \bibinfo{person}{A. Fabri}.}
  \bibinfo{year}{2020}\natexlab{}.
\newblock \showarticletitle{Surface Mesh}.
\newblock In \bibinfo{booktitle}{\emph{{CGAL} User and Reference Manual}
  (\bibinfo{edition}{{5.2}} ed.)}. \bibinfo{publisher}{{CGAL Editorial Board}}.
\newblock
\urldef\tempurl%
\url{https://doc.cgal.org/5.2/Manual/packages.html#PkgSurfaceMesh}
\showURL{%
\tempurl}


\bibitem[\protect\citeauthoryear{Br{\"o}nnimann, Fabri, Giezeman, Hert,
  Hoffmann, Kettner, Pion, and Schirra}{Br{\"o}nnimann et~al\mbox{.}}{2020}]%
        {bronnimann:2020:kernel}
\bibfield{author}{\bibinfo{person}{H. Br{\"o}nnimann}, \bibinfo{person}{A.
  Fabri}, \bibinfo{person}{G.-J. Giezeman}, \bibinfo{person}{S. Hert},
  \bibinfo{person}{M. Hoffmann}, \bibinfo{person}{L. Kettner},
  \bibinfo{person}{S. Pion}, {and} \bibinfo{person}{S. Schirra}.}
  \bibinfo{year}{2020}\natexlab{}.
\newblock \showarticletitle{{2D} and {3D} Linear Geometry Kernel}.
\newblock In \bibinfo{booktitle}{\emph{{CGAL} User and Reference Manual}
  (\bibinfo{edition}{{5.2}} ed.)}. \bibinfo{publisher}{{CGAL Editorial Board}}.
\newblock
\urldef\tempurl%
\url{https://doc.cgal.org/5.2/Manual/packages.html#PkgKernel23}
\showURL{%
\tempurl}


\bibitem[\protect\citeauthoryear{Brunton, Wuhrer, Shu, Bose, and
  Demaine}{Brunton et~al\mbox{.}}{2009}]%
        {brunton:2009:filling}
\bibfield{author}{\bibinfo{person}{A. Brunton}, \bibinfo{person}{S. Wuhrer},
  \bibinfo{person}{C. Shu}, \bibinfo{person}{P. Bose}, {and}
  \bibinfo{person}{E. Demaine}.} \bibinfo{year}{2009}\natexlab{}.
\newblock \showarticletitle{Filling holes in triangular meshes by curve
  unfolding}. In \bibinfo{booktitle}{\emph{2009 IEEE International Conference
  on Shape Modeling and Applications}}. \bibinfo{pages}{66--72}.
\newblock
\urldef\tempurl%
\url{https://doi.org/10.1109/SMI.2009.5170165}
\showDOI{\tempurl}


\bibitem[\protect\citeauthoryear{Cong, Bao, E, Bhat, and Fedkiw}{Cong
  et~al\mbox{.}}{2015}]%
        {cong:2015:fully}
\bibfield{author}{\bibinfo{person}{M. Cong}, \bibinfo{person}{M. Bao},
  \bibinfo{person}{J. E}, \bibinfo{person}{K. Bhat}, {and} \bibinfo{person}{R.
  Fedkiw}.} \bibinfo{year}{2015}\natexlab{}.
\newblock \showarticletitle{Fully automatic generation of anatomical face
  simulation models}. In \bibinfo{booktitle}{\emph{Proc ACM
  SIGGRAPH/Eurographics Symp Comp Anim}}. \bibinfo{pages}{175--183}.
\newblock


\bibitem[\protect\citeauthoryear{Cong, Bhat, and Fedkiw}{Cong
  et~al\mbox{.}}{2016}]%
        {cong:2016:face}
\bibfield{author}{\bibinfo{person}{M. Cong}, \bibinfo{person}{L. Bhat}, {and}
  \bibinfo{person}{R. Fedkiw}.} \bibinfo{year}{2016}\natexlab{}.
\newblock \showarticletitle{Art-Directed Muscle Simulation for High-End Facial
  Animation}. In \bibinfo{booktitle}{\emph{Proc 2016 ACM SIGGRAPH/Eurographics
  Symp Comp Anim}}. Eurographics Association, \bibinfo{pages}{119--127}.
\newblock


\bibitem[\protect\citeauthoryear{Doran, Chang, and Bridson}{Doran
  et~al\mbox{.}}{2013}]%
        {doran:2013:isosurface}
\bibfield{author}{\bibinfo{person}{C. Doran}, \bibinfo{person}{A. Chang}, {and}
  \bibinfo{person}{R. Bridson}.} \bibinfo{year}{2013}\natexlab{}.
\newblock \showarticletitle{Isosurface stuffing improved: acute lattices and
  feature matching}.
\newblock In \bibinfo{booktitle}{\emph{ACM SIGGRAPH 2013 Talks}}.
\newblock


\bibitem[\protect\citeauthoryear{Edwards and Bridson}{Edwards and
  Bridson}{2014}]%
        {edwards:2014:adaptive-dg}
\bibfield{author}{\bibinfo{person}{E. Edwards} {and} \bibinfo{person}{R.
  Bridson}.} \bibinfo{year}{2014}\natexlab{}.
\newblock \showarticletitle{Detailed water with coarse grids: combining surface
  meshes and adaptive Discontinuous Galerkin}.
\newblock \bibinfo{journal}{\emph{ACM Trans Graph}} \bibinfo{volume}{33},
  \bibinfo{number}{4} (\bibinfo{year}{2014}), \bibinfo{pages}{136:1--136:9}.
\newblock


\bibitem[\protect\citeauthoryear{Eppstein and Mumford}{Eppstein and
  Mumford}{2009}]%
        {eppstein:2009:self}
\bibfield{author}{\bibinfo{person}{D. Eppstein} {and} \bibinfo{person}{E.
  Mumford}.} \bibinfo{year}{2009}\natexlab{}.
\newblock \showarticletitle{Self-overlapping curves revisited}. In
  \bibinfo{booktitle}{\emph{Proceedings of the Twentieth Annual ACM-SIAM
  Symposium on Discrete Algorithms}}. SIAM, \bibinfo{pages}{160--169}.
\newblock


\bibitem[\protect\citeauthoryear{Evans, Fasy, and Wenk}{Evans
  et~al\mbox{.}}{2020}]%
        {evans:2020:combinatorial}
\bibfield{author}{\bibinfo{person}{P. Evans}, \bibinfo{person}{B. Fasy}, {and}
  \bibinfo{person}{C. Wenk}.} \bibinfo{year}{2020}\natexlab{}.
\newblock \showarticletitle{{Combinatorial Properties of Self-Overlapping
  Curves and Interior Boundaries}}. In \bibinfo{booktitle}{\emph{36th
  International Symposium on Computational Geometry (SoCG 2020)}}
  \emph{(\bibinfo{series}{Leibniz International Proceedings in Informatics
  (LIPIcs)})}, \bibfield{editor}{\bibinfo{person}{Sergio Cabello} {and}
  \bibinfo{person}{Danny~Z. Chen}} (Eds.), Vol.~\bibinfo{volume}{164}.
  \bibinfo{publisher}{Schloss Dagstuhl--Leibniz-Zentrum f{\"u}r Informatik},
  \bibinfo{address}{Dagstuhl, Germany}, \bibinfo{pages}{41:1--41:17}.
\newblock
\showISBNx{978-3-95977-143-6}
\showISSN{1868-8969}
\urldef\tempurl%
\url{https://doi.org/10.4230/LIPIcs.SoCG.2020.41}
\showDOI{\tempurl}


\bibitem[\protect\citeauthoryear{Funck, Theisel, and Seidel}{Funck
  et~al\mbox{.}}{2006}]%
        {vonfunck:2006:vector}
\bibfield{author}{\bibinfo{person}{W.~Von Funck}, \bibinfo{person}{H. Theisel},
  {and} \bibinfo{person}{H.-P. Seidel}.} \bibinfo{year}{2006}\natexlab{}.
\newblock \showarticletitle{Vector field based shape deformations}.
\newblock \bibinfo{journal}{\emph{ACM Trans. Graph.}} \bibinfo{volume}{25},
  \bibinfo{number}{3} (\bibinfo{year}{2006}), \bibinfo{pages}{1118--1125}.
\newblock


\bibitem[\protect\citeauthoryear{Gain and Dodgson}{Gain and Dodgson}{2001}]%
        {gain:2001:preventing}
\bibfield{author}{\bibinfo{person}{J. Gain} {and} \bibinfo{person}{N.
  Dodgson}.} \bibinfo{year}{2001}\natexlab{}.
\newblock \showarticletitle{Preventing self-intersection under free-form
  deformation}.
\newblock \bibinfo{journal}{\emph{IEEE Trans Viz Comp Grap}}
  \bibinfo{volume}{7}, \bibinfo{number}{4} (\bibinfo{year}{2001}),
  \bibinfo{pages}{289--298}.
\newblock


\bibitem[\protect\citeauthoryear{Gao, Chen, Xiang, Jacobson, McGuire, and
  Fidler}{Gao et~al\mbox{.}}{2020}]%
        {gao:2020:learning}
\bibfield{author}{\bibinfo{person}{J. Gao}, \bibinfo{person}{W. Chen},
  \bibinfo{person}{T. Xiang}, \bibinfo{person}{A. Jacobson},
  \bibinfo{person}{M. McGuire}, {and} \bibinfo{person}{S. Fidler}.}
  \bibinfo{year}{2020}\natexlab{}.
\newblock \showarticletitle{Learning Deformable Tetrahedral Meshes for 3D
  Reconstruction}. In \bibinfo{booktitle}{\emph{Advances in Neural Information
  Processing Systems}}, \bibfield{editor}{\bibinfo{person}{H.~Larochelle},
  \bibinfo{person}{M.~Ranzato}, \bibinfo{person}{R.~Hadsell},
  \bibinfo{person}{M.~F. Balcan}, {and} \bibinfo{person}{H.~Lin}} (Eds.),
  Vol.~\bibinfo{volume}{33}. \bibinfo{publisher}{Curran Associates, Inc.},
  \bibinfo{pages}{9936--9947}.
\newblock
\urldef\tempurl%
\url{https://proceedings.neurips.cc/paper/2020/file/7137debd45ae4d0ab9aa953017286b20-Paper.pdf}
\showURL{%
\tempurl}


\bibitem[\protect\citeauthoryear{Graver and Cargo}{Graver and Cargo}{2011}]%
        {graver:2011:does}
\bibfield{author}{\bibinfo{person}{J. Graver} {and} \bibinfo{person}{G.
  Cargo}.} \bibinfo{year}{2011}\natexlab{}.
\newblock \showarticletitle{When Does a Curve Bound a Distorted Disk?}
\newblock \bibinfo{journal}{\emph{SIAM Journal on Discrete Mathematics}}
  \bibinfo{volume}{25}, \bibinfo{number}{1} (\bibinfo{year}{2011}),
  \bibinfo{pages}{280--305}.
\newblock


\bibitem[\protect\citeauthoryear{Harmon, Panozzo, Sorkine, and Zorin}{Harmon
  et~al\mbox{.}}{2011}]%
        {harmon:2011:interference}
\bibfield{author}{\bibinfo{person}{D. Harmon}, \bibinfo{person}{D. Panozzo},
  \bibinfo{person}{O. Sorkine}, {and} \bibinfo{person}{D. Zorin}.}
  \bibinfo{year}{2011}\natexlab{}.
\newblock \showarticletitle{Interference-aware geometric modeling}.
\newblock \bibinfo{journal}{\emph{ACM Transactions on Graphics (TOG)}}
  \bibinfo{volume}{30}, \bibinfo{number}{6} (\bibinfo{year}{2011}),
  \bibinfo{pages}{1--10}.
\newblock


\bibitem[\protect\citeauthoryear{Horn and Taylor}{Horn and Taylor}{1989}]%
        {horn:1989:theorem}
\bibfield{author}{\bibinfo{person}{W. Horn} {and} \bibinfo{person}{D. Taylor}.}
  \bibinfo{year}{1989}\natexlab{}.
\newblock \showarticletitle{A theorem to determine the spatial containment of a
  point in a planar polyhedron}.
\newblock \bibinfo{journal}{\emph{Comp Vis Graph Imag Proc}}
  \bibinfo{volume}{45}, \bibinfo{number}{1} (\bibinfo{year}{1989}),
  \bibinfo{pages}{106--116}.
\newblock


\bibitem[\protect\citeauthoryear{Hu, Schneider, Wang, Zorin, and Panozzo}{Hu
  et~al\mbox{.}}{2020}]%
        {hu:2020:fast}
\bibfield{author}{\bibinfo{person}{Y. Hu}, \bibinfo{person}{T. Schneider},
  \bibinfo{person}{B. Wang}, \bibinfo{person}{D. Zorin}, {and}
  \bibinfo{person}{D. Panozzo}.} \bibinfo{year}{2020}\natexlab{}.
\newblock \showarticletitle{Fast tetrahedral meshing in the wild}.
\newblock \bibinfo{journal}{\emph{ACM Trans. Graph.}} \bibinfo{volume}{39},
  \bibinfo{number}{4} (\bibinfo{year}{2020}), \bibinfo{pages}{117--1}.
\newblock


\bibitem[\protect\citeauthoryear{Hu, Zhou, Gao, Jacobson, Zorin, and
  Panozzo}{Hu et~al\mbox{.}}{2018}]%
        {hu:2018:tetwild}
\bibfield{author}{\bibinfo{person}{Y. Hu}, \bibinfo{person}{Q. Zhou},
  \bibinfo{person}{X. Gao}, \bibinfo{person}{A. Jacobson}, \bibinfo{person}{D.
  Zorin}, {and} \bibinfo{person}{D. Panozzo}.} \bibinfo{year}{2018}\natexlab{}.
\newblock \showarticletitle{Tetrahedral Meshing in the Wild}.
\newblock \bibinfo{journal}{\emph{ACM Trans. Graph.}} \bibinfo{volume}{37},
  \bibinfo{number}{4}, Article \bibinfo{articleno}{60} (\bibinfo{date}{July}
  \bibinfo{year}{2018}), \bibinfo{numpages}{14}~pages.
\newblock
\showISSN{0730-0301}
\urldef\tempurl%
\url{https://doi.org/10.1145/3197517.3201353}
\showDOI{\tempurl}


\bibitem[\protect\citeauthoryear{Hu and Ling}{Hu and Ling}{1995}]%
        {hu:1995:geometric}
\bibfield{author}{\bibinfo{person}{Z.-J. Hu} {and} \bibinfo{person}{Z.-K.
  Ling}.} \bibinfo{year}{1995}\natexlab{}.
\newblock \showarticletitle{Geometric modeling of a moving object with
  self-intersection}. In \bibinfo{booktitle}{\emph{International Design
  Engineering Technical Conferences and Computers and Information in
  Engineering Conference}}, Vol.~\bibinfo{volume}{17162}. American Society of
  Mechanical Engineers, \bibinfo{pages}{141--148}.
\newblock


\bibitem[\protect\citeauthoryear{Jacobson, Kavan, and Sorkine-Hornung}{Jacobson
  et~al\mbox{.}}{2013}]%
        {jacobson:2013:robust}
\bibfield{author}{\bibinfo{person}{A. Jacobson}, \bibinfo{person}{L. Kavan},
  {and} \bibinfo{person}{O. Sorkine-Hornung}.} \bibinfo{year}{2013}\natexlab{}.
\newblock \showarticletitle{Robust inside-outside segmentation using
  generalized winding numbers}.
\newblock \bibinfo{journal}{\emph{ACM Trans. Graph.}} \bibinfo{volume}{32},
  \bibinfo{number}{4} (\bibinfo{year}{2013}), \bibinfo{pages}{1--12}.
\newblock


\bibitem[\protect\citeauthoryear{Jamin, Alliez, Yvinec, and Boissonnat}{Jamin
  et~al\mbox{.}}{2015}]%
        {jamin:2015:cgal}
\bibfield{author}{\bibinfo{person}{C. Jamin}, \bibinfo{person}{P. Alliez},
  \bibinfo{person}{M. Yvinec}, {and} \bibinfo{person}{J.-D. Boissonnat}.}
  \bibinfo{year}{2015}\natexlab{}.
\newblock \showarticletitle{{CGALmesh}: a generic framework for delaunay mesh
  generation}.
\newblock \bibinfo{journal}{\emph{ACM Trans. Math. Soft.}}
  \bibinfo{volume}{41}, \bibinfo{number}{4} (\bibinfo{year}{2015}),
  \bibinfo{pages}{1--24}.
\newblock


\bibitem[\protect\citeauthoryear{Kazhdan, Solomon, and Ben-Chen}{Kazhdan
  et~al\mbox{.}}{2012}]%
        {kazhdan:2012:flow}
\bibfield{author}{\bibinfo{person}{M. Kazhdan}, \bibinfo{person}{J. Solomon},
  {and} \bibinfo{person}{M. Ben-Chen}.} \bibinfo{year}{2012}\natexlab{}.
\newblock \showarticletitle{Can Mean-Curvature Flow Be Modified to Be
  Non-Singular?}
\newblock \bibinfo{journal}{\emph{Comput. Graph. Forum}} \bibinfo{volume}{31},
  \bibinfo{number}{5} (\bibinfo{date}{Aug.} \bibinfo{year}{2012}),
  \bibinfo{pages}{1745–1754}.
\newblock
\showISSN{0167-7055}
\urldef\tempurl%
\url{https://doi.org/10.1111/j.1467-8659.2012.03179.x}
\showDOI{\tempurl}


\bibitem[\protect\citeauthoryear{Kim and Tautges}{Kim and Tautges}{2010}]%
        {kim:2010:ebmesh}
\bibfield{author}{\bibinfo{person}{H.-J. Kim} {and} \bibinfo{person}{T.
  Tautges}.} \bibinfo{year}{2010}\natexlab{}.
\newblock \showarticletitle{EBMesh: An Embedded Boundary Meshing Tool}. In
  \bibinfo{booktitle}{\emph{Proceedings of the 19th International Meshing
  Roundtable}}, \bibfield{editor}{\bibinfo{person}{Suzanne Shontz}} (Ed.).
  \bibinfo{publisher}{Springer Berlin Heidelberg}, \bibinfo{address}{Berlin,
  Heidelberg}, \bibinfo{pages}{227--242}.
\newblock
\showISBNx{978-3-642-15414-0}


\bibitem[\protect\citeauthoryear{Koschier, Bender, and Thuerey}{Koschier
  et~al\mbox{.}}{2017}]%
        {koschier:2017:xfem}
\bibfield{author}{\bibinfo{person}{D. Koschier}, \bibinfo{person}{J. Bender},
  {and} \bibinfo{person}{N. Thuerey}.} \bibinfo{year}{2017}\natexlab{}.
\newblock \showarticletitle{Robust eXtended Finite Elements for complex cutting
  of deformables}.
\newblock \bibinfo{journal}{\emph{ACM Trans Graph}} \bibinfo{volume}{36},
  \bibinfo{number}{4} (\bibinfo{year}{2017}), \bibinfo{pages}{55:1--55:13}.
\newblock
\urldef\tempurl%
\url{https://doi.org/10.1145/3072959.3073666}
\showDOI{\tempurl}


\bibitem[\protect\citeauthoryear{Labelle and Shewchuk}{Labelle and
  Shewchuk}{2007}]%
        {labelle:2007:mesh}
\bibfield{author}{\bibinfo{person}{F. Labelle} {and} \bibinfo{person}{J.
  Shewchuk}.} \bibinfo{year}{2007}\natexlab{}.
\newblock \showarticletitle{Isosurface Stuffing: Fast Tetrahedral Meshes with
  Good Dihedral Angles}. In \bibinfo{booktitle}{\emph{ACM SIGGRAPH 2007}} (San
  Diego, California) \emph{(\bibinfo{series}{SIGGRAPH '07})}.
  \bibinfo{publisher}{ACM}, \bibinfo{address}{New York, NY, USA},
  \bibinfo{pages}{57--es}.
\newblock
\showISBNx{9781450378369}
\urldef\tempurl%
\url{https://doi.org/10.1145/1275808.1276448}
\showDOI{\tempurl}


\bibitem[\protect\citeauthoryear{Li}{Li}{2011}]%
        {li:2011:detecting}
\bibfield{author}{\bibinfo{person}{W. Li}.} \bibinfo{year}{2011}\natexlab{}.
\newblock \showarticletitle{Detecting Ambiguities in 3D Polygons with
  Self-Intersecting Projections}. In \bibinfo{booktitle}{\emph{2011 12th
  International Conference on Computer-Aided Design and Computer Graphics}}.
  \bibinfo{pages}{11--16}.
\newblock
\urldef\tempurl%
\url{https://doi.org/10.1109/CAD/Graphics.2011.31}
\showDOI{\tempurl}


\bibitem[\protect\citeauthoryear{Li and Barbi\v{c}}{Li and Barbi\v{c}}{2018}]%
        {li:2018:immersion}
\bibfield{author}{\bibinfo{person}{Y. Li} {and} \bibinfo{person}{J.
  Barbi\v{c}}.} \bibinfo{year}{2018}\natexlab{}.
\newblock \showarticletitle{Immersion of Self-Intersecting Solids and
  Surfaces}.
\newblock \bibinfo{journal}{\emph{ACM Trans. Graph.}} \bibinfo{volume}{37},
  \bibinfo{number}{4}, Article \bibinfo{articleno}{45} (\bibinfo{date}{July}
  \bibinfo{year}{2018}), \bibinfo{numpages}{14}~pages.
\newblock
\showISSN{0730-0301}
\urldef\tempurl%
\url{https://doi.org/10.1145/3197517.3201327}
\showDOI{\tempurl}


\bibitem[\protect\citeauthoryear{Loriot, Rouxel-Labb{\'e}, Tournois, and
  Yaz}{Loriot et~al\mbox{.}}{2020}]%
        {loriot:2020:polygon}
\bibfield{author}{\bibinfo{person}{S. Loriot}, \bibinfo{person}{M.
  Rouxel-Labb{\'e}}, \bibinfo{person}{J. Tournois}, {and} \bibinfo{person}{I.
  Yaz}.} \bibinfo{year}{2020}\natexlab{}.
\newblock \showarticletitle{Polygon Mesh Processing}.
\newblock In \bibinfo{booktitle}{\emph{{CGAL} User and Reference Manual}
  (\bibinfo{edition}{{5.2}} ed.)}. \bibinfo{publisher}{{CGAL Editorial Board}}.
\newblock
\urldef\tempurl%
\url{https://doc.cgal.org/5.2/Manual/packages.html#PkgPolygonMeshProcessing}
\showURL{%
\tempurl}


\bibitem[\protect\citeauthoryear{Marx}{Marx}{1974}]%
        {marx:1974:extensions}
\bibfield{author}{\bibinfo{person}{M. Marx}.} \bibinfo{year}{1974}\natexlab{}.
\newblock \showarticletitle{Extensions of normal immersions of $\mathcal{S}^1$
  into $\mathcal{R}^2$}.
\newblock \bibinfo{journal}{\emph{Trans. Amer. Math. Soc.}}
  \bibinfo{volume}{187} (\bibinfo{year}{1974}), \bibinfo{pages}{309--326}.
\newblock


\bibitem[\protect\citeauthoryear{Mitchell, Aanjaneya, Setaluri, and
  Sifakis}{Mitchell et~al\mbox{.}}{2015}]%
        {mitchell:2015:nonmanifold}
\bibfield{author}{\bibinfo{person}{N. Mitchell}, \bibinfo{person}{M.
  Aanjaneya}, \bibinfo{person}{R. Setaluri}, {and} \bibinfo{person}{E.
  Sifakis}.} \bibinfo{year}{2015}\natexlab{}.
\newblock \showarticletitle{Non-Manifold Level Sets: A Multivalued Implicit
  Surface Representation with Applications to Self-Collision Processing}.
\newblock \bibinfo{journal}{\emph{ACM Trans. Graph.}} \bibinfo{volume}{34},
  \bibinfo{number}{6}, Article \bibinfo{articleno}{247} (\bibinfo{date}{Oct.}
  \bibinfo{year}{2015}), \bibinfo{numpages}{9}~pages.
\newblock
\showISSN{0730-0301}
\urldef\tempurl%
\url{https://doi.org/10.1145/2816795.2818100}
\showDOI{\tempurl}


\bibitem[\protect\citeauthoryear{Molino, Bao, and Fedkiw}{Molino
  et~al\mbox{.}}{2004}]%
        {molino:2004:vna}
\bibfield{author}{\bibinfo{person}{N. Molino}, \bibinfo{person}{Z. Bao}, {and}
  \bibinfo{person}{R. Fedkiw}.} \bibinfo{year}{2004}\natexlab{}.
\newblock \showarticletitle{A virtual node algorithm for changing mesh topology
  during simulation}.
\newblock \bibinfo{journal}{\emph{ACM Trans Graph}} \bibinfo{volume}{23},
  \bibinfo{number}{3} (\bibinfo{year}{2004}), \bibinfo{pages}{385--392}.
\newblock
\urldef\tempurl%
\url{https://doi.org/10.1145/1015706.1015734}
\showDOI{\tempurl}


\bibitem[\protect\citeauthoryear{Molino, Bridson, and Fedkiw}{Molino
  et~al\mbox{.}}{2003a}]%
        {molino:2003:tetrahedral}
\bibfield{author}{\bibinfo{person}{N. Molino}, \bibinfo{person}{R. Bridson},
  {and} \bibinfo{person}{R. Fedkiw}.} \bibinfo{year}{2003}\natexlab{a}.
\newblock \showarticletitle{Tetrahedral mesh generation for deformable bodies}.
  In \bibinfo{booktitle}{\emph{Proc. Symposium on Computer Animation}}.
  \bibinfo{pages}{8}.
\newblock


\bibitem[\protect\citeauthoryear{Molino, Bridson, Teran, and Fedkiw}{Molino
  et~al\mbox{.}}{2003b}]%
        {molino:2003:mesh}
\bibfield{author}{\bibinfo{person}{N. Molino}, \bibinfo{person}{R. Bridson},
  \bibinfo{person}{J. Teran}, {and} \bibinfo{person}{R. Fedkiw}.}
  \bibinfo{year}{2003}\natexlab{b}.
\newblock \showarticletitle{A Crystalline, Red Green Strategy for Meshing
  Highly Deformable Objects with Tetrahedra.}. In \bibinfo{booktitle}{\emph{Int
  Mesh Round}}. Citeseer, \bibinfo{pages}{103--114}.
\newblock


\bibitem[\protect\citeauthoryear{Mukherjee}{Mukherjee}{2014}]%
        {mukherjee:2014:self}
\bibfield{author}{\bibinfo{person}{U. Mukherjee}.}
  \bibinfo{year}{2014}\natexlab{}.
\newblock \showarticletitle{Self-overlapping curves: Analysis and
  applications}.
\newblock \bibinfo{journal}{\emph{Computer-Aided Design}}  \bibinfo{volume}{46}
  (\bibinfo{year}{2014}), \bibinfo{pages}{227--232}.
\newblock


\bibitem[\protect\citeauthoryear{Osher and Fedkiw}{Osher and Fedkiw}{2003}]%
        {osher:2003:LSMDIS}
\bibfield{author}{\bibinfo{person}{S. Osher} {and} \bibinfo{person}{R.
  Fedkiw}.} \bibinfo{year}{2003}\natexlab{}.
\newblock \bibinfo{booktitle}{\emph{Level set methods and dynamic implicit
  surfaces}}.
\newblock \bibinfo{publisher}{Springer}, \bibinfo{address}{New York, N.Y.}
\newblock
\showISBNx{0-387-95482-1}


\bibitem[\protect\citeauthoryear{Sacht, Jacobson, Panozzo, Sch\"{u}ller, and
  Sorkine-Hornung}{Sacht et~al\mbox{.}}{2013}]%
        {sacht:2013:consistent}
\bibfield{author}{\bibinfo{person}{L. Sacht}, \bibinfo{person}{A. Jacobson},
  \bibinfo{person}{D. Panozzo}, \bibinfo{person}{C. Sch\"{u}ller}, {and}
  \bibinfo{person}{O. Sorkine-Hornung}.} \bibinfo{year}{2013}\natexlab{}.
\newblock \showarticletitle{Consistent Volumetric Discretizations inside
  Self-Intersecting Surfaces}. In \bibinfo{booktitle}{\emph{Proceedings of the
  Eleventh Eurographics/ACMSIGGRAPH Symposium on Geometry Processing}} (Genova,
  Italy) \emph{(\bibinfo{series}{SGP '13})}. \bibinfo{publisher}{Eurographics
  Association}, \bibinfo{address}{Goslar, DEU}, \bibinfo{pages}{147–156}.
\newblock
\urldef\tempurl%
\url{https://doi.org/10.1111/cgf.12181}
\showDOI{\tempurl}


\bibitem[\protect\citeauthoryear{Sederberg and Parry}{Sederberg and
  Parry}{1986}]%
        {sederberg:1986:free}
\bibfield{author}{\bibinfo{person}{T. Sederberg} {and} \bibinfo{person}{S.
  Parry}.} \bibinfo{year}{1986}\natexlab{}.
\newblock \showarticletitle{Free-form deformation of solid geometric models}.
  In \bibinfo{booktitle}{\emph{Proc. 13th Ann. Conf. Comp. Graph. Interactive
  Techniques}}. \bibinfo{pages}{151--160}.
\newblock


\bibitem[\protect\citeauthoryear{Shen, O'Brien, and Shewchuk}{Shen
  et~al\mbox{.}}{2004}]%
        {shen:2004:interpolating}
\bibfield{author}{\bibinfo{person}{C. Shen}, \bibinfo{person}{J. O'Brien},
  {and} \bibinfo{person}{J. Shewchuk}.} \bibinfo{year}{2004}\natexlab{}.
\newblock \showarticletitle{Interpolating and Approximating Implicit Surfaces
  from Polygon Soup}. In \bibinfo{booktitle}{\emph{ACM SIGGRAPH 2004 Papers}}
  (Los Angeles, California) \emph{(\bibinfo{series}{SIGGRAPH '04})}.
  \bibinfo{publisher}{Association for Computing Machinery},
  \bibinfo{address}{New York, NY, USA}, \bibinfo{pages}{896–904}.
\newblock
\showISBNx{9781450378239}
\urldef\tempurl%
\url{https://doi.org/10.1145/1186562.1015816}
\showDOI{\tempurl}


\bibitem[\protect\citeauthoryear{Shor and {Van Wyk}}{Shor and {Van
  Wyk}}{1992}]%
        {shor:1992:detecting}
\bibfield{author}{\bibinfo{person}{P. Shor} {and} \bibinfo{person}{C. {Van
  Wyk}}.} \bibinfo{year}{1992}\natexlab{}.
\newblock \showarticletitle{Detecting and decomposing self-overlapping curves}.
\newblock \bibinfo{journal}{\emph{Computational Geometry}} \bibinfo{volume}{2},
  \bibinfo{number}{1} (\bibinfo{year}{1992}), \bibinfo{pages}{31--50}.
\newblock
\showISSN{0925-7721}
\urldef\tempurl%
\url{https://doi.org/10.1016/0925-7721(92)90019-O}
\showDOI{\tempurl}


\bibitem[\protect\citeauthoryear{Si}{Si}{2015}]%
        {si:2015:tetgen}
\bibfield{author}{\bibinfo{person}{H. Si}.} \bibinfo{year}{2015}\natexlab{}.
\newblock \showarticletitle{{TetGen}, a {Delaunay}-Based Quality Tetrahedral
  Mesh Generator}.
\newblock \bibinfo{journal}{\emph{ACM Trans. Math. Softw.}}
  \bibinfo{volume}{41}, \bibinfo{number}{2}, Article \bibinfo{articleno}{11}
  (\bibinfo{date}{Feb.} \bibinfo{year}{2015}), \bibinfo{numpages}{36}~pages.
\newblock
\showISSN{0098-3500}
\urldef\tempurl%
\url{https://doi.org/10.1145/2629697}
\showDOI{\tempurl}


\bibitem[\protect\citeauthoryear{Sifakis and Barbic}{Sifakis and
  Barbic}{2012}]%
        {sifakis:2012:course}
\bibfield{author}{\bibinfo{person}{E. Sifakis} {and} \bibinfo{person}{J.
  Barbic}.} \bibinfo{year}{2012}\natexlab{}.
\newblock \showarticletitle{FEM simulation of 3D deformable solids: a
  practitioner's guide to theory, discretization and model reduction}. In
  \bibinfo{booktitle}{\emph{ACM SIGGRAPH 2012 Courses}} (Los Angeles,
  California) \emph{(\bibinfo{series}{SIGGRAPH '12})}.
  \bibinfo{publisher}{ACM}, \bibinfo{address}{New York, NY, USA},
  \bibinfo{pages}{20:1--20:50}.
\newblock
\urldef\tempurl%
\url{https://doi.org/10.1145/2343483.2343501}
\showDOI{\tempurl}


\bibitem[\protect\citeauthoryear{Sifakis, Der, and Fedkiw}{Sifakis
  et~al\mbox{.}}{2007}]%
        {sifakis:2007:arbitrary}
\bibfield{author}{\bibinfo{person}{E. Sifakis}, \bibinfo{person}{K. Der}, {and}
  \bibinfo{person}{R. Fedkiw}.} \bibinfo{year}{2007}\natexlab{}.
\newblock \showarticletitle{Arbitrary cutting of deformable tetrahedralized
  objects}. In \bibinfo{booktitle}{\emph{Proc ACM SIGGRAPH/Eurograph Symp Comp
  Anim}}. \bibinfo{pages}{73--80}.
\newblock


\bibitem[\protect\citeauthoryear{Song and Belytschko}{Song and
  Belytschko}{2009}]%
        {song:2009:cracking}
\bibfield{author}{\bibinfo{person}{J.-H. Song} {and} \bibinfo{person}{T.
  Belytschko}.} \bibinfo{year}{2009}\natexlab{}.
\newblock \showarticletitle{Cracking node method for dynamic fracture with
  finite elements}.
\newblock \bibinfo{journal}{\emph{Int. J. Num. Meth. Engr.}}
  \bibinfo{volume}{77}, \bibinfo{number}{3} (\bibinfo{year}{2009}),
  \bibinfo{pages}{360--385}.
\newblock


\bibitem[\protect\citeauthoryear{Tao, Batty, Fiume, and Levin}{Tao
  et~al\mbox{.}}{2019}]%
        {tao:2019:mandoline}
\bibfield{author}{\bibinfo{person}{M. Tao}, \bibinfo{person}{C. Batty},
  \bibinfo{person}{E. Fiume}, {and} \bibinfo{person}{D. Levin}.}
  \bibinfo{year}{2019}\natexlab{}.
\newblock \showarticletitle{Mandoline: Robust Cut-Cell Generation for Arbitrary
  Triangle Meshes}.
\newblock \bibinfo{journal}{\emph{ACM Trans. Graph.}} \bibinfo{volume}{38},
  \bibinfo{number}{6}, Article \bibinfo{articleno}{179} (\bibinfo{date}{Nov.}
  \bibinfo{year}{2019}), \bibinfo{numpages}{17}~pages.
\newblock
\showISSN{0730-0301}
\urldef\tempurl%
\url{https://doi.org/10.1145/3355089.3356543}
\showDOI{\tempurl}


\bibitem[\protect\citeauthoryear{Teran, Sifakis, Blemker, Ng-Thow-Hing, Lau,
  and Fedkiw}{Teran et~al\mbox{.}}{2005}]%
        {teran:2005:muscle}
\bibfield{author}{\bibinfo{person}{J. Teran}, \bibinfo{person}{E. Sifakis},
  \bibinfo{person}{S. Blemker}, \bibinfo{person}{V. Ng-Thow-Hing},
  \bibinfo{person}{C. Lau}, {and} \bibinfo{person}{R. Fedkiw}.}
  \bibinfo{year}{2005}\natexlab{}.
\newblock \showarticletitle{Creating and simulating skeletal muscle from the
  visible human data set}.
\newblock \bibinfo{journal}{\emph{IEEE Trans Vis Comp Graph}}
  \bibinfo{volume}{11}, \bibinfo{number}{3} (\bibinfo{year}{2005}),
  \bibinfo{pages}{317--328}.
\newblock


\bibitem[\protect\citeauthoryear{{The CGAL Project}}{{The CGAL
  Project}}{2020}]%
        {cgal:2020:cgal}
\bibfield{author}{\bibinfo{person}{{The CGAL Project}}.}
  \bibinfo{year}{2020}\natexlab{}.
\newblock \bibinfo{booktitle}{\emph{{CGAL} User and Reference Manual}
  (\bibinfo{edition}{{5.2}} ed.)}.
\newblock \bibinfo{publisher}{{CGAL Editorial Board}}.
\newblock
\urldef\tempurl%
\url{https://doc.cgal.org/5.2/Manual/packages.html}
\showURL{%
\tempurl}


\bibitem[\protect\citeauthoryear{Titus}{Titus}{1961}]%
        {titus:1961:combinatorial}
\bibfield{author}{\bibinfo{person}{C. Titus}.} \bibinfo{year}{1961}\natexlab{}.
\newblock \showarticletitle{The combinatorial topology of analytic functions of
  the boundary of a disk}.
\newblock \bibinfo{journal}{\emph{Acta Mathematica}} \bibinfo{volume}{106},
  \bibinfo{number}{1-2} (\bibinfo{year}{1961}), \bibinfo{pages}{45--64}.
\newblock


\bibitem[\protect\citeauthoryear{Wang, Ding, Gast, Zhu, Gagniere, Jiang, and
  Teran}{Wang et~al\mbox{.}}{2019}]%
        {wang:2019:fracture}
\bibfield{author}{\bibinfo{person}{S. Wang}, \bibinfo{person}{M. Ding},
  \bibinfo{person}{T. Gast}, \bibinfo{person}{L. Zhu}, \bibinfo{person}{S.
  Gagniere}, \bibinfo{person}{C. Jiang}, {and} \bibinfo{person}{J. Teran}.}
  \bibinfo{year}{2019}\natexlab{}.
\newblock \showarticletitle{Simulation and Visualization of Ductile Fracture
  with the Material Point Method}.
\newblock \bibinfo{journal}{\emph{Proceedings of the ACM on Computer Graphics
  and Interactive Techniques}} \bibinfo{volume}{2}, \bibinfo{number}{2},
  \bibinfo{pages}{18}.
\newblock


\bibitem[\protect\citeauthoryear{Wang, Jiang, Schroeder, and Teran}{Wang
  et~al\mbox{.}}{2014}]%
        {wang:2015:cutting}
\bibfield{author}{\bibinfo{person}{Y. Wang}, \bibinfo{person}{C. Jiang},
  \bibinfo{person}{C. Schroeder}, {and} \bibinfo{person}{J. Teran}.}
  \bibinfo{year}{2014}\natexlab{}.
\newblock \showarticletitle{An adaptive virtual node algorithm with robust mesh
  cutting}. In \bibinfo{booktitle}{\emph{Proc ACM SIGGRAPH/Eurograph Symp Comp
  Anim}}. \bibinfo{publisher}{Eurographics Association},
  \bibinfo{pages}{77--85}.
\newblock


\bibitem[\protect\citeauthoryear{Wu, Westermann, and Dick}{Wu
  et~al\mbox{.}}{2015}]%
        {wu:2015:cutting}
\bibfield{author}{\bibinfo{person}{J. Wu}, \bibinfo{person}{R. Westermann},
  {and} \bibinfo{person}{C. Dick}.} \bibinfo{year}{2015}\natexlab{}.
\newblock \showarticletitle{A survey of physically based simulation of cuts in
  deformable bodies}.
\newblock \bibinfo{journal}{\emph{Comp Graph Forum}} \bibinfo{volume}{34},
  \bibinfo{number}{6} (\bibinfo{year}{2015}), \bibinfo{pages}{161--187}.
\newblock
\urldef\tempurl%
\url{https://doi.org/10.1111/cgf.12528}
\showDOI{\tempurl}


\bibitem[\protect\citeauthoryear{Zhang, Duan, Zhou, Jiang, Wang, Wu, Huang, Du,
  Liu, Zhou, and Shang}{Zhang et~al\mbox{.}}{2018}]%
        {zhang:2018:stable}
\bibfield{author}{\bibinfo{person}{J. Zhang}, \bibinfo{person}{F. Duan},
  \bibinfo{person}{M. Zhou}, \bibinfo{person}{D. Jiang}, \bibinfo{person}{X.
  Wang}, \bibinfo{person}{Z. Wu}, \bibinfo{person}{Y. Huang},
  \bibinfo{person}{G. Du}, \bibinfo{person}{S. Liu}, \bibinfo{person}{P. Zhou},
  {and} \bibinfo{person}{X. Shang}.} \bibinfo{year}{2018}\natexlab{}.
\newblock \showarticletitle{Stable and realistic crack pattern generation using
  a cracking node method}.
\newblock \bibinfo{journal}{\emph{Front. Comp. Sci.}} \bibinfo{volume}{12},
  \bibinfo{number}{4} (\bibinfo{year}{2018}), \bibinfo{pages}{777--797}.
\newblock


\end{thebibliography}
